\documentclass[12pt]{article}
\usepackage{jheppub}
\usepackage{comment}
\usepackage[x11names,dvipsnames]{xcolor}
\usepackage{amsmath}

\usepackage{tikz}
\usetikzlibrary{decorations.pathreplacing}
\usetikzlibrary{automata,topaths}
\usetikzlibrary{arrows.meta}
\usetikzlibrary{decorations.pathmorphing}
\tikzset{snake it/.style={decorate, decoration=snake}}
\usetikzlibrary{arrows.meta,decorations.pathmorphing,patterns}
\usetikzlibrary{patterns}
\usepackage{mathtools,slashed}
\usepackage[normalem]{ulem}

\usepackage{amssymb}

\usepackage{subcaption}

\makeatletter
\renewcommand{\@fnsymbol}[1]{%
  \ifcase#1\or 
  *\or         
  \dagger\or   
  \ddagger\or  
  \S\or        
  \P\or        
  \spadesuit\or
  \heartsuit\or
  \diamondsuit\or
  \clubsuit    
  \else\@ctrerr\fi}
\makeatother

\def\bomega{{\boldsymbol{\omega}}}
\def\P{{\mathcal{P}}}
\def\r{{\mathfrak{r}}}

\newcommand{\eg}{{\it e.g.,}\ }
\newcommand{\ie}{{\it i.e.,}\ }
\newcommand{\mt}[1]{\textrm{\tiny #1}}

\newcommand{\mK}{\mathcal{K}}
\newcommand{\tigma}{\tilde{\alpha}}

\author[1]{Damián A. Galante,}
\affiliation[1]{Department of Mathematics, King’s College London,\\ Strand, London WC2R 2LS, UK}
\author[2]{Robert C. Myers}
\author[2]{and Themistocles Zikopoulos}
\affiliation[2]{Perimeter Institute for Theoretical Physics,\\ 31 Caroline Street North, Waterloo, ON N2L2Y5, Canada}
\emailAdd{damian.galante@kcl.ac.uk}
\emailAdd{rmyers.perimeter@gmail.com}
\emailAdd{tzikopoulos@perimeterinstitute.ca}
\title{Conformal Boundary Conditions and Higher Curvature Gravity}

\abstract{We initiate a systematic study of Einstein-Gauss-Bonnet gravity in the presence of boundaries subject to conformal boundary conditions, in which the conformal class of the boundary metric is kept fixed. In Einstein gravity, the trace of the extrinsic curvature is also fixed at the boundary. Here we generalize this boundary condition with the appropriate higher curvature correction. We study the problem both in Euclidean and Lorentzian signature. In Euclidean signature, we show that, similarly to the Einstein gravity case, the entropy at large temperatures exhibits the behavior of a conformal field theory in one lower dimension. We also show that in the flat space limit, the higher curvature corrections do not contribute to the leading behavior at high temperatures. We conjecture that this result is a universal feature of the flat space limit in the presence of conformal boundaries. We test our conjecture by analyzing charged black holes. In Lorentzian signature, we analyze the dynamics of the boundary Weyl factor in black hole backgrounds at the linearized level.

}

\begin{document}

\maketitle

\newpage

\section{Introduction}

In Einstein gravity, an AdS$_{d+1}$ black brane has a tree-level entropy given by the Bekenstein-Hawking formula
\begin{equation}\label{eq:BB_S_BH}
    \mathcal{S}_{\text{BH}}=\frac{A}{4G_N}=\left(\frac{4\pi }{d}\right)^{d-1}\frac{\ell^{d-1}_{\mt{AdS}}}{4G_N}\,\frac{V_{d-1}}{\beta^{d-1}} \ ,
\end{equation}
with $V_{d-1}$ being the formally infinite volume of the spatial boundary directions, $\ell_{\mt{AdS}}$, the AdS curvature scale and $\beta$, the inverse Hawking temperature. The temperature scaling of the entropy precisely matches that of a thermal conformal field theory in one dimension lower (CFT$_{d}$) and, within the AdS/CFT correspondence, one can interpret the coefficient in eq.~\eqref{eq:BB_S_BH} as a measure of the degrees of freedom of the dual boundary theory, \ie
\begin{equation}    \mathcal{N}_{\text{dof}}\sim\left(\frac{4\pi }{d}\right)^{d-1}\frac{\ell^{d-1}_{\mt{AdS}}}{4G_N} \ .
\end{equation}
This intuition can be made precise in various top-down constructions of AdS/CFT from string theory, the prototypical example being the duality\footnote{At large-$N$ and large 't Hooft coupling $\lambda$.} between a type IIB supergravity AdS black brane and $\mathcal{N}=4$ super Yang-Mills with $SU(N)$ gauge group at finite temperature~\cite{Maldacena:1997re,Gubser:1998bc,Witten:1998qj,Witten:1998zw}.~In this example, one finds $N^2\propto \ell_{\mt{AdS}}^{d-1}/G_N$.

From the bulk point of view, the thermodynamic behavior of black holes can be analyzed with the Euclidean gravitational path integral via the standard Gibbons-Hawking prescription~\cite{Gibbons:1976ue}.~This prescription can also accommodate situations where one is interested in (quasi-local) thermodynamics of finite spacetime regions, by for instance placing the black hole inside a finite boundary as in the classic work of York~\cite{York:1986it}. In that particular case, spherical Dirichlet boundary conditions were imposed at the boundary.

Surprisingly enough, however, Dirichlet boundary conditions, where one fixes the profile of the induced metric along the boundary $\partial\mathcal{M}$, do not generically yield a well-posed (Initial) Boundary Value Problem (IBVP) for general relativity (GR) in (Lorentzian) Euclidean signature~\cite{anderson2008boundary,anderson2010extension,Anninos:2022ujl}. Furthermore, in Euclidean signature, they also fail to be elliptic, meaning that quantum perturbation theory around a classical solution is ill-defined~\cite{Avramidi:1997sh,Witten:2018lgb}.\footnote{Note, however, that certain classes of boundary geometries exhibit better behavior under Dirichlet boundary conditions, provided the Brown–York stress tensor defines a Lorentzian metric on the boundary of the same signature as the bulk~\cite{An:2025gvr}. An analogue statement holds in Euclidean signature \cite{anderson2010extension, Witten:2018lgb}. These include \eg spherical boundaries in de Sitter space \cite{Coleman:2021nor,Silverstein:2024xnr}.} 

An interesting choice of boundary conditions are the conformal boundary conditions (CBC) \cite{York:1972sj}, whereby the conformal structure of the boundary $[h_{mn}]$ along with the trace of the extrinsic curvature $\mathcal{K}$ are held fixed,
\begin{equation}\label{eq:CBCs}
    \text{Conformal boundary conditions}: \ \lbrace [h_{mn}]|_{\partial\mathcal{M}},\, \mathcal{K}|_{\partial\mathcal{M}}\rbrace \ .
\end{equation}
Conformal boundary conditions yield a well-posed and elliptic Euclidean boundary problem~\cite{anderson2008boundary,anderson2010extension}.~The Lorentzian problem is more intricate, but has better existence and uniqueness properties than the Dirichlet case. See, \eg \cite{An:2020nfw,An:2021fcq,An:2025rlw,An:2025gvr,An:2025cbs,Liu:2025xij} for recent developments. 

The thermodynamic properties of black holes in Einstein gravity subject to CBC have been investigated with $d=2$ and 3 in \cite{Anninos:2023epi,Anninos:2024wpy,Galante:2025tnt} and the analysis was then generalized to arbitrary dimension $d$ in \cite{Banihashemi:2024yye,Banihashemi:2025qqi}.~When discussing Euclidean thermodynamics subject to CBC, the canonical thermal ensemble is defined by a conformally invariant temperature $\tilde{\beta}$ and the trace of the extrinsic curvature $\mathcal{K}$ and hence the Euclidean problem gets modified, compared to its Dirichlet counterpart.~One of the main outcomes of \cite{Anninos:2023epi,Anninos:2024wpy,Galante:2025tnt,Banihashemi:2024yye,Banihashemi:2025qqi} was that, in the small-$\tilde{\beta}$ limit, the conformal entropy has a structure that coincides with the expectations for the thermal effective action of a local CFT in one lower dimension -- in contrast with Dirichlet boundary conditions.~More specifically, the leading contribution to the black hole conformal entropy in the high-temperature expansion  was found to be 
\begin{equation}\label{eq:S_conf}
    \mathcal{S}_{\text{conf}}=\frac{\mathcal{N}_{\text{dof}}(\mathcal{K})}{\tilde{\beta}^{d-1}}+\mathcal{O}\left(\frac{1}{\tilde{\beta}^{d-3}}\right)\ ,
\end{equation}
with
\begin{equation}\label{eq:Nodf_intro}
    \mathcal{N}_{\text{dof}}^{(\text{GR})}(\mathcal{K})=\begin{cases}
        \dfrac{\Sigma_{k,d-1}\ell_{\mt{dS}}^{d-1}}{4G_N}\left(\dfrac{4\pi}{d^2}\right)^{d-1}\left(\sqrt{\mathcal{K}^2\ell_{\mt{dS}}^2+d^2}-\mathcal{K}\ell_{\mt{dS}}\right)^{d-1}  &\quad\Lambda>0 \ , \\[2ex]
        \dfrac{\Sigma_{k,d-1}\left(2\pi\right)^{d-1}}{4G_N \mathcal{K}^{d-1}} \  &\quad \Lambda=0 \ ,\\[2ex]
        \dfrac{\Sigma_{k,d-1}\ell_{\mt{AdS}}^{d-1}}{4G_N}\left(\dfrac{4\pi}{d^2}\right)^{d-1}\left(\mathcal{K}\ell_{\mt{AdS}}-\sqrt{\mathcal{K}^2\ell_{\mt{AdS}}^2-d^2}\right)^{d-1} \  &\quad \Lambda<0 \ .
    \end{cases}
\end{equation}
where $\Lambda=\pm \tfrac{d(d-1)}{2\ell_{\mt{(A)dS}}^2}$  and $\Sigma_{k,d-1}$ is the volume of the $(d-1)$-dimensional maximally symmetric black hole horizon. Note that in $d=2$, and for black branes, these formulas are exact. This observation, drawing inspiration from the traditional AdS/CFT examples~\eqref{eq:BB_S_BH}, could possibly provide hints towards a version of holography for finite portions of spacetime, even beyond $\Lambda<0$.~A tree-level proposal for the dual field theoretic description of these timelike boundaries in AdS$_3$, that reproduces eq.~\eqref{eq:Nodf_intro}, was recently proposed in \cite{Allameh:2025gsa}.

Given the above, a well-motivated question is to study the universality of the extensive behavior of the conformal entropy~\eqref{eq:S_conf}, when higher curvature contributions to the gravitational effective action are taken into account.~In this work, we take a step towards addressing this question by considering Einstein-Maxwell-Gauss-Bonnet (EMGB) gravity. Further, we provide arguments regarding the robustness of our conclusions against further higher curvature terms of Lanczos-Lovelock (LL) type~\cite{Lanczos:1938sf,Lovelock:1971yv}.

More specifically, we find that the CBC~\eqref{eq:CBCs} of Einstein gravity have to be generalized to what we will call \textit{extended conformal boundary conditions} (ECBC), whereby along the conformal class of the induced metric on $\partial\mathcal{M}$, a suitable higher curvature generalization of $\mK$ has to be fixed, given by
\begin{equation}
    \mathcal{P}:=\mK +\frac{\alpha}{2(d-1)(d-2)}\,\mathcal{Q} \ ,
\end{equation}
with $\alpha$ being the Gauss-Bonnet (GB) coupling (see eq.~\eqref{eq:full_action}), and $\mathcal{Q}$ a function of the extrinsic curvature $\mathcal{K}_{\mu\nu}$ and the intrinsic curvature $\tilde{R}_{\mu\nu}$, with the overhead tilde denoting curvature tensors evaluated for the boundary metric. The explicit expression for $\mathcal{Q}$ is given in eq.~\eqref{GBbound}.

Studying the thermodynamics of EMGB black holes with spherical maximally symmetric horizons subject to the ECBC, we find that the extensive form of the conformal entropy in the high-temperature limit~\eqref{eq:S_conf} persists, with \eqref{eq:Nodf_intro} generalizing to
\begin{equation}
\mathcal{N}_{\mathrm{dof}}(\mathcal{P}) =
\begin{cases}
\begin{aligned}
&\frac{\Sigma_{k,d-1}\ell_{\mt{dS}}^{d-1}}{4G_N}\left(\frac{4\pi}{d^2}\right)^{d-1}\left(\tilde{\Delta}-\mathcal{P}\ell_{\mt{dS}}\right)^{d-1} \\
    & \quad\times\Bigg[1 - \frac{\alpha}{\ell_\mt{dS}^2}\,\frac{(d-1)\big(d^2+\mathcal{P}\ell_{\mt{dS}}(\mathcal{P}\ell_{\mt{dS}}+2\tilde{\Delta})\big)\left(\tilde{\Delta}-\mathcal{P}\ell_{\mt{dS}}\right)^{2}}{3d^2\,(\mathcal{P}^2\ell_{\mt{dS}}^2+d^2)}
 \Bigg]
\end{aligned}
&\qquad \Lambda>0 \ ,
\\[1.2em]\\
\begin{aligned}
&\frac{\Sigma_{k,d-1}\ell_{\mt{AdS}}^{d-1}}{4G_N}\left(\frac{4\pi}{d^2}\right)^{d-1}\left(\mathcal{P}\ell_{\mt{AdS}}-\Delta\right)^{d-1} \\
    & \quad\times\Bigg[1 + \frac{\alpha}{\ell_\mt{AdS}^2}\,\frac{(d-1)\big(d^2-\mathcal{P}\ell_{\mt{AdS}}(\mathcal{P}\ell_{\mt{AdS}}+2\Delta)\big)\left(\mathcal{P}\ell_{\mt{AdS}}-\Delta\right)^{2}}{3d^2\,(\mathcal{P}^2\ell_{\mt{AdS}}^2-d^2)}
 \Bigg]
\end{aligned}
&\qquad \Lambda<0 \ ,
\end{cases}
\label{eq:house_on_fire}
\end{equation}
where we defined $\tilde{\Delta}\equiv\sqrt{\mathcal{P}^2\ell_{\mt{dS}}^2+d^2}$ and $\Delta\equiv\sqrt{\mathcal{P}^2\ell_{\mt{AdS}}^2-d^2}$. The (A)dS expressions are obtained perturbatively in the coupling $\alpha$ and they are the analytic continuation of each other, under $\ell_{\mt{dS}}\to i \ell_{\mt{AdS}}$.~We will also provide numerical results for these expressions for finite $\alpha$.

Remarkably, it is possible to take the flat space limit, $\P \ell_\mt{AdS} \to \infty$, of the $\Lambda<0$ expression (or similarly for $\Lambda>0$) to obtain that for vanishing cosmological constant this quantity does not receive corrections from the higher curvature terms,
\begin{equation}
\mathcal{N}_{\mathrm{dof}}(\mathcal{P}) =
\dfrac{\Sigma_{k,d-1}(2\pi)^{d-1}}{4 G_N\, \mathcal{P}^{d-1}} \ , \qquad \ \ \Lambda = 0 \ .
\end{equation}
As we demonstrate, this expression is in fact valid non-perturbatively in the Gauss-Bonnet coupling $\alpha$.~We further show that it does not get modified when one considers charged black holes. 

We conjecture that the leading coefficient in the high-temperature expansion of the conformal entropy~\eqref{eq:S_conf} in flat space is universal, and does not receive further contributions in Lanczos-Lovelock gravity. The existence of a dimensionless, universal new constant for semiclassical gravity with vanishing cosmological constant may play an important role for flat space holography.

We also study linearized dynamics of EMGB gravity subject to ECBC in Lorentzian signature for spherical, planar and hyperbolic horizon topology.~We demonstrate that the exponentially growing modes featured in Einstein gravity~\cite{Anninos:2023epi,Anninos:2024wpy,Liu:2024ymn,Anninos:2024xhc, Anninos:2025zgr} persist, and the GB correction does not stabilize them.~Nonetheless, we observe that planar and hyperbolic black hole horizons enjoy better stability properties than the spherical ones, even within Einstein-Maxwell (EM) theory.~In particular, for black branes of sufficiently large charge and for hyperbolic black holes with any value of the charge, all linearized modes are stable, except when the timelike boundary is placed very close to the horizon.

Before concluding the introduction, we would like to point out two important caveats. First, we comment on how the Gauss-Bonnet (GB) interaction in the action~\eqref{eq:full_action}
should be interpreted. When gravity is treated as an effective field theory, higher-curvature
corrections to the Einstein-Hilbert action are organized in powers of the Planck length
$\ell_p$, which acts as the natural ultraviolet cutoff of the low-energy theory. Therefore, the effect of these higher curvature interactions is suppressed at long distances. From this perspective, the GB coupling is expected to
scale as $\alpha = k\, \ell_p^2$, where $k$ is an $\mathcal{O}(1)$ coefficient.

One may alternatively view the action~\eqref{eq:full_action} as arising from the low-energy
limit of a weakly coupled string theory, in which case it is possible that
$\alpha \gg \ell_p^2$. In this regime, ref.~\cite{Camanho:2014apa} showed that tree-level
three-point graviton scattering in Einstein-Gauss-Bonnet gravity alone exhibits acausal behavior.
The proposed resolution is the appearance of an infinite tower of massive higher-spin states
at the scale $\alpha^{1/2}$. The suggested resolution is that EMGB gravity is only consistent when embedded
in a string-theoretic completion, and that one must proceed with caution, as stringy effects may
become important in unexpected ways.

In the following, we present analytic (and some numerical) results for small but finite
$\alpha$, treating EMGB gravity as a useful theoretical laboratory for testing ideas related to
conformal boundary conditions in gravity. This perspective has proven fruitful in the past, \eg see \cite{Myers:2010ru,Myers:2010tj}.

Secondly, even though the formulas presented in this paper will be for arbitrary dimension $d$, we would like to emphasize that it is not, even in principle, clear how these could correspond to the large-$N$ limit of some CFT when the boundary dimension $d\geq7$.~This is because Nahm's classification~\cite{Nahm:1977tg,Minwalla:1997ka,Cordova:2016emh} prohibits the existence of superconformal algebras in $d\geq7$, a necessary ingredient in all the known top down constructions of AdS/CFT from string theory.

The rest of this paper is organized as follows. In section~\ref{sec:gen_setup}, we discuss the conventions we adopt in this paper, as well as the general setup underlying the black hole analysis that follows, including a brief review of the thermal effective action.~In section~\ref{sec:entropies}, we review the conformal thermodynamics of GR and then provide the corresponding analysis in EMGB gravity, with an emphasis on the high-temperature limit of the conformal entropy.~We then turn to the study of linearized dynamics in Lorentzian signature for EMGB in section~\ref{sec:Lornetzian_dynamics}.~We close by offering some concluding remarks and perspectives on future directions in section~\ref{sec:discuss}. Various technical details are deferred to the appendices: appendix~\ref{app:action_variations} contains the variations of the EMGB action subject to various boundary conditions;
appendix \ref{app:no_K_bdy_term}
demonstrates that there is no local and covariant boundary term that produces a well-defined variational principle for the EMGB action subject to the standard CBC, which fixes $\mathcal{K}$;
appendix~\ref{app:on-shell_actions} gives the computation of the on-shell EMGB action; appendix~\ref{app: flat EGB bh}, the details of the conformal entropies of black holes in Minkowski space; and appendix~\ref{app:GB_AdS_bdy}, the derivation of the value of $\mathcal{P}$ at the AdS boundary. Finally, appendix~\ref{ap:entropy_univ} contains a review of the universal derivation of the Wald formula in LL gravity and appendix~\ref{app:pert_GR} describes the behavior of linearized Lorentzian dynamics for planar and hyperbolic horizons in EM theory.

\section{General Framework}\label{sec:gen_setup}

In the following, we would like to evaluate thermodynamic quantities stemming from the Euclidean gravitational path integral in the saddle-point approximation.~We consider static black hole solutions for the theory of Einstein-Maxwell-Gauss-Bonnet (EMGB) gravity, whose Euclidean action on a Riemannian manifold $(\mathcal{M},g)$ with boundary $(\partial\mathcal{M},h)$ is given by
\begin{equation}\label{eq:full_action}
    \begin{aligned}
        \mathcal{I}_{\text{EMGB}}=&-\frac{1}{16\pi G_N}\int_{\mathcal{M}} d^{d+1}x \,\sqrt{g}\,\bigg(-2\Lambda+R -\frac{1}{4}F^{\mu\nu}F_{\mu\nu}\\
        &\qquad\qquad\qquad\qquad +
        \frac{\alpha}{(d-2)(d-3)}\left(R^{\mu\nu\rho\sigma}R_{\mu\nu\rho\sigma}-4R^{\mu\nu}R_{\mu\nu}+R^2\right)\bigg)\\
        &\qquad\  -\frac{1}{8\pi G_N}\int_{\partial\mathcal{M}}d^{d}y\,\sqrt{h}\,\left(a_d\,\mathcal{K}+\frac{\alpha\ b_d}{(d-2)(d-3)}\mathcal{Q} +\frac{1}{2}F^{\mu\nu}n_\mu A_\nu\right) \ .
    \end{aligned}
\end{equation}
In the first line, we have the usual Einstein-Maxwell action with a cosmological constant.~The next line is the Gauss-Bonnet (GB) term with a coupling $\alpha$, which has dimensions of length$^2$.~In four dimensions (\ie $d=3$),  the GB term is topological and so we work in $d\ge4$ in order to have non-trivial (thermo)dynamics.\footnote{One can have nontrivial thermodynamics with four bulk dimensions, since the topological GB term will contribute an additive constant to the free energy or the entropy in this case, see \eg \cite{Bunch1981, Liko:2007vi, Sarkar:2010xp}.}~The last line contains various boundary terms.~In particular, the third term involving the Maxwell field ensures that we are working in the canonical ensemble, where the charge of the black hole is fixed -- see appendix \ref{MaxApp} for details. 

Of course, the gravitational boundary terms and conditions are the focus of our paper.~In the third line of eq.~\eqref{eq:full_action}, the first term (\ie the trace of the extrinsic curvature $\mathcal{K}_{\mu\nu}$) is associated with the Einstein term, while
\begin{equation}
\mathcal{Q} = 4\left(\mathcal{K}\mathcal{K}^{\mu\nu}\mathcal{K}_{\mu\nu}-\tfrac{2}{3}\mathcal{K}^{\mu\nu}\mathcal{K}_{\nu\rho}\mathcal{K}^{\rho}_{\mu}-\tfrac{1}{3}\mathcal{K}^3-2\tilde{G}^{\mu\nu}\mathcal{K}_{\mu\nu}\right)\,,
  \label{GBbound}  
\end{equation}
is associated with the GB term.~Note that above, $\tilde{G}^{\mu\nu} \equiv \tilde{R}^{\mu\nu}-\tfrac{1}{2}h^{\mu\nu}\tilde{R}$ is the Einstein tensor evaluated for the induced metric $h_{\mu\nu}$ on the boundary, and $n_\mu$ is the outward-pointing normal vector to $\partial\mathcal{M}$.~The dimensionless coefficients $a_d, b_d$ are  fixed by requiring the variational principle is well-defined for various boundary conditions -- see appendix \ref{app:action_variations} for details.

For instance, in the case of the Dirichlet problem, in which we fix the full boundary metric $h_{\mu\nu}$, the coefficients are 
\begin{equation}
    \text{Dirichlet boundary conditions:}  \, (a_d = 1,\ b_d = \tfrac{1}{2}) \,.
\end{equation}
In this case, the $\mathcal{K}$ boundary term is the Gibbons-Hawking-York (GHY) term~\cite{York:1972sj,Gibbons:1976ue} and the $\mathcal{Q}$ term is the standard boundary term derived for Gauss-Bonnet gravity \cite{Bunch1981,  Myers:1987yn}. Other types of boundary conditions, such as Neumann or Umbilic \cite{Fournodavlos:2021eye} can also be generalized for GB gravity, as explained in appendix \ref{app:action_variations}.

Our primary interest is conformal boundary conditions, where we only fix the conformal class of the boundary metric. One can ask what is the natural Gauss-Bonnet generalization of the conformal boundary conditions \eqref{eq:CBCs} applied in Einstein gravity. One finds that it is impossible to obtain a local and covariant boundary action that fixes the conformal class and $\mathcal{K}$ as in eq.~\eqref{eq:CBCs} -- see appendix \ref{app:no_K_bdy_term}. Instead, we define \textit{extended} conformal boundary conditions (ECBC), that fix 
\begin{equation}\label{eq:ECBCs}
    \text{Extended conformal boundary conditions}: \ \left\lbrace [h_{mn}]|_{\partial\mathcal{M}},\,\mathcal{P}|_{\partial\mathcal{M}}\right\rbrace \ ,
\end{equation}
where 
\begin{equation}
\mathcal{P}:=\mathcal{K}+\,\frac{\alpha}{2(d-1)(d-2)}\,\mathcal{Q}\,.    \label{eq:paste}
\end{equation}
Of course, these extended conformal boundary conditions reduce to the CBC \eqref{eq:CBCs} of Einstein gravity when $\alpha=0$.  As described in appendix \ref{app:action_variations}, for the ECBC, we find
\begin{equation}
    \text{Extended conformal boundary conditions:}  \, \left(a_d = \frac{1}{d}, b_d =  \frac{3}{2d} \right) \,.
\end{equation}

\subsection{Regular Euclidean solutions} 
We focus our analysis on Euclidean black hole solutions of the form
\begin{equation}\label{eq:line_element}
    ds^2=f(r)d\tau^2+\frac{dr^2}{f(r)}+r^2d\Sigma_{k,d-1}^2 \ ,
\end{equation}
where
\begin{equation}
    d\Sigma_{k,d-1}^2=\begin{cases}
        \ d\Omega_{d-1}^2 \ , \qquad  k=1 \ ,\\
        \ \frac{1}{\ell^2}\,d\vec{x}^{\,2}_{d-2} \ , \quad \  k=0 \ ,\\
        \ dH^2_{d-1} \ , \qquad k = -1 \ .
    \end{cases}\label{eq:kmetric}
\end{equation}
That is, with $k=\pm1$, the above line element corresponds to the metric of a ($d$--1)-sphere or ($d$--1)-dimensional hyperbolic plane with unit curvature radius. With $k=0$, we have a flat metric and we have introduced a length scale $\ell$ because the $x$-coordinates are dimensionful.\footnote{In the case of AdS, this $\ell$ coincides with the one appearing in the cosmological constant term of the action~\eqref{eq:full_action}.} The metric~\eqref{eq:line_element} is a solution of the EMGB field equations provided \cite{Boulware:1985wk,Padmanabhan:2013xyr,Cai:2001dz,WILTSHIRE198636}
\begin{equation}\label{eq:f(r)_EMGB}
    f(r)=k+\frac{r^2}{2{\alpha}}\left(1-\sqrt{1+4{\alpha}\left(\frac{\mu}{r^{d}}-\frac{q^2}{r^{2(d-1)}}+\frac{2\Lambda}{d(d-1)}\right)}\right) \,.
\end{equation}
In the following, we will mostly focus on $\Lambda<0$, in which case we replace $\tfrac{2\Lambda}{d(d-1)}=-\tfrac1{\ell^2}$ in the above expression.~Of course, for positive $\Lambda$, we can replace
$\tfrac{2\Lambda}{d(d-1)}=\tfrac1{\ell^2}$.~In what follows, it will also be useful to introduce $r_h$ as the horizon radius, \ie the outermost solution of $f(r=r_h)=0$.\footnote{Note that when the blackening factor admits multiple horizons, these formulas will also be valid for inner patches (up to a sign in the normal unit vector).}

The full solution also includes the Maxwell $U(1)$ connection\footnote{Recall that upon analytically continuing the charged black hole to Euclidean signature, the gauge field acquires an imaginary time component, \eg see \cite{Gibbons:1976ue}.}
\begin{equation}
    A_\mu dx^\mu=i\sqrt{\frac{d-1}{2(d-2)}}\left(\frac{q}{r_h^{d-2}}-\frac{q}{r^{d-2}}\right)d\tau \,.
\end{equation}
Here we have chosen a gauge where $A_\tau(r=r_h)=0$, which ensures that the connection remains a well-defined one-form at the horizon.
In the absence of finite boundaries (and working in Lorentzian signature), the integration constants $\mu \ \text{and} \ q$ in eq.~\eqref{eq:f(r)_EMGB} are related to the ADM mass $M$ and charge $Q$ of the black hole via
\begin{equation}
    M=\frac{(d-1)\Sigma_{k,d-1}}{16\pi \,G_N}\,\mu\,,\ \quad  \quad \ Q=\sqrt{2(d-1)(d-2)}\,\frac{\Sigma_{k,d-1}}{8\pi\, G_N}\,q \ ,
\end{equation}
where $\Sigma_{k,d-1}$ is the volume of maximally symmetric horizon~\eqref{eq:kmetric}. In the presence of a finite timelike boundary these will be instead related to quasi-local charges defined at the finite boundary \cite{Odak:2021axr,Anninos:2024xhc}.

In eq.~\eqref{eq:f(r)_EMGB}, $f(r)$ in fact admits two branches, corresponding to choosing either the plus or minus sign in front of the square root. However, perturbing around the plus branch, one finds that the gravitons are ghosts \cite{Boulware:1985wk,Myers:2010ru}. Further, it is only the minus branch that has a smooth limit to Einstein gravity, when $\alpha \to 0$. In this limit, eq.~\eqref{eq:f(r)_EMGB} reduces to the familiar solution
\begin{equation}\label{eq:f(r)_GR}
    f(r)=k-\frac{\mu}{r^{d-2}}+\frac{q^2}{r^{2(d-2)}}-\frac{2\Lambda}{d(d-1)}\,.
\end{equation}
Therefore, in the following, we will only consider the solution with the minus branch given in eq.~\eqref{eq:f(r)_EMGB}.

Before proceeding, let us comment on the GB coupling $\alpha$.~In low-energy effective actions coming from string theory, the GB coupling is always positive definite~\cite{Boulware:1985wk}.~Note also that if ${\alpha}<0$, the metric~\eqref{eq:f(r)_EMGB} has branch cut singularities, but these are always shielded by event horizons (except for small spherical black holes, where the branch cut is at the horizon)~\cite{Wiltshire:1988uq}.~Further, with negative $\Lambda$, proper AdS vacua only appear when $4\alpha\leq\ell^2$, since for larger values the square root in the blackening factor~\eqref{eq:f(r)_EMGB} becomes imaginary. Holographic studies of EGB gravity~\cite{Camanho:2009vw,Buchel:2009sk}  yielded further constraints on ${\alpha}$, by requiring consistency with the unitarity, causality and positivity of various energy fluxes in the dual CFT. These constraints, take the form~\cite{Buchel:2009sk}
\begin{equation}\label{eq:GB_coupling_boundD}
    -\frac{(3d+2)(d-2)}{4(d+2)^2}\leq  \frac{{\alpha}}{\ell^2} \leq \frac{(d-2)(d-3)(d^2-d+6)}{4(d^2-3d+6)^2} \ .
\end{equation}
However, as noted in the introduction, the consistency of high-energy scattering processes indicates that EMGB theory is incomplete for any non-vanishing ${\alpha}$ \cite{Camanho:2014apa}.~With these caveats in mind, in the following we focus on perturbative expansions for small ${\alpha}$ but also take advantage of the exact solution \eqref{eq:f(r)_EMGB} to examine results for small but finite values of the coupling.

\subsection{On-shell action and conformal thermodynamics}

To examine conformal spacetime thermodynamics in this new setting, we are interested in boundaries that are conformal to $\mathbb{S}^1 \times \Sigma_{k,d-1}$,
\begin{equation}
\label{eq: conformal class}
    ds^2|_{\partial\mathcal{M}}=e^{2\bomega}\left(du^2+ \mathfrak{r}^2\, d\Sigma_{k,d-1}^2\right) \, ,
\end{equation}
where, in principle, the conformal factor $\bomega$ can depend on the boundary coordinates and is not fixed by boundary data.

Given a bulk solution $(\mathcal{M},g)$ with line element~\eqref{eq:line_element}, such boundaries correspond to simply placing $\partial\mathcal{M}$ at some $r=e^{\bomega} \mathfrak{r}$, so that the induced metric $h$ takes the form
\begin{equation}
    ds^2|_{\partial\mathcal{M}}=f(e^\bomega \mathfrak{r})\,d\tau^2+(e^\bomega \mathfrak{r})^2\,d\Sigma_{k,d-1}^2 \ . 
\end{equation}
This can be immediately recognized to be in the desired form \eqref{eq: conformal class}, after identifying $d\tau\rightarrow e^\bomega du/\sqrt{f(e^{\bomega} \mathfrak{r})}$.

As usual, the Euclidean time is periodically identified as $\tau \sim \tau + \beta$.~Following the classic work of York~\cite{York:1986it}, the canonical ensemble is defined by fixing the proper boundary temperature $\beta_{\partial\mathcal{M}}$ along with any other charges. 
For (extended) conformal boundary conditions, $\beta_{\partial\mathcal{M}}$ can in principle fluctuate, so instead we define the (dimensionless) inverse temperature
\begin{equation}
    \tilde{\beta} \equiv \frac{\beta}{\mathfrak{r}}\,,
\end{equation}
which is conformally invariant, \ie $\tilde{\beta}$ is fixed when the conformal class of $h$ is fixed. Then, the conformal canonical ensemble is defined by fixing $\tilde{\beta}$ along with $\mathcal{P}$ in eq.~\eqref{eq:paste} and any other charges (the electric charge $Q$, in our case) along $\partial\mathcal{M}$. 

In other words, the gravitational partition function in the saddle point approximation, with the saddle obeying ECBC, is given by
\begin{equation}
    \mathcal{Z}_{\text{grav}}[\tilde{\beta},\mathcal{P},Q]
    \simeq e^{-\mathcal{\hat{I}}_{\text{EMGB}}} \,,
\end{equation} 
where the overhat indicates that this is the on-shell action of the solution in question. If there is more than one saddle satisfying the boundary conditions, then we are instructed to sum over saddles.

Given $\mathcal{Z}_{\text{grav}}[\tilde{\beta},\mathcal{P},Q]$, we follow the Gibbons-Hawking prescription~\cite{Gibbons:1976ue} to compute the entropy, energy and specific heat of this conformal canonical ensemble as
    \begin{align} 
        \mathcal{S}_{\text{conf}}&=\left(1-\tilde{\beta}\partial_{\tilde{\beta}}\right)\log{\mathcal{Z}_{\text{grav}}[\tilde{\beta},\mathcal{P},Q]} \ , \nonumber \\
        \mathcal{E}_{\text{conf}}&=-\partial_{\tilde{\beta}}\log{\mathcal{Z}_{\text{grav}}[\tilde{\beta},\mathcal{P},Q]} \ , \label{eq:thermod_relations} \\
\mathcal{C}_{\mathcal{P}}&=\tilde{\beta}^2\partial^2_{\tilde{\beta}}\log{\mathcal{Z}_{\text{grav}}[\tilde{\beta},\mathcal{P},Q]} \ , \nonumber
    \end{align}
with $\mathcal{P}$ and $Q$ kept fixed in the various derivatives. For the class of solutions~\eqref{eq:line_element}, the boundary data can be written in terms of the bulk parameters as 
\begin{align}
        \tilde{\beta}&=\frac{4\pi \sqrt{f(e^\bomega \mathfrak{r})}}{e^\bomega\mathfrak{r}|f'(r_h)|} \ , \nonumber\\
        \mathcal{P}&=\frac{1}{2e^\bomega \mathfrak{r}\sqrt{f(e^\bomega\mathfrak{r}})}\big(e^\bomega\mathfrak{r}f'(e^\bomega\mathfrak{r})+2(d-1)f(e^\bomega\mathfrak{r})\big)
        \label{eq:invert_devil_b,K}\\
        &\qquad+{\alpha}\,\frac{3e^\bomega\mathfrak{r}\big(k-f(e^\bomega\mathfrak{r})\big)f'(e^\bomega\mathfrak{r})+2(d-3)\big(3k-f(e^\bomega\mathfrak{r})\big)f(e^\bomega\mathfrak{r})}{3 e^{3\bomega}\mathfrak{r}^3\sqrt{f(e^\bomega\mathfrak{r})}} \ .
        \nonumber
\end{align}
As mentioned above, $\bomega$ is a possibly fluctuating conformal factor. In our Euclidean analysis in section~\ref{sec:entropies}, $\bomega$ will be constant, while we will allow for a time-dependent conformal factor when analyzing Lorentzian dynamics in section~\ref{sec:Lornetzian_dynamics}.

\paragraph{Thermal effective action}
Before actually computing thermodynamic quantities in our setup, let us briefly comment on some expectations coming from thermal effective field theory. As emphasized in \cite{Banihashemi:2024yye,Banihashemi:2025qqi}, one can organize the high-temperature limit of the conformal thermodynamics (for GR) in terms of an effective action of a thermal CFT in one dimension lower. 

Take a $d$-dimensional thermal CFT living on the manifold $\partial\mathcal{M}\cong \mathbb{S}_{\tilde{\beta}}^1\times X$, whose metric can be written as
\begin{equation}
    ds^2=e^{2\bomega}\left(d\tau^2+\gamma_{ij}dx^idx^j\right) \ .
\end{equation}

Assuming that the theory is trivially gapped at finite temperature,
in the $\tilde{\beta}\rightarrow0$ limit, one can dimensionally reduce over the thermal circle, yielding an effective action of the form \cite{Banerjee:2012iz,Benjamin:2023qsc}\footnote{Note that $c_0$ is proportional to the thermal one-point function of the stress tensor $\langle T^{\mu\nu}(t,\vec{x})\rangle_{\tilde{\beta}}$ on $\mathbb{R}^{d-1}$, or alternatively to the Casimir energy on $\mathbb{S}^1\times\mathbb{R}^{d-2}$ \cite{Benjamin:2023qsc}. Similarly, $c_1$ is related to the subleading contribution of the Casimir energy when the latter is compactified to $\mathbb{S}^1\times\mathbb{S}^{d-2}$ \cite{Allameh:2024qqp}.}
\begin{equation}\label{eq:I_EFT}
\mathcal{I}_{\text{EFT}}^{\tilde{\beta}}=\int_X  d^{d-1}x\, \sqrt{\gamma} \left(-\frac{c_0}{\tilde{\beta}^{d-1}}+\frac{c_1}{\tilde{\beta}^{d-3}}R_\gamma+\cdots\right) \ ,
\end{equation}
where $R_\gamma$ is the Ricci scalar of the `spatial' metric $\gamma$, \eg see section $1.3$ of ref.~\cite{Banihashemi:2024yye}. From the thermal effective action, we can compute the thermal entropy as usual,
\begin{equation}\label{eq:S_EFT}  \mathcal{S}_{\text{EFT}}=\left(\tilde{\beta}\partial_{\tilde{\beta}}-1\right)\mathcal{I}_{\text{EFT}}^{\tilde{\beta}}=\Sigma_{k,d-1}\left(d\,\frac{c_0}{\tilde{\beta}^{d-1}}-\frac{k\,(d-1)(d-2)^2}{\mathfrak{r}^2}\frac{c_1}{\tilde{\beta}^{d-3}}+\dots\right) \ ,
\end{equation}
where we have introduced $\Sigma_{k,d-1}$ as the volume and
substituted $R_\gamma=\frac{k\,(d-1)(d-2)}{\mathfrak{r}^2}$ for the constant curvature metrics of curvature radius $\mathfrak{r}$ in eq.~\eqref{eq:kmetric}.~Moreover, it was conjectured in~\cite{Allameh:2024qqp} that the subleading coefficient $c_1>0$. This conjecture is supported by examining free field theories and thermal CFTs with an Einstein gravity dual. Nevertheless, it was observed in~\cite{Banihashemi:2024yye,Banihashemi:2025qqi} that thermodynamics of CBC consistently violate this bound. In turn, this violation was interpreted as a sign that gravity does not completely decouple in this holographic setup. Instead, it was argued that the boundary conformal mode fluctuates and must be integrated over in a corresponding CFT path integral.

\section{Euclidean Conformal Thermodynamics}\label{sec:entropies}

We now turn to discuss Euclidean thermodynamics in the conformal ensemble. The objective is to compute the partition function $\mathcal{Z}_{\text{grav}}[\tilde{\beta},\mathcal{P},Q]$, that by definition, is a function of the boundary data $\lbrace\tilde\beta, \mathcal{P}, Q\rbrace$.

The problem is that the bulk metric is usually expressed in other variables (like $\mu$, $q$, $r_h$), so one needs to write the bulk parameters in terms of boundary data to actually evaluate the partition function in the conformal canonical ensemble. A first step to do this, is to conveniently write the blackening factor \eqref{eq:f(r)_EMGB} in terms of the radius of the horizon $r_h$ as 
\begin{equation}
    f(r)=k+\frac{r^2}{2{\alpha}}\left(1-\sqrt{1+4{\alpha}\left(\frac{ky^{d}}{r_h^2}\left(1+\frac{k{\alpha}}{r_h^2}\right)-\frac{q^2}{r^{2(d-1)}}\left(1-y^{2-d}\right)-\frac{1}{\ell^2}\left(1-y^{d}\right)\right)}\right) \ ,
\end{equation}
where $y \equiv r_h/r$. For the Einstein case, this reduces to 
\begin{equation}
    f(r)=k\left(1-y^{d-2}\right)+\frac{q^2}{r^{2(d-2)}}\left(1-y^{2-d}\right)+\frac{r^2}{\ell^2}\left(1-y^{d}\right)  \ .
\end{equation}
Note that at the boundary $y = r_h/e^\bomega \mathfrak{r}$. In principle, the relations defining the (E)CBC \eqref{eq:invert_devil_b,K} could be inverted to express $r_h$ and $e^\bomega\mathfrak{r}$ as functions of the boundary data $\lbrace\tilde{\beta},\mathcal{P},Q\rbrace$. However this becomes analytically intractable even in pure Einstein gravity (with the exception of the $d=2$ case, that can be solved exactly \cite{Anninos:2024wpy}).

One alternative is to proceed by solving the full problem numerically. This was done, for instance, in \cite{Anninos:2024wpy} for the positive cosmological constant case in $d=3$, and in \cite{Banihashemi:2025qqi} for charged black holes in $d=3$. Here we will be mostly interested in the high-temperature behavior of the partition function, so, as we will show, we only need to invert the expressions perturbatively in this limit. 

Recall that the conformal temperature is given by $\tilde \beta = \beta_{\partial\mathcal{M}}/e^\bomega \mathfrak{r}$, where $\beta_{\partial\mathcal{M}}$ is the proper inverse temperature at the timelike boundary $\partial\mathcal{M}$. Then, taking the high-temperature limit (\ie $\tilde\beta \to 0$) implies that one must take both the horizon radius $r_h$ and the size of the boundary $e^\bomega\mathfrak{r}$ to infinity, while keeping their ratio $y$ fixed as determined by the requirement that $\mathcal{P}$ is fixed along the boundary. Thus, from the bulk perspective, the high-temperature limit corresponds to a large-volume limit.~Note this limit is not accessible with Dirichlet boundary conditions, since in that case, the size of the boundary is kept fixed.

In section \ref{subsec:CBCs_GR}, we review how to carefully take this limit in the Einstein case, before proceeding to EMGB in section \ref{sec: emgb thermo}.

\subsection{Conformal Thermodynamics in Einstein Gravity}\label{subsec:CBCs_GR}

In this subsection, we briefly review Euclidean conformal thermodynamics in Einstein gravity with an emphasis on the conformal entropy, to set the stage for our subsequent EMGB analysis.~This problem has been analyzed previously in \cite{Anninos:2023epi,Anninos:2024wpy,Banihashemi:2024yye,Banihashemi:2025qqi,Liu:2024ymn}.

Consider pure Einstein gravity with negative cosmological constant in $d+1$ dimensions.\footnote{Note that in Einstein gravity, in contrast to more general Lovelock theories, the length scale $\ell$ appearing in the action~\eqref{eq:full_action} consides with the (A)dS curvature radius $\ell_{\mt{\text{(A)dS}}}$ and thus we do not distinguish between the two in this subsection.}~If we only consider stationary solutions, by Birkhoff’s theorem we know that the line element~\eqref{eq:f(r)_GR} with $q=0$ covers the full solution space of the problem, up to solutions that differ in their boundary Weyl factor.\footnote{\label{foot:Weyl_factor_solns}This is because the bulk Killing vector fields act as conformal Killing vectors at the boundary.~Hence, Birkhoff's theorem only guarantees the uniqueness for the solution in the bulk, up to possibly a finite number of isolated solutions with non-trivial Weyl factor.~We will return to this point in Section~\ref{sec:Lornetzian_dynamics}, when we discuss dynamics in Lorentzian signature.} For the spherical black hole case (\ie $k=+1$),\footnote{The results readily generalize to all maximally symmetric horizon topologies, as we discuss in the EMGB analysis below.} the CBC~\eqref{eq:CBCs} for the line element~\eqref{eq:f(r)_GR} read
\begin{align}\label{eq:GR_devil_relations}
        \tilde{\beta}&= \frac{4\pi y}{(d-2)+d\frac{r_h^2}{\ell^2}}\sqrt{1- y^{d-3}+\frac{r_h^2}{\ell^2} y(1-y^{d-1})} \ ,\\
        \mathcal{K}\ell &= \dfrac{\ell
        }{2\mathfrak{r}\sqrt{f(e^\omega \mathfrak{r})}}\left(2(d-1)+2d\left(\frac{e^\omega \mathfrak{r}}{\ell}\right)^2-d\left(1+\frac{r_h^2}{\ell^2}\right)y^{d-2}\right)\ .
        \nonumber
        \end{align}
One can compute the on-shell action for these solutions and evaluate the various thermodynamic quantities using eq.~\eqref{eq:thermod_relations}.~Instead, what we observe is that irrespective of the boundary conditions, the entropy is always given by the Bekenstein-Hawking area formula,
\begin{equation}\label{eq:BH_SGR} 
    \mathcal{S}_{\text{conf}}^{(\text{GR})}=\frac{\Omega_{d-1}r_h^{d-1}}{4G_N} \ .
\end{equation}
One way of explaining this fact is that black hole entropy can be computed using Wald's formula~\cite{Wald:1993nt} (or the generalizations in \cite{PhysRevLett.70.3684,Jacobson:1993vj,Jacobson:1994qe,Dong:2013qoa,Camps:2013zua}). Alternatively, it is possible to derive this universal formula only assuming the existence of a local, covariant boundary term \cite{Carlip:1993sa,Banados:1993qp}. We review this derivation in appendix~\ref{ap:entropy_univ}. We will later generalize these arguments for general Lovelock theories.

For now, we write \eqref{eq:BH_SGR} in terms of the boundary data $\lbrace\tilde{\beta},\mathcal{K}\rbrace$. In the high-temperature limit, the relations~\eqref{eq:GR_devil_relations} become 
\begin{align}\label{eq:GR_AdS_data_invert}
\tilde{\beta}&= \dfrac{4\pi\ell}{dr_h}u+\mathcal{O}(r_h^{-3}) \ , \\
    \mathcal{K}\ell&=\dfrac{d}{2u}\left(1+u^2\right)+\mathcal{O}(r_h^{-2}) \ ,\nonumber
\end{align}
where we defined $u\equiv\sqrt{1-y^{d}}$. Recall the goal is to express $r_h$ and $u$ in terms of boundary data. Note that eq.~\eqref{eq:GR_AdS_data_invert} simply implies that,
\begin{align}
     &r_h= \dfrac{4\pi\ell}{d\tilde{\beta}}u +\mathcal{O}(\tilde{\beta}) \ ,\\
     &u^2-\dfrac{2\mathcal{K}\ell}{d}u+1=0 \ .\nonumber
\end{align}
Solving the quadratic equation in the second line, we find $u$ as a function of $K\ell$,\footnote{We discard the solution with the plus sign as it leads to $u>1$, which is prohibited by $u=\sqrt{1-y^{d}}$ when $y\in[0,1]$.}
\begin{equation}\label{eq:sln_u0}
    u=\frac{\mathcal{K}\ell-\sqrt{\mathcal{K}^2\ell^2-d^2}}{d} \ ,
\end{equation}
which in turn implies that
\begin{equation}
    r_h=\frac{4\pi\ell}{d\tilde{\beta}}\,\frac{\mathcal{K}\ell-\sqrt{\mathcal{K}^2\ell^2-d^2}}{d} +\mathcal{O}(\tilde{\beta}) \ .
\end{equation}
Therefore, the conformal entropy~\eqref{eq:BH_SGR} reads
\begin{equation}
    \mathcal{S}_{\text{conf}}^{(\text{GR})}=\frac{\Omega_{d-1}\ell^{d-1}}{4G_N}\left(\frac{4\pi}{d^2}\right)^{d-1}\left(\mathcal{K}\ell-\sqrt{\mathcal{K}^2\ell^2-d^2}\right)^{d-1}\frac{1}{\tilde{\beta}^{d-1}} +\mathcal{O}\left(\frac{1}{\tilde{\beta}^{d-3}}\right)\ ,
    \label{eq:house2}
\end{equation}
where $\Omega_{d-1}=\tfrac{2\pi^{d/2}}{\Gamma({d}/{2})}$ is the area of a unit $(d{-}1)$-sphere. Here, we see that this conformal entropy agrees with the structure of \eqref{eq:S_EFT}, which corresponds to the thermal entropy of a $d$-dimensional CFT~.~Subleading corrections of this high-temperature asymptotic expansion can also be matched  \cite{Banihashemi:2024yye,Banihashemi:2025qqi}. The results for positive and vanishing cosmological constant can be easily obtained from this one. For the flat space limit, we just take the dimensionless combination $\mathcal{K} \ell \to \infty$, while to obtain the positive cosmological constant answer, we analytically continue $\ell\rightarrow i\ell$.\footnote{\label{ft:analytic_cont}Throughout, we choose the principal branch in defining $\mathcal{N}_\text{dof}$ as an analytic function of $\ell$.}
In summary, in all cases we obtain
\begin{equation} \label{eq: conf entropy GR}
\mathcal{S}_{\text{conf}}^{(\text{GR})}=\frac{\mathcal{N}_{\text{dof}}^{(\text{GR})}(\mathcal{K})}{\tilde{\beta}^{d-1}}+\mathcal{O}\left(\frac{1}{\tilde{\beta}^{d-3}}\right)\ ,
\end{equation}
where
\begin{equation}\label{eq:N_dof(K)_GR_cases}
    \mathcal{N}_{\text{dof}}^{(\text{GR})}(\mathcal{K})=\begin{cases}
        \dfrac{\Sigma_{k,d-1}\ell^{d-1}}{4G_N}\left(\dfrac{4\pi}{d^2}\right)^{d-1}\left(\sqrt{\mathcal{K}^2\ell^2+d^2}-\mathcal{K}\ell\right)^{d-1}  &\quad\Lambda>0 \ , \\[2ex]
        \dfrac{\Sigma_{k,d-1}\left(2\pi\right)^{d-1}}{4G_N \mathcal{K}^{d-1}} \  &\quad \Lambda=0 \ ,\\[2ex]
        \dfrac{\Sigma_{k,d-1}\ell^{d-1}}{4G_N}\left(\dfrac{4\pi}{d^2}\right)^{d-1}\left(\mathcal{K}\ell-\sqrt{\mathcal{K}^2\ell^2-d^2}\right)^{d-1} \  &\quad \Lambda<0 \ .
    \end{cases}
\end{equation}

This is one of the main results of the Euclidean analysis of conformal boundary conditions in Einstein gravity.~Moreover, in the case of vanishing cosmological constant, the trace of the extrinsic curvature provides an extra parameter from which a dimensionless $\mathcal{N}_{\text{dof}}$ can be obtained. This is usually absent when studying gravity with vanishing cosmological constant.~Let us also stress that for $d=2$, the result \eqref{eq: conf entropy GR} is exact and the conformal entropy does not receive any $\tilde \beta$ corrections.

An interesting observation is that $\mathcal{N}^{(\text{GR})}_{\text{dof}}(\mathcal{K})$ is a monotonic function.~For negative cosmological constant, it decreases from the conformal boundary of AdS, where $\mK\ell = d$, all the way to the black hole horizon, where $\mK\ell \to \infty$. Of course, when $\mK\ell = d$, we recover the standard AdS/CFT result. For instance, when $d=2$, we obtain that $\mathcal{N}_{\text{dof}}(\mathcal{K})$ at the boundary is the Brown-Henneaux central charge. This monotonic behavior is somewhat reminiscent of the holographic $c$-functions~\cite{Freedman:1999gp,Myers:2010tj,Myers:2012ed}. It would be interesting to have a microscopic interpretation of this number of degrees of freedom\footnote{A specific proposal for $d=2$ is given in \cite{Allameh:2025gsa}.} -- see section~\ref{sec:discuss} for further discussion.

With positive cosmological constant, $\mathcal{N}^{(\text{GR})}_{\text{dof}}(\mathcal{K})$ is again monotonic from the observer's worldline in the center of the static patch ($K\ell \to - \infty$) to the stretched cosmological horizon ($K\ell \to \infty$).

Before considering the case of Gauss-Bonnet gravity, let us briefly review the thermodynamics of the canonical ensemble with Dirichlet boundary conditions.

For simplicity, we restrict to the case of flat space, though the discussion readily generalizes to all values of $\Lambda$. In the Dirichlet problem, we consider a boundary $\mathbb{S}^1\times \mathbb{S}^{d-1}$, where the size of the $\mathbb{S}^{d-1}$ is fixed at $r=\mathfrak{r}$, while the size of the thermal circle is fixed to $\beta_{\partial\mathcal{M}}$. In terms of the bulk data, the latter is given by
\begin{equation}
    \beta_{\partial \mathcal{M}}=\frac{4\pi \sqrt{f(\mathfrak{r})}}{f'(r_h)}  \ .
\end{equation}
For the spherical black hole solution of Einstein gravity~\eqref{eq:f(r)_GR}, it is convenient to introduce $x\equiv(r_h/\mathfrak{r})^{d-2}$, such that in the $\beta_{\partial\mathcal{M}} \to 0$ high-temperature expansion, we obtain,
\begin{equation}
    \beta_{\partial\mathcal{M}}=\frac{4\pi\mathfrak{r}}{d-2}\sqrt{1-x} \Rightarrow x = 1-\frac{(d-2)^2\beta_{\partial\mathcal{M}}^2}{(4\pi \mathfrak{r})^2} +\mathcal{O}(\beta_{\partial\mathcal{M}}^4) \ .
\end{equation}
Computing the on-shell action for this solution and evaluating the entropy through the usual formula~\eqref{eq:thermod_relations}, we find that
\begin{equation}\label{eq:Dirichlet_entropy}
    \mathcal{S}_D^{(\text{GR})}=\frac{\Omega_{d-1}r_h^{d-1}}{4G_N} =\frac{\Omega_{d-1}\mathfrak{r}^{d-1}}{4G_N}-\frac{(d-1)(d-2)\Omega_{d-1}\mathfrak{r}^{d-3}}{64\pi^2 G_N}\beta_{\partial\mathcal{M}}^2+\mathcal{O}(\beta_{\partial\mathcal{M}}^4) \ .
\end{equation}
 Notice that
in contrast to the conformal case discussed previously, the entropy here approaches to a constant value at high temperatures. Further, the leading correction away from this constant goes to zero in a way that does not depend on the spacetime 
dimension. In AdS/CFT duality, these finite boundaries appear to have an interpretation as irrelevant deformations of the boundary theory~\cite{Freidel:2008sh,McGough:2016lol,Zamolodchikov:2004ce,Smirnov:2016lqw,Taylor:2018xcy, Hartman:2018tkw}. For instance, in $d=2$ this corresponds to the $T\bar{T}$ deformation of Zamolodchikov~\cite{Zamolodchikov:2004ce}.

Even though both the canonical and the conformal entropy are given in terms of the same (bulk) area formula, their behavior in terms of boundary data is strikingly different. This is due to the fact that the conformal boundary allows for a fluctuating Weyl factor that can accommodate many black hole solutions as we take the horizon radius to be large.

\subsection{Conformal Thermodynamics in Einstein-Gauss-Bonnet Gravity} \label{sec: emgb thermo}
We now turn to discuss conformal thermodynamics within EGB gravity, for all values of the cosmological constant. As in the Einstein case, one could start by computing the on-shell action \eqref{eq:full_action} for solutions given by blackening factor \eqref{eq:f(r)_EMGB}. We do that in appendix \ref{app:on-shell_actions}.

As we are interested mainly in the behavior of the conformal entropy, here we adopt an alternative approach.~Generalizing the Einstein case, it can be shown that the entropy in EMGB gravity universally takes the form,
\begin{equation}\label{eq:Swald}
    \mathcal{S}^{\text{(EMGB)}}_{\text{conf}}=\frac{\Sigma_{k,d-1} r_h^{d-1}}{4 G_{N}} \left(1 + 2k\frac{d-1}{d-3}\frac{{\alpha}}{r_h^2}\right) \ ,
\end{equation}
which is the well-known result derived in \cite{Jacobson:1994qe} that corresponds to the Wald entropy for GB black holes~\cite{Wald:1993nt}.  This bulk formula holds independently of the boundary conditions, the sign of the cosmological constant and the matter content of the theory. The derivation only requires the existence of a local, covariant boundary term.~We review this derivation in appendix \ref{ap:entropy_univ}.

The objective then is to write \eqref{eq:Swald} in terms of the conformal boundary data.~In the following, we will do that, first for the uncharged $Q=0$ case in (A)dS.~We will further show that the flat space result can be obtained in the limit of $\P \ell \to \infty$. We will finally consider the case with finite charge in section~\ref{subsec:charged_conformal_thermo}.

\subsubsection{EGB AdS Black Hole} 
We start the evaluation of the conformal entropy by considering the AdS uncharged black hole in EGB. That is, setting $q=0$ and $\Lambda = -\tfrac{d(d-1)}{2\ell^2}$ in eq.~\eqref{eq:f(r)_EMGB}, to obtain the following line element,
\begin{equation}\label{eq:blacken_AdS_EGB}
    ds^2_{\text{BD}}=f(r)d\tau^2+\frac{dr^2}{f(r)}+r^2d\Sigma^2_{d-1} \ , \  \ f(r)=k+\frac{r^2}{2\alpha}\left(1-\sqrt{1+\frac{4\alpha\mu}{r^{d}}-\frac{4\alpha}{\ell^2}}\right) \ .
\end{equation}
The boundary is located at $r=e^\bomega \mathfrak{r}$, and we consider the region between the boundary and the (Euclidean) horizon, $r\in(e^\bomega\mathfrak{r},r_h)$.

First note that in GB gravity, the AdS curvature scale $\ell_{\mt{AdS}}$ is related to the scale $\ell$ appearing in the action~\eqref{eq:full_action} through 
\begin{equation}\label{eq:f(sigma)}
    \ell_{\mt{AdS}}=\ell\,g(\tigma) \ , \quad \text{with} \ \ g(\tigma) := \ \sqrt{\frac{2\tigma}{1-\sqrt{1-4\tigma}}}  = 1-\frac{\tigma }{2}+\mathcal{O}\left(\tigma ^2\right) \, ,
\end{equation} 
as can be seen by taking the $r\rightarrow\infty$ limit of eq.~\eqref{eq:blacken_AdS_EGB}.~Note that we defined $\tigma:= \tfrac{\alpha}{\ell^2}$.~The distinction between $\ell$ and $\ell_\mt{AdS}$ will become important when trying to compare to the pure Einstein gravity results.

Now recall that for Einstein gravity, the conformal boundary of AdS has $\mathcal{K}\ell \rightarrow d$. In EGB gravity, this result generalizes to 
\begin{equation}
    \mathcal{P}\,\ell_{\mt{AdS}}\rightarrow \frac{d}{3}\left(2+\sqrt{1-4\tigma}\right) = d\left( 1-\frac{2 \tigma }{3}+\mathcal{O}\left(\tigma ^2\right) \right)\ ,
\end{equation}
as derived in appendix \ref{app:GB_AdS_bdy}. 

With this information, we are ready to compute the conformal entropy. We first obtain analytic expressions perturbatively in $\tigma$. Further, as we are interested in the high-temperature expansion, we consider the bulk large-volume limit by taking $r_h,e^{\bomega \mathfrak{r}}\rightarrow\infty$ with $y$ kept fixed.~In this double limit and for the AdS black holes under consideration, ~eq.~\eqref{eq:invert_devil_b,K} becomes 
\begin{align}\label{eq:leading_bKl_spheAdS}
        \tilde{\beta}&= \dfrac{4\pi \ell}{dr_h}\sqrt{1-y^{d}}+\tigma\dfrac{2\pi\ell\left(1-y^{d}\right)^\frac{3}{2}}{d r_h} + \mathcal{O}(r_h^{-3},\tigma^2) \ ,\\
        \mathcal{P}\ell &= \dfrac{d}{2\sqrt{1-y^{d}}}\left(2-y^{d}\right)-\tigma\dfrac{d}{12}\sqrt{1-y^{d}}\left(2+y^{d}\right)+\mathcal{O}(r_h^{-2},\tigma^2)\ .
        \nonumber
        \end{align}
Notice that the boundary conditions are independent of the horizon topology, \ie independent of $k$.~As in the Einstein case, it is convenient to re-write these expressions in terms of the variable $u=\sqrt{1-y^{d}}$. Similarly, the second boundary condition can be now written as a quartic equation,
\begin{equation}\label{dep_quartic_u}
    \tigma u^4 -3\left(\tigma-2\right)u^2-\frac{12\mathcal{P}\ell}{d}u+6=0 \ .
\end{equation}
As before, given a solution $u$, we can immediately use the first line in eq.~\eqref{eq:leading_bKl_spheAdS} to find $r_h$ as
\begin{equation}\label{eq:r_h_from_u}
    r_h=\frac{2\pi\ell u}{d\tilde{\beta}}\left(2+\tigma u^2\right) \ .
\end{equation}
Once we have $r_h$, it is immediate to evaluate the conformal entropy using eq.~\eqref{eq:Swald}.\\[-1.5ex]

\noindent \textbf{The quartic equation} One then needs to solve the depressed quartic \eqref{dep_quartic_u}. That can be achieved by implementing Ferrari's method to obtain an analytic result, however the expressions even for $d=4$ are quite complicated and rather unilluminating.

Given that, for now, we are interested in the leading $\tigma$ contributions, we instead solve the quartic perturbatively.~Assume a solution of the form $u=u_0+\tigma \,u_1+\mathcal{O}(\tigma^2)$. The leading order term $u_0$ is the Einstein solution \eqref{eq:sln_u0} with $\mathcal{K}\rightarrow\mathcal{P}$, \ie
\begin{equation}
     u_0=\frac{\mathcal{P}\ell-\sqrt{\mathcal{P}^2\ell^2-d^2}}{d} \ .
\label{eq:housecoat}
\end{equation}
To get the first correction $u_1$, it is convenient to define,
\begin{equation}\label{eq:depressed}
    F(u,\tigma) := \tigma u^4 -3\left(\tigma-2\right)u^2-\frac{12\mathcal{P}\ell}{d}u+6 \, .
\end{equation}
Then eq.~\eqref{dep_quartic_u} can be written to leading order as
\begin{equation}\label{eq:perturb_quartic}
    F(u_0,\tigma=0)+\tigma\left(\frac{\partial F}{\partial \tigma}(u_0,\tigma=0)+u_1 \frac{\partial F}{\partial u}(u_0,\tigma=0)\right)+\mathcal{O}(\tigma^2) = 0 \ .
\end{equation}
Solving eq.~\eqref{eq:perturb_quartic} order by order in $\tigma$, we find
\begin{equation}\label{eq:IFThm}
    u_1=-\frac{\frac{\partial F}{\partial \tigma}(u_0,\tigma=0)}{\frac{\partial F}{\partial u}(u_0,\tigma=0)}=\frac{du_0^2(u_0^2-3)}{12(\mathcal{P}\ell-du_0)} \ .
\end{equation}
The Inverse Function Theorem guarantees the uniqueness of this solution in an open neighborhood of $\tigma \ll 1$, provided $\frac{\partial F}{\partial u}(u_0,\tigma=0)\neq0$. Note that this means that we can only trust the perturbative solution for $\mathcal{P}\ell>d$.\footnote{Note that using eq.~\eqref{eq:housecoat}, one finds $\mathcal{P}\ell-du_0=\sqrt{\mathcal{P}^2\ell^2-d^2}$.}~To obtain the right correction close to the AdS boundary, we will solve the equation exactly but numerically, momentarily. For now, the perturbative leading order solution reads
\begin{equation}\label{eq:pert_soln_wrong}
    u=u_0+\tigma\frac{du_0^2(u_0^2-3)}{12(\mathcal{P}\ell-du_0)}+\mathcal{O}(\tigma^2) \,,
\end{equation}
with $u_0$ given in eq.~\eqref{eq:housecoat}.\\[-1ex]

\noindent \textbf{Conformal entropy} Combining eqs.~\eqref{eq:Swald} and \eqref{eq:r_h_from_u} with our perturbative solution~\eqref{eq:pert_soln_wrong}, we finally obtain that
\begin{equation}\label{eq:leading_a_SconfAdS}
\begin{aligned}  \mathcal{S}_\text{conf}^{(\text{EGB})}&=\frac{\Sigma_{k,d-1}\ell_{\mt{AdS}}^{d-1}}{4G_N}\left(\frac{4\pi}{d^2}\right)^{d-1}\,\frac{\left(\mathcal{P}\ell_{\mt{AdS}}-\Delta\right)^{d-1}}{\tilde{\beta}^{d-1}} \\
    &\ \quad\times\Bigg[1 + \tigma\frac{(d-1)\big(d^2-\mathcal{P}\ell_{\mt{AdS}}(\mathcal{P}\ell_{\mt{AdS}}+2\Delta)\big)\left(\mathcal{P}\ell_{\mt{AdS}}-\Delta\right)^{2}}{3d^2\,(\mathcal{P}^2\ell_{\mt{AdS}}^2-d^2)}
 \Bigg]+\mathcal{O}\left(\frac{1}{\tilde{\beta}^{d-3}}\right)\ ,
\end{aligned}    
\end{equation}
where $\Delta =\sqrt{\mathcal{P}^2\ell_{\mt{AdS}}^2-d^2}$. From this expression, one can read off the effective number of degrees of freedom to be
\begin{equation}\label{eq:d=5_Sconf_EGB_AdS}
\begin{aligned}
\mathcal{N}_\text{dof}^{(\text{EGB})}(\mathcal{P}\ell_{\mt{AdS}})&=\frac{\Sigma_{k,d-1}\ell_{\mt{AdS}}^{d-1}}{4G_N}\left(\frac{4\pi}{d^2}\right)^{d-1}\left(\mathcal{P}\ell_{\mt{AdS}}-\Delta\right)^{d-1} \\
    & \quad\times\Bigg[1 + \tigma\frac{(d-1)\big(d^2-\mathcal{P}\ell_{\mt{AdS}}(\mathcal{P}\ell_{\mt{AdS}}+2\Delta)\big)\left(\mathcal{P}\ell_{\mt{AdS}}-\Delta\right)^{2}}{3d^2\,(\mathcal{P}^2\ell_{\mt{AdS}}^2-d^2)}
 \Bigg] \ .
\end{aligned} 
\end{equation}
Eqs.~\eqref{eq:leading_a_SconfAdS} and \eqref{eq:d=5_Sconf_EGB_AdS} generalize the Einstein gravity results in eqs.~\eqref{eq:house2} and \eqref{eq:N_dof(K)_GR_cases}, which are recovered in the limit $\tigma\rightarrow0$.\footnote{\label{foot:P_depends_on_a}Note that $\mathcal{P}$ implicitly depends on $\alpha$, by its definition~\eqref{eq:ECBCs}.~In the GR $\alpha\rightarrow0$ limit it reduces to the trace of the extrinsic curvature $\mathcal{K}$.~With this understanding, in all our expressions we keep $\mathcal{P}$.}

In the limit where we approach the AdS boundary, the GB contribution diverges as can be seen by inspecting the second term in in the square brackets in eq.~\eqref{eq:leading_a_SconfAdS}.~Here, this simply signals the breakdown of our perturbative expansion in $\tilde\alpha$ -- see the comments below eq.~\eqref{eq:IFThm}. Indeed, we will show that the non-perturbative result is monotonic and finite.\\[-1.7ex]

\begin{figure}[t]
  \centering
  \includegraphics[width=.8\textwidth,height=8cm]{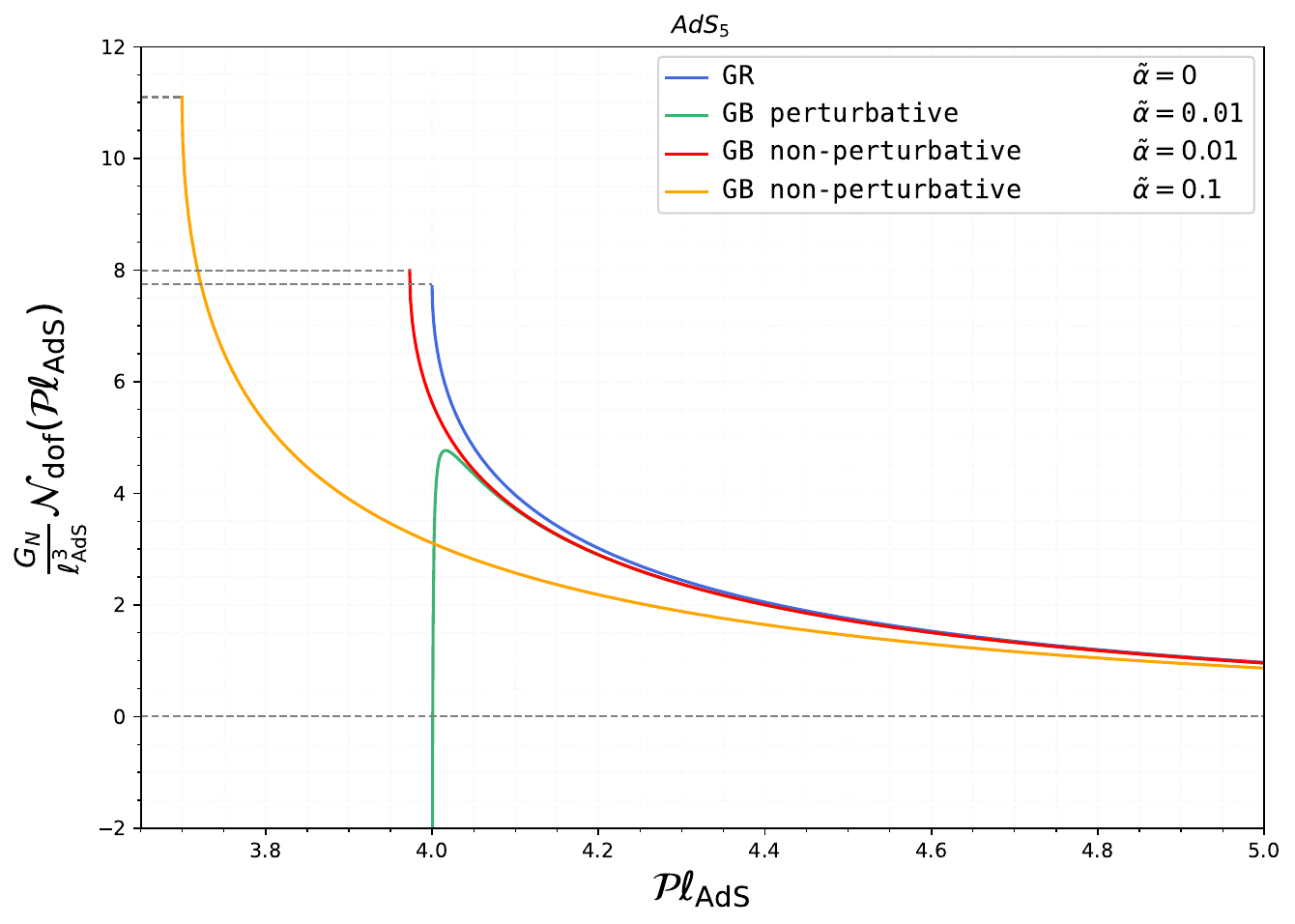}
  \caption{The function $\frac{G_N}{\ell_\mt{AdS}^{d-1}}\mathcal{N}_{\text{dof}}(\mathcal{P}\ell_\mt{AdS})$ for Einstein gravity (GR)~\eqref{eq:N_dof(K)_GR_cases}, the leading Gauss-Bonnet (GB perturbative) correction in perturbation theory~\eqref{eq:leading_a_SconfAdS} and the numerical solution for finite $\tigma=0.01,\,0.1$ (GB non-perturbative).~The horizontal dashed semi-lines correspond to the value of each function at the AdS boundary.~The numerical GB solution exhibits similar monotonic behavior for any finite value of  the coupling $\tigma$ and in any dimension $d$.}
  \label{fig:non-pert_AdS}
\end{figure}

\noindent \textbf{Non-perturbative conformal entropy} Now we consider the problem for finite $\tigma$, but we are still interested in the high-temperature expansion of the entropy.~The ECBC~\eqref{eq:ECBCs} for the metric~\eqref{eq:blacken_AdS_EGB} now take the form
\begin{align}\label{eq:non-pert_AdS_invert}
    \tilde{\beta}&= \dfrac{2^{\frac{3}{2}}\pi\ell}{d\sqrt{\tigma}r_h}w+\dots  \ ,\\
        \mathcal{P}\,\ell& = \dfrac{d(-w^4+6w^2+12\tigma)}{12\sqrt{2\tigma}\ell w} +\cdots \ ,
        \nonumber
\end{align}
where $w := \sqrt{1-\sqrt{1-4\tigma(1-y^{d})}}$ and the dots indicate subleading corrections in the high-temperature limit.~Even though inverting these relations for finite $\tigma$ is again intractable via analytic methods, we proceed numerically. The method is the same as before. Namely, from the second equation in~\eqref{eq:non-pert_AdS_invert}, we derive a fourth order algebraic equation for $w$ in terms $\mathcal{P}\ell_\mt{AdS}$,
\begin{equation}\label{eq:algebraic_numerics}
w^4-6w^2+\frac{12\sqrt{1-\sqrt{1-4\tigma}}\mathcal{P}\ell_\mt{AdS}}{d-1}w-12\tigma=0 \ .
\end{equation}
Given a solution to this equation,\footnote{The quartic eq.~\eqref{eq:algebraic_numerics} has four roots, and we choose the one that is consistent with the definition of $w$ when $y\in[0,1]$.} then it is straightforward to use the first line of eq.~\eqref{eq:non-pert_AdS_invert} to write the horizon radius as a function of the boundary data as
\begin{equation}
    r_h=\frac{2\pi \sqrt{1-\sqrt{1-4\tigma}}\ \ell_\mt{AdS}}{d\,\tigma\,\tilde{\beta}}\,w \ .
\end{equation}
From $r_h$, it is direct to obtain the conformal entropy via eq.~\eqref{eq:Swald}.~The algorithm is straightforward and only demands solving a quartic equation numerically. The result for two different values of $\tigma$ can be seen in figure~\ref{fig:non-pert_AdS}, where we also show the naive perturbative result.~It is clear from this figure that the divergence close to the AdS boundary was an artifact of the perturbative expansion, \ie the non-perturbative result is a smooth function of $\P \ell_\mt{AdS}$ and yields a finite result at $\P \ell_\mt{AdS}=d$.

Indeed, the non-perturbative GB contribution takes a finite value at the AdS boundary.~This value coincides with the one expected from AdS/CFT, which for finite GB coupling takes the form~\cite{Buchel:2009sk}
\begin{equation}
    \mathcal{S}_{\text{EGB}}=\frac{\Sigma_{k,d-1}\ell_{\mt{AdS}}^{d-1}}{4G_N}\left(\frac{4\pi}{d\,g^2(\tigma)}\right)^{d-1}\frac{1}{\tilde{\beta}^{d-1}} +\mathcal{O}\left(\frac{1}{\tilde{\beta}^{d-3}}\right) \ ,
\end{equation}
where $g(\tigma)$ was defined in eq.~\eqref{eq:f(sigma)}.
When $\tigma>0$ the value of $\mathcal{N}_\text{dof}^{(\text{EGB})}$ at the AdS boundary is greater than that in pure GR, while it is smaller when $\tigma<0$.~The deviation of the two values at the AdS boundary might be interpreted as  a modification of the microscopic ultraviolet degrees of freedom via the addition of a Gauss-Bonnet term in the Einstein-Hilbert action~\eqref{eq:full_action}.

As we increase $\P\ell_\mt{AdS}$, the boundary is driven into the interior of the bulk, and $\mathcal{N}_{\text{dof}}(\mathcal{P}\ell_\mt{AdS})$ decreases monotonically towards the black hole horizon, which has $\P \ell_\mt{AdS}\to \infty$. For $\P\ell_\mt{AdS}$ away from the boundary, $\mathcal{N}_{\text{dof}}(\mathcal{P}\ell_\mt{AdS})$  becomes indistinguishable from the Einstein gravity result. As in the Einstein case, this behavior is reminiscent of that for holographic $c$-functions~\cite{Freedman:1999gp,Myers:2010tj,Myers:2012ed}.

\subsubsection{EGB dS Black Hole} 

Similarly, we can study the problem in EGB with a positive cosmological constant. Given the existence of both black hole and cosmological horizons, in this case, there are different types of patches that can be considered. Ref.~\cite{Anninos:2024wpy} presents an exhaustive analysis of spherically symmetric static configurations in Einstein gravity. Here, we will just restrict ourselves to what are called \textit{cosmic patches}, that is the bulk region in between the boundary and the cosmological horizon.

The bulk blackening factor is given by eq.~\eqref{eq:f(r)_EMGB} with $q=0$, $\Lambda=\tfrac{d(d-1)}{2\ell^2}$ and $k=+1$,
\begin{equation}
    ds^2_{\text{BD}}=f(r)d\tau^2+\frac{dr^2}{f(r)}+r^2d\Omega^2_{d-1} \ , \  \ f(r)=1+\frac{r^2}{2\alpha}\left(1-\sqrt{1+\frac{4\alpha\mu}{r^{d}}+\frac{4\alpha}{\ell^2}}\right) \ .
\end{equation}
Performing a similar analysis as in the AdS case above, one arrives at expressions for the conformal entropy that are related to the AdS ones via a simple analytic continuation $\ell\rightarrow i\ell$ in eq.~\eqref{eq:leading_a_SconfAdS}, recall footnote~\ref{ft:analytic_cont}. More specifically, the dS conformal entropy reads
\begin{equation}\label{eq:leading_a_SconfdS}
\begin{aligned}
    \mathcal{S}_\text{conf}^{(\text{EGB})}&=\frac{\Omega_{d-1}\ell_{\mt{dS}}^{d-1}}{4G_N}\left(\frac{4\pi}{d^2}\right)^{d-1}\,\frac{\left(\tilde{\Delta}-\mathcal{P}\ell_{\mt{dS}}\right)^{d-1}}{\tilde{\beta}^{d-1}} \\
    & \quad\times\Bigg[1 - \tigma\frac{(d-1)\big(d^2+\mathcal{P}\ell_{\mt{dS}}(\mathcal{P}\ell_{\mt{dS}}+2\tilde{\Delta})\big)\left(\tilde{\Delta}-\mathcal{P}\ell_{\mt{dS}}\right)^{2}}{3d^2\,(\mathcal{P}^2\ell_{\mt{dS}}^2+d^2)}
 \Bigg]+\mathcal{O}\left(\frac{1}{\tilde{\beta}^{d-3}}\right)\ ,
\end{aligned} 
\end{equation}
\begin{figure}[t] 
  \centering
  \includegraphics[width=.7\textwidth,height=7cm]{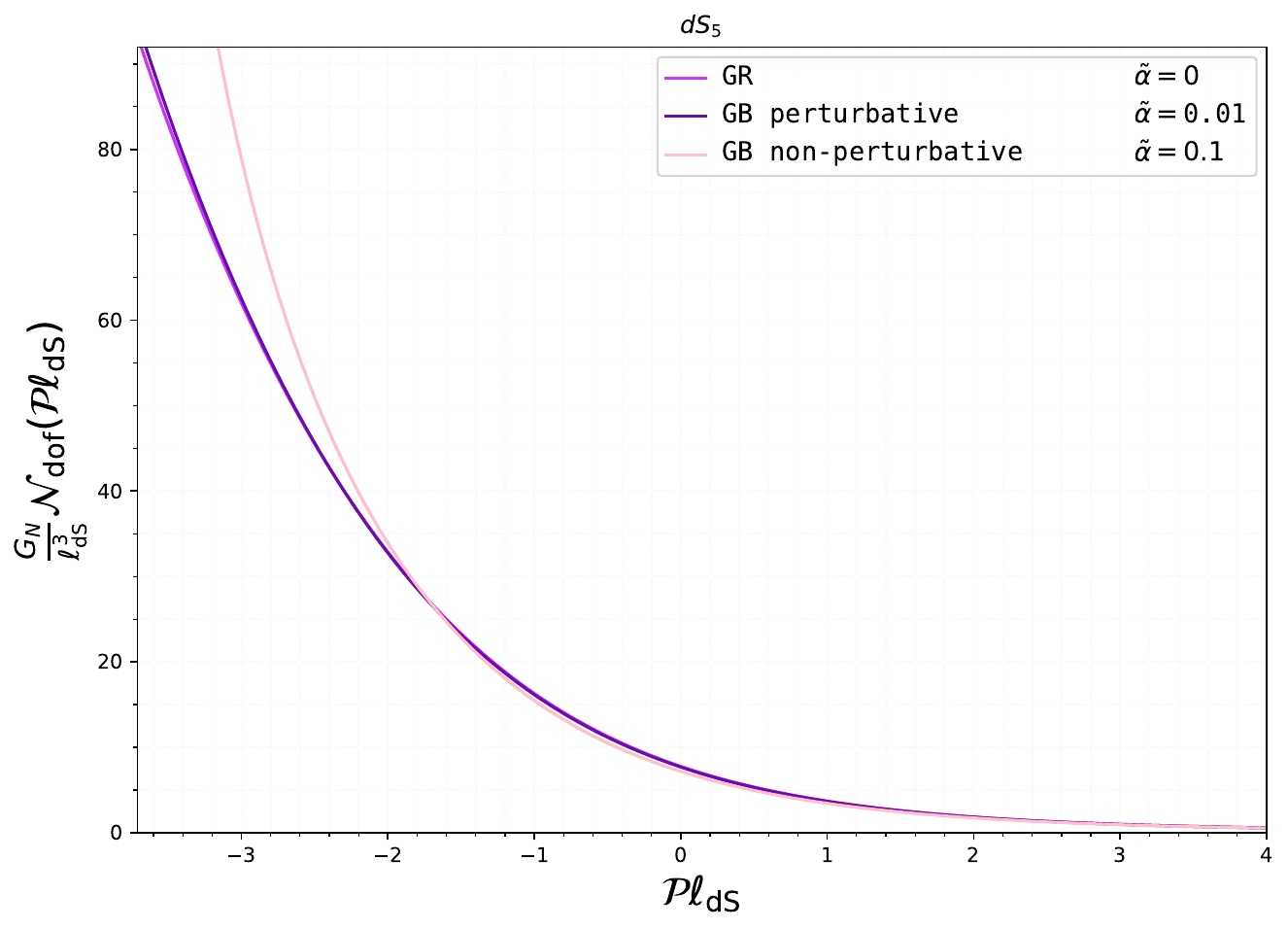}
  \caption{Plot of the function $\frac{G_N}{\ell_\mt{dS}^{d-1}}\mathcal{N}_{\text{dof}}(\mathcal{P}\ell_\mt{dS})$ of the number of degrees of freedom in the putative boundary dual of dS$_5$, for the leading Gauss-Bonnet (GB) correction of eq.~\eqref{eq:leading_a_SconfdS}.~Both the perturbative~\eqref{eq:Ndof_dS} and non-perturbative EGB curves as shown, along with the GR one.~The general trend is that $\mathcal{N}_{\text{dof}}(\mathcal{P}\ell_\mt{dS})$ is larger for greater values of $\tigma$.~Again, the story is the same in higher dimensions.}
  \label{fig:fN_dof(K)_EGB_dS}
\end{figure}
perturbatively in $\tigma$,\footnote{Notice that again $\tigma=\tfrac{\alpha}{\ell^2}$, but in the analytic continuation $\ell\to i\ell$ one sends $\tigma\to -\tigma$ to arrive at eq.~\eqref{eq:leading_a_SconfdS}.} with $\tilde{\Delta}=\sqrt{\mathcal{P}^2\ell_{\mt{dS}}^2+d^2}$. This gives,
\begin{equation}\label{eq:Ndof_dS}
\begin{aligned}
    \mathcal{N}_{\text{dof}}(\mathcal{P}\ell_\mt{dS})  &=\frac{\Omega_{d-1}\ell_{\mt{dS}}^{d-1}}{4G_N}\left(\frac{4\pi}{d^2}\right)^{d-1}\left(\tilde{\Delta}-\mathcal{P}\ell_{\mt{dS}}\right)^{d-1} \\
    & \quad\times\Bigg[1 - \tigma\frac{(d-1)\big(d^2+\mathcal{P}\ell_{\mt{dS}}(\mathcal{P}\ell_{\mt{dS}}+2\tilde{\Delta})\big)\left(\tilde{\Delta}-\mathcal{P}\ell_{\mt{dS}}\right)^{2}}{3d^2\,(\mathcal{P}^2\ell_{\mt{dS}}^2+d^2)}
 \Bigg]\ .
 \end{aligned}
\end{equation}
Note that, perturbatively in $\tilde{\alpha}$, in dS there is no asymptotic timelike boundary, so $\P\ell_\mt{dS}$ is allowed to take values from $-\infty$, close to the observer's worldline (or a black hole horizon), to $+\infty$, near the cosmological horizon. In each case we obtain that,
\begin{equation}
    \mathcal{N}_\text{dof}(\mathcal{P}\ell_\mt{dS})  =
    \begin{cases}
   \begin{aligned}
       \dfrac{2^{d-5} \pi ^{d-1} \Omega_{d-1} }  {G_N \mathcal{P}^{d+1}\ell_{\mt{dS}}^2} &\Big(\left(4 \mathcal{P}\ell_{\mt{dS}}^2-(d-1) d^2\right)-\tigma(d-1)d^2\Big)\\
       &\qquad\qquad\qquad+\mathcal{O}\left(\left(\mathcal{P}\ell_{\mt{dS}}\right)^{-d-2}\right)  ,  \ \mathcal{P}\ell\to +\infty \ ,
   \end{aligned}\\ \\
   \begin{aligned}
\dfrac{\Omega_{d-1}\ell_{\mt{dS}}^{d-1}}{4G_N}\left(-\dfrac{8\pi\mathcal{P}\ell_{\mt{dS}}}{d^2}\right)^{d-1} &\left(1+\tigma\dfrac{4(d-1)d^2}{3\ell_{\mt{dS}}^2}\left(-\dfrac{8\pi\mathcal{P}\ell_{\mt{dS}}}{d^2}\right)^2\right)\\
& \qquad+ \mathcal{O}\left((\mathcal{P}\ell_{\mt{dS}})^{d-3}\right)  , \ \mathcal{P}\ell\to -\infty \ .
\end{aligned}
    \end{cases}
\end{equation}
Non-perturbatively in $\tigma$, however, there is a subtlety.~In particular, for the analytically continued version of eq.~\eqref{eq:algebraic_numerics} under $\ell\rightarrow-i\ell$ and $\tilde{\alpha}\to-\tilde{\alpha}$, one can find real positive solutions for $w$ only when $\mathcal{P}\ell_{\mt{dS}}\in [-p_\text{crit}(\tilde{\alpha}),+\infty)$.~Here, the lower bound is negative, \ie $p_\text{crit}(\tilde{\alpha})>0$, and depends on the value of the coupling $\tilde{\alpha}$.~The dependence is such that $p_\text{crit}(\tilde{\alpha})$ increases when $\tilde{\alpha}$ decreases.~In the $\tilde{\alpha}\to 0$ limit we recover the Einstein behavior, and $\P\ell_{\mt{dS}}$ is valued over the entire real line.~In EGB, this can be interpreted as saying that there do not exist cosmic patches consistent with the ECBC in the high-temperature regime, for values of $\mathcal{P}\ell_{\mt{dS}}<p_\text{crit}(\tilde{\alpha})$.

We plot eq.~\eqref{eq:Ndof_dS} in figure~\ref{fig:fN_dof(K)_EGB_dS}. We observe that as in Einstein gravity, $\mathcal{N}_{\text{dof}}(\mathcal{P}\ell_\mt{dS})$ is a monotonically decreasing function of $\P \ell_\mt{dS}$.~As opposed to the AdS case, the perturbative expansion does not break down, and it is a good approximation to the non-perturbative result for small $\tigma$, for all allowed values of $\P\ell_{\mt{dS}}$.

\subsubsection{Flat EGB Black Hole}

An appealing feature of considering gravity in the presence of a boundary is that it provides an additional dimensionful constant, in our case $\P$. Provided $\P$ and a non-vanishing cosmological constant, it becomes physically meaningful to consider the flat space limit, in which $\P\ell$ goes to infinity. This is true both for positive and negative values of the cosmological constant.

Taking the large-$\P\ell$ limit of, \eg the conformal entropy for the AdS black hole~\eqref{eq:leading_a_SconfAdS}, we obtain\footnote{Note the solutions with maximally symmetric horizons for $k=0$ and --1 are not conventional black holes, but correspond to what are  known as flat space cosmologies, see \eg ref~\cite{Banihashemi:2025qqi}.}
\begin{equation}\label{eq:Ndof_Mink}
    \mathcal{S}_\text{conf}^{(\text{EGB})}=\frac{(2\pi)^{d-1}\Sigma_{k,d-1}}{4G_N\mathcal{P}^{d-1}}\frac{1}{\tilde{\beta}^{d-1}}+\mathcal{O}\left(\frac{1}{\tilde{\beta}^{d-3}}\right) \ .
\end{equation}
Note that the leading term in the high-temperature expansion does \textit{not} receive a correction in the GB coupling (recall footnote~\ref{foot:P_depends_on_a}).~The first non-trivial correction comes at order $\mathcal{O}(\tilde\beta^{3-d})$. This is unlike the AdS and dS cases, and might be a distinct feature of quantum gravity in flat space. The same result can be obtained directly from the flat space black hole, but then the GB coupling needs to scale with $G_N$ -- see appendix \ref{app: flat EGB bh}.

As a matter of fact, in flat space we can do better.~In particular, in the high-temperature limit one can analytically invert the boundary data relations~\eqref{eq:invert_devil_b,K} to obtain $r_h$, as follows.~In the high-temperature limit\footnote{This limit is subtle in flat space due to the absence of a background scale.~See appendix~\ref{app: flat EGB bh} for a more careful discussion of this limit.} the boundary data for the line element of eq.~\eqref{eq:f(r)_EMGB} with $k=1,\Lambda=q=0$, read
\begin{align}
        \tilde{\beta}&=\dfrac{4 \pi(1-y)\sqrt{1+2\rho}}{\sqrt{ \left(d-3+(d-5) \rho\right)(1-y)}}+\mathcal{O}(1-y) \ ,\\
        \mathcal{P}\ell &= \frac{\sqrt{ \left(1+2\rho\right) \left(d-3+(d-5) \rho\right)(1-y)}}{2 r_h (1-y)} +\mathcal{O}(\sqrt{1-y}) \ ,
        \nonumber
\end{align}
with $\rho:=\tfrac{\alpha}{r_h^2}$.~It is then straightforward to combine these relations and obtain the following expression
\begin{equation}
    r_h=\frac{2\pi}{\mathcal{P}\tilde{\beta}}\left(1+2\frac{\alpha}{r_h^2}\right) \ .
\end{equation}
Solving for $r_h$ and expanding around $\tilde{\beta}\to 0$, we arrive at
\begin{equation}\label{eq:rh_flat_non-pert}
    r_h=\frac{2\pi}{ \mathcal{P}\tilde{\beta}}+\alpha\frac{\mathcal{P} \tilde{\beta}}{\pi }+\mathcal{O}(\alpha^2,\tilde{\beta}^3) \ .
\end{equation}
Eq.~\eqref{eq:rh_flat_non-pert} illustrates that the leading high-temperature contribution to the conformal entropy in Minkowski space is not affected by the GB coupling $\alpha$, coinciding with the Einstein gravity result~\eqref{eq:Nodf_intro}, \ie
\begin{equation}
    \mathcal{N}^{(\text{EGB})}_\text{dof}(\mathcal{P})=\frac{(2\pi)^{d-1}\Omega_{d-1}}{4G_N\mathcal{P}^{d-1}} \ .
\end{equation}
The fact that \textit{non-perturbatively} $\mathcal{N}_\text{dof}(\P)$ is robust against higher derivative contributions is noteworthy.~We will comment further on this in section~\ref{sec:discuss}, where we conjecture that this is true in flat space regardless of the type of higher curvature correction one considers.

\subsection{Adding Charge}\label{subsec:charged_conformal_thermo}

In an attempt to test our conjecture on the universality of the effective number of degrees of freedom $\mathcal{N}_\text{dof}$ in theories of gravity with vanishing cosmological constant, we consider the case with charge. We first study the high-temperature limit of the AdS charged black hole, and then take a suitable flat space limit, to obtain that also in this case, $\mathcal{N}_\text{dof}$ does not receive higher curvature corrections.

\subsubsection{EMGB AdS Black Hole}

Consider a spherical and charged black hole in AdS. The line element is given by eq.~\eqref{eq:f(r)_EMGB} with $\Lambda=-\tfrac{d(d-1)}{2\ell^2}$ and $k=+1$, 
\begin{equation}
    ds^2_{\text{BD}}=f(r)d\tau^2+\frac{dr^2}{f(r)}+r^2d\Omega^2_{d-1} \ , \  \  f(r)=1+\frac{r^2}{2\alpha}\left(1-\sqrt{1+\frac{4\alpha\mu}{r^{d}}-\frac{4\alpha q^2}{r^{2(d-1)}}-\frac{4\alpha}{\ell^2}}\right) \ .
\end{equation}
As usual, we are interested in the conformal thermodynamics of the black hole patch $r\in(e^\omega\mathfrak{r},r_h)$.~Using the dimensionless variables $\tigma=\alpha/\ell^2, \ \tau:=q^2/r_h^{2(d-2)}$, we can repeat the preceding analysis.~Working to leading order in~$\tigma$~and taking the large-volume limit, we arrive at
\begin{align}
        \tilde{\beta}&= \dfrac{4\pi \ell}{dr_h}\sqrt{1-y^{d}}+\tigma\dfrac{2\pi\ell\left(1-y^{d}\right)^\frac{3}{2}}{d r_h} + \mathcal{O}(r_h^{-3},\tigma^2) \ ,\\
        \mathcal{P}\ell &= \dfrac{d}{2\sqrt{1-y^{d}}}\left(2-y^{d}\right)-\tigma\dfrac{d}{12}\sqrt{1-y^{d}}\left(2+y^{d}\right)+\mathcal{O}(r_h^{-2},\tigma^2)\ .
        \nonumber
\end{align}
which are precisely the leading expressions \eqref{eq:leading_bKl_spheAdS} for the uncharged case. Therefore, we conclude that the AdS expressions remain unaffected to leading order by the presence of (electric) charge, and they are given by eq.~\eqref{eq:leading_a_SconfAdS}.

The fact that the charge does not enter the leading order behavior of the entropy is neither endemic to (extended) conformal boundary conditions nor to higher curvature gravity. For comparison, we can consider the case of charged AdS black branes (\ie $k=0$) within Einstein gravity.~The blackening factor appearing in the metric is given by eq.~\eqref{eq:f(r)_GR},
\begin{equation}
    f(r)=\frac{r^2}{\ell^2}-\frac{\mu}{r^{d-2}}+\frac{q^2}{r^{2(d-2)}} \ .
\end{equation}
The Hawking temperature is given by 
\begin{equation}\label{eq:GR_AdS_Q_BB_beta}
    \beta=\frac{4\pi}{d\frac{r_h}{\ell^2}-(d-2)\frac{q^2}{r_h^{2d-3}}} \ .
\end{equation}
Taking $\beta\rightarrow0$, to leading order gives that the first term in the denominator dominates and thus to leading order, the Bekenstein-Hawking entropy is the same as in the zero-charge case, \eg see \cite{Chamblin:1999tk}.~We will provide further remarks on this in section~\ref{sec:discuss}.

\subsubsection{Flat EMGB Black Hole} 

As already remarked, since, perturbatively in the GB coupling, the leading high-temperature contribution in the black hole entropy in Minkowski space~\eqref{eq:Ndof_Mink} is unaffected, one might wonder whether the presence of other parameters (such as electric charge) can modify this behavior.~Similarly as before, we can arrive at the corresponding expressions by taking the flat $\P\ell\to \infty$ limit of the AdS expressions.~But since, to leading order in the high-temperature expansions, these were unaffected, one is led to the conclusion that the effective number of degrees of freedom is the same as in eq.~\eqref{eq:Ndof_Mink}, \ie
\begin{equation}
    \mathcal{N}_\text{dof}(\P)=\frac{(2\pi)^{d-1}\Omega_{d-1}}{4G_N \P^{d-1}} \ .
\end{equation}
One can arrive at this result also from a direct analysis of charged spherical black holes in Minkowski space, as we demonstrate in appendix~\ref{app: flat EGB bh}.

\section{Lorentzian Linearized Dynamics}\label{sec:Lornetzian_dynamics}

In this section, we initiate the discussion of real-time dynamics of Einstein-Maxwell-Gauss-Bonnet theory in the presence of a timelike boundary subject to extended conformal boundary conditions. In this setting, the advantage of working with EGB (or in general with LL) gravity, instead of any other generic higher curvature theory, is that the equations of motion for the metric remain second order.

The IBVP, compared to the Euclidean case, is much more subtle.~For instance, the (Euclidean) ellipticity of conformal boundary conditions in Einstein gravity was already proven in ref.~\cite{anderson2008boundary}, while results on the initial boundary corner value problem in Lorentzian signature are quite recent~\cite{An:2021fcq,Liu:2025xij, An:2025cbs}.~Adding higher-curvature corrections can only make these proofs more involved.~In fact, for EGB, the initial value problem has only recently been understood \cite{Kovacs:2020pns,Kovacs:2020ywu,Figueras:2024bba},\footnote{This recent progress built on earlier work by Choquet-Bruhat~\cite{Choquet-Bruhat:1988jdt,choquet1989gravitation}, who showed that the IVP of EGB gravity is well-posed in the restrictive case of \textit{analytic} initial data.} and we are not aware of any mathematical results regarding the problem in the presence of boundaries.

With that word of caution, we proceed to study the problem of Lorentzian dynamics inside an EGB worldtube with a finite boundary subject to ECBC. We consider the following induced metric at the boundary 
\begin{equation}\label{eq:L_bdy_metric}
    ds^2|_{\partial\mathcal{M}} = e^{2\bomega} [h]_{\mu\nu} dx^\mu dx^\nu \,,
\end{equation} 
where $[h]_{\mu\nu}$ is one particular representative of the conformal class of the induced metric, which is fixed at the boundary. The boundary mode $\bomega(y^m)$ (Latin indices are intrinsic to the boundary), instead, is not fixed in this problem, and it should be considered as a boundary degree of freedom. Additionally, we have the standard bulk gravitational degrees of freedom.

The interaction between bulk and boundary degrees of freedom is encoded in the generalized (conformal) version of the Brown-York stress tensor, that in EGB is found to be\footnote{In defining the conformal stress tensor we followed the conventions of ref.~\cite{Anninos:2024xhc}.}
\begin{equation}\label{eq:conformal_EGB_BY}
   T_{\mu\nu}^{(\text{EGB})}=-\dfrac{e^{(d-2)\bomega}}{8\pi G_N}\left(\mathcal{K}_{\mu\nu}-\frac{1}{d}h_{\mu\nu}\mathcal{K}+\frac{\alpha}{(d-2)(d-3)}\Big(\mathcal{B}_{\mu\nu}-\dfrac{1}{d}h_{\mu\nu}\mathcal{B}\Big)\right) \ ,
\end{equation}
where $\mathcal{B}=h^{\mu\nu}\mathcal{B}_{\mu\nu}$, and the (long) explicit form of the tensor $\mathcal{B}_{\mu\nu}$ is given in eq.~\eqref{eq:GB_relations}. For the Einstein case, this conformal stress tensor was derived in refs.~\cite{An:2021fcq, Odak:2021axr}. For a derivation of the EGB generalization, see appendix~\ref{app:action_variations}, \eg eqs.~\eqref{eq:def_EH} and~\eqref{eq:def_GB}.

Note that, as for the case of CBC in Einstein gravity, this conformal stress tensor~\eqref{eq:conformal_EGB_BY} is traceless. Furthermore, the divergence of the conformal stress tensor can be obtained through the analogue of the momentum constraint at the boundary,
\begin{equation}
    D^\mu T^{(\text{EGB})}_{\mu\nu}=-\frac{(d-1)e^{d\bomega}}{d8\pi G_N} D_{\nu} \mathcal{P} \ ,
\end{equation}
where the divergence is taken with respect to the conformal representative of the metric at the boundary $[g]_{\mu\nu}$. In Einstein gravity, the stress tensor is transverse provided $\mathcal{K}$ is fixed to a constant at the boundary \cite{Anninos:2024xhc}. In EGB, we observe that this generalizes and the conformal stress tensor is covariantly conserved when $\mathcal{P}$ is constant along $\partial\mathcal{M}$.

The equivalent of the Hamiltonian constraint, for a boundary metric of the form \eqref{eq:L_bdy_metric}, becomes a dynamical equation for the boundary mode $\bomega(y^m)$.~In Einstein gravity, this can be nicely written in terms of the conformal stress tensor, see \eg eq.~(2.13) of ref.~\cite{Anninos:2024xhc}.~It would be interesting to see whether a similar formula can be obtained for EGB gravity. Given it is cumbersome, we will not attempt to write here the Hamiltonian constraint as a function of the $\bomega(y^m)$ and $[g]_{\mu\nu}$ in full generality.

Instead, we will solve a simplified version of the problem \cite{Anninos:2024wpy,Liu:2024ymn,Anninos:2024xhc,Galante:2025tnt}. We will consider the bulk solution to be the Lorentzian version of the $f(r)$ metrics described in section~\ref{sec:gen_setup}, see eq.~\eqref{eq:f(r)_EMGB}.~Given the previous discussion, in the presence of a finite boundary subject to (extended) conformal boundary conditions, this does not fully solve the gravitational problem, as the Weyl factor at the boundary is still allowed to fluctuate. We will provide the form of the non-linear equation governing the dynamics of the Weyl factor when the conformal representative of the metric at the boundary is chosen to be the standard metric on $\mathbb{R}\times \Sigma_{k,d-1}$ (up to the overall Weyl factor which is only allowed to depend on the boundary time). Then we will consider the linearized version of that equation, expanded around the static solution with $\bomega =0$.~For Einstein gravity with spherical boundaries, the solutions to the linearized problem are always exponentially increasing (and decreasing) with time \cite{Galante:2025tnt}. In EGB, we show that this is still true for the spherical case.~We also show that planar and  hyperbolic boundaries allow for oscillatory behavior, even in Einstein gravity. We will consider both the charged and uncharged cases.

Before moving into that, let us make a brief comment about the bulk dynamics problem.~Of course, in the same way we are linearizing the boundary dynamics equation, we could consider the bulk linearized problem. Around spherically symmetric backgrounds in Einstein gravity, this problem has been analyzed in refs.~\cite{Anninos:2023epi,Anninos:2024wpy,Liu:2024ymn,Anninos:2024xhc}.~Namely, the linearized perturbations admit a Kodama-Ishibashi decomposition~\cite{Kodama:2000fa,Kodama:2003jz}, which results in dynamics governed by a master field, whose equation of motion is solvable for maximally symmetric spacetimes.~In a Fourier basis, imposing conformal boundary conditions selects the allowed mode frequencies.~It was observed that this problem in Einstein gravity admits solutions which grow exponentially with time at the linearized level~\cite{Anninos:2023epi,Anninos:2024wpy,Liu:2024ymn,Anninos:2024xhc, Liu:2025xij}.

Turning to EGB, given that the bulk linearized equations of motion about maximally symmetric backgrounds will be the same, the only difference would be that we are now fixing $\mathcal{P}$, instead of $\mathcal{K}$ at the boundary $\partial\mathcal{M}$.~It would be interesting to see whether our extended conformal boundary conditions change the qualitative behavior of the bulk dynamics.~See section~\ref{sec:discuss} for further discussion on this.

\subsection{Boundary mode dynamics}

For now, we restrict ourselves to the boundary dynamics problem just described. Here we only consider the dynamics of the Weyl factor in the case where it is a function of the conformal time $u$ (and not of the other boundary coordinates).

Let us begin by discussing the general setup, before specializing to a particular metric and horizon topology.~Looking at bulk solutions of the form,
\begin{equation}\label{eq:f(r)_gen_dynamics}
    ds^2=-f(r)dt^2+\frac{dr^2}{f(r)}+r^2d\Sigma_{k,d-1}^2 \ ,
\end{equation}
we place our boundary at $r=e^{\bomega(u)}\mathfrak{r}$ and we are going to impose the extended conformal boundary conditions~\eqref{eq:ECBCs}, so that
\begin{equation}\label{eq:dynamics_sphere_bcs}
    \begin{cases}
        ds^2|_{\partial\mathcal{M}}=e^{2\bomega(u)}\left(-du^2+\mathfrak{r}^2d\Sigma_{k,d-1}^2\right) \ ,\\
        \mathcal{P}|_{\partial\mathcal{M}}=\text{constant} \ .
    \end{cases}
\end{equation}
For the charged case, we also fix the charge $q$ at the boundary. Requiring $\mathcal{P}$ being constant, yields the following differential equation governing the dynamics of the Weyl factor:
\begin{equation}\label{eq:f(r)_dynamics}
\small
\begin{aligned}
  \mathfrak{r}^2 \partial_u^2\bomega(u)=&\frac{1}{6\mathfrak{r}^2 \left(\mathfrak{r} ^2 (e^{2\bomega(u)}+4\alpha(\partial_u\bomega(u))^2)+
  2 \alpha\big(k+ f(e^{\bomega(u)}\mathfrak{r})\big) \right)}\Big[6 \alpha e^{\bomega(u)}\mathfrak{r}  \left(f(e^{\bomega(u)}\mathfrak{r})-k\right) f'(e^{\bomega(u)}\mathfrak{r})\\
  &-3 e^{3\bomega(u)}\mathfrak{r} ^3  f'(e^{\bomega(u)}\mathfrak{r})+4(d-3) \alpha \left(f(e^{\bomega(u)}\mathfrak{r})+(\mathfrak{r}\partial_u\bomega(u))^2 \right) \left(f(e^{\bomega(u)}\mathfrak{r})-2(\mathfrak{r}\partial_u\bomega(u))^2 -3k\right)\\
  &-6(d-1) e^{2\bomega(u)}\mathfrak{r} ^2  \left(f(e^{\bomega(u)}\mathfrak{r} )+(\mathfrak{r}\partial_u\bomega(u))^2 \right)+6 \mathcal{P} e^{3\bomega(u)}\mathfrak{r} ^3 \sqrt{f(e^{\bomega(u)}\mathfrak{r})+(\mathfrak{r}\partial_u\bomega(u))^2 }\Big]  \ .
\end{aligned}
\end{equation}
In the limit $\alpha\rightarrow0$ we recover the corresponding expression from Einstein gravity~\cite{Liu:2024ymn,Galante:2025tnt}, namely eqs.~\eqref{eq:GR_Weyl_black_brane dynamics} and~\eqref{eq:GR_Weyl_brane dynamics} for planar and hyperbolic horizons, respectively. Note that a solution to eq.~\eqref{eq:f(r)_dynamics} leaves the bulk solution~\eqref{eq:f(r)_gen_dynamics} unchanged, so these solutions can all be interpreted as physical diffeomorphims.~Moreover, as a second order (non-linear) differential equation, eq.~\eqref{eq:f(r)_dynamics} needs to be supplemented by \textit{corner} data, $\lbrace\bomega(u)|_{\mathcal{C}},\partial_u\bomega(u)|_{\mathcal{C}}\rbrace$. These data is neither part of the standard initial data of the initial value problem nor of the  extended conformal boundary conditions and resides at what is usually termed the corner, \ie the intersection between the Cauchy slice and the timelike boundary.~After imposing this corner data, together with the standard initial data (at least within Einstein gravity), solutions to the initial boundary corner value problem (IBCVP) exist and are unique \cite{Anninos:2022ujl,An:2025rlw}.~We expect that something similar holds for the IBCVP in EGB gravity, subject to the ECBC. 

Solutions to eq.~\eqref{eq:f(r)_dynamics} are still difficult to find.~However, note that there always exists a static solution with $\bomega (u) = 0$.~In particular, plugging $\bomega (u) = 0$ in eq.~\eqref{eq:f(r)_dynamics} retrieves the form of $\mathcal{P}$ at a constant $r=\mathfrak{r}$ surface, \ie eq.~\eqref{eq:invert_devil_b,K}.~One can further linearize eq.~\eqref{eq:f(r)_dynamics} around this solution, so that $\bomega(u)= \delta\bomega(u) + \mathcal{O}(\delta \bomega^2)$, to obtain,
\begin{equation}\label{eq:BH_linearized}
   \mathfrak{r}^2\partial^2_u\delta\bomega(u)= \lambda^2\,\delta\bomega(u) \ ,
\end{equation}
with
\begin{equation}\label{eq:BH_lambda2}
\small
\begin{aligned}
\lambda^2
&=
\dfrac{1}{4 f(\mathfrak{r})\bigl(\mathfrak{r}^2+ 2 \alpha (k+f(\mathfrak{r})) \bigr)}
\begin{multlined}[t]
\Bigl[
4 f^2(\mathfrak{r}) \bigl(
      \alpha \mathfrak{r} \bigl( (d-5) f'(\mathfrak{r}) + \mathfrak{r} f''(\mathfrak{r}) \bigr)
    + 6 k \alpha(d-3)  
    + (d-1) \mathfrak{r}^2
  \bigr) \\
  {}+ 2 \mathfrak{r} f(\mathfrak{r}) \bigl(
      f'(\mathfrak{r}) \bigl(
        -2 k \alpha (d-5)  
        - (d-1) \mathfrak{r}^2
        + \alpha \mathfrak{r} f'(\mathfrak{r})
      \bigr)
      - \mathfrak{r} f''(\mathfrak{r}) \bigl( 2 k \alpha + \mathfrak{r}^2 \bigr)
  \bigr) \\
  {}- 8 (d-3) \alpha f^3(\mathfrak{r})
  + \mathfrak{r}^2 (f'(\mathfrak{r}))^2 \bigl( 2 k \alpha + \mathfrak{r}^2 \bigr)
\Bigr]
\end{multlined}\\
&\equiv
- \dfrac{\mathfrak{r}^2 \sqrt{f(\mathfrak{r})}}
        {\mathfrak{r}^2 + 2 \alpha\bigl(k+f(\mathfrak{r}) \bigr)}\,
  \partial_r \mathcal{P}\big|_{r=\mathfrak{r}}\ .
\end{aligned}
\end{equation}
for $\mathcal{P}$ as in eq.~\eqref{eq:invert_devil_b,K}.~Thus eq.~\eqref{eq:BH_linearized} admits solutions of the form,
\begin{equation}\label{eq:f(r)_linear_dynamics}
    \delta\bomega(u)\propto e^{\pm \lambda u} \ .
\end{equation}
Note that in principle, $\lambda\in \mathbb{C}$. We now turn to study the behavior of these modes for the EMGB black hole line elements~\eqref{eq:f(r)_EMGB}.

\paragraph{Spherical horizons}

\begin{figure}[t] 
  \centering
  \includegraphics[width=.65\textwidth,height=7cm]{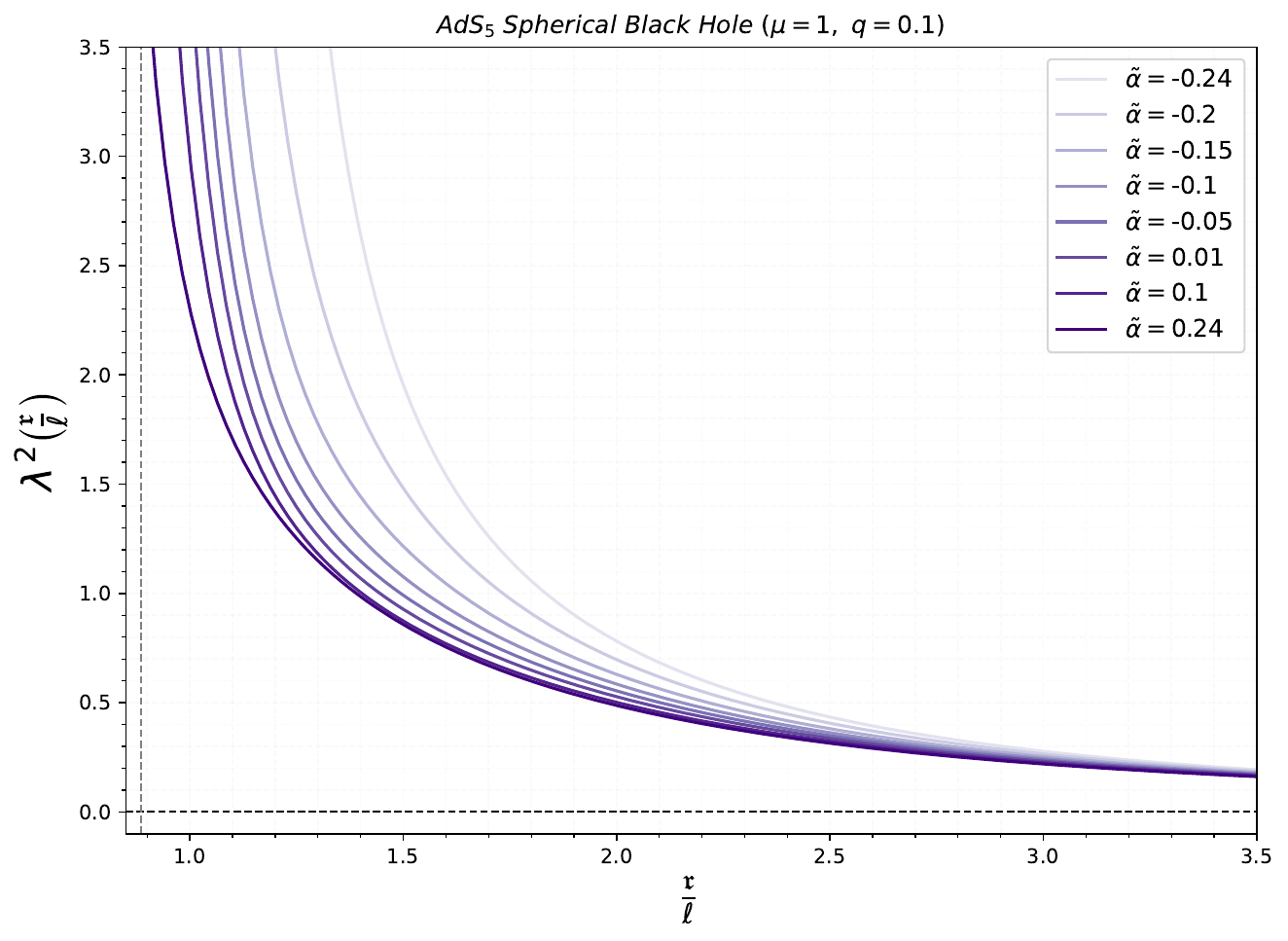}
  \caption{Plot of $\lambda^2(\tfrac{\mathfrak{r}}{\ell})$~\eqref{eq:BH_lambda2} for the charged AdS$_5$ black hole for $\mu=1, \ q=0.1 $, and for various negative and positive values of the coupling $\tigma$.~The gray vertical dashed line indicates the radius of the outer event horizon for $\tigma=0.24$.~The qualitative features of the curves do not depend on the precise value of the charge $q$ or the dimension $d$.}
  \label{fiG:dynamics_AdS}
\end{figure}

\begin{figure}[t]
  \centering
  \includegraphics[width=.65\textwidth,height=7cm]{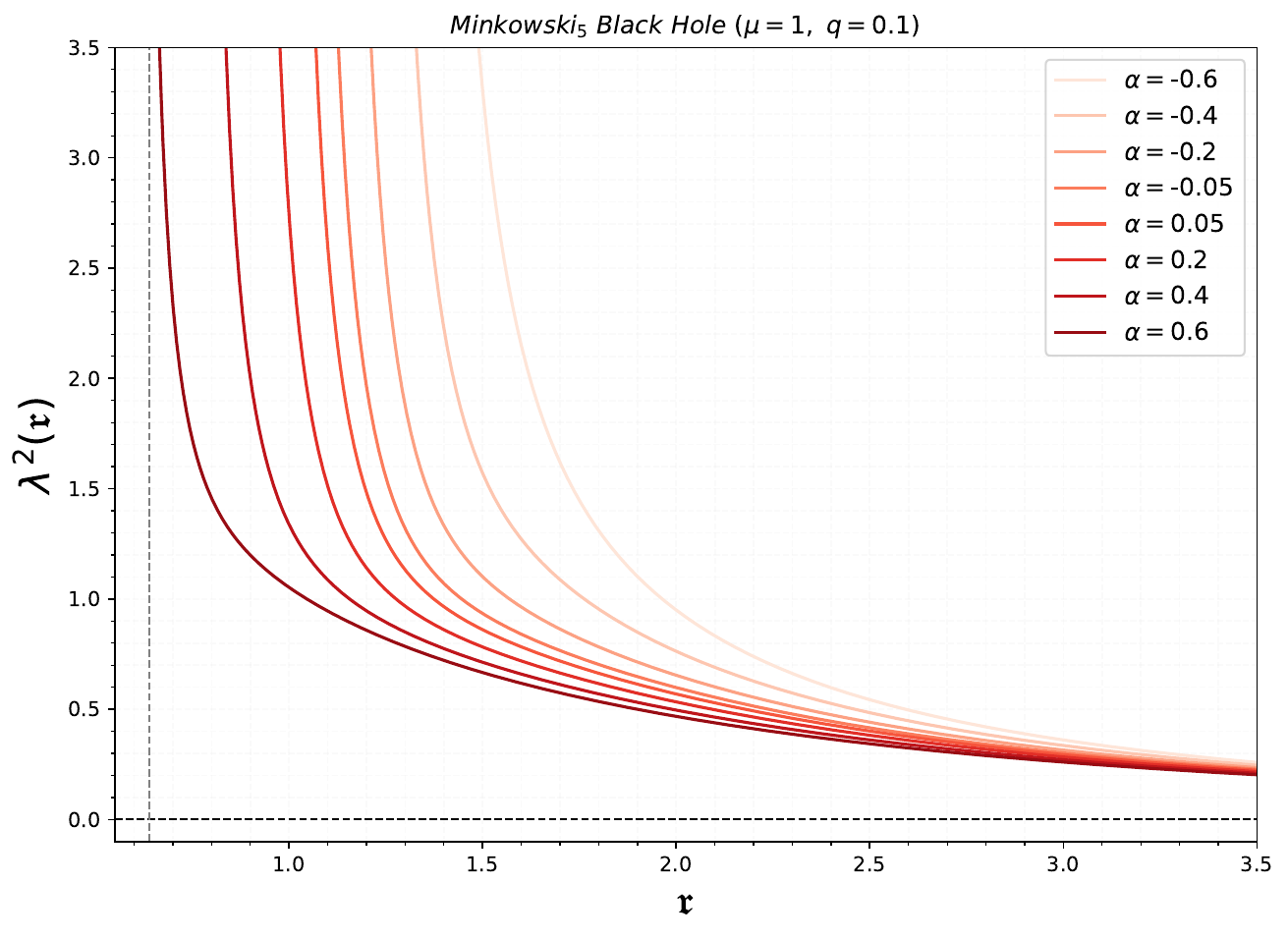}
  \caption{Plot of $\lambda^2(\mathfrak{r})$~\eqref{eq:BH_lambda2} for a charged black hole in $5$-dimensional flat space, where $\mu=1, \ q=0.1 $ (the values of $d,q$ do not affect the qualitative aspects of the curves), for various negative and positive values of the coupling $\alpha$.~The gray vertical dashed line on the left plot corresponds to the horizon position when $\alpha=0.6$.}
  \label{fiG:RN_dynamics_flat}
\end{figure}

Let us first examine the behavior of these modes by considering the case of a charged spherical EMGB black hole in AdS, with blackening factor~\eqref{eq:f(r)_EMGB},
\begin{equation}\label{eq:arguments}
    f(r)=1+\frac{r^2}{2\alpha}\left(1-\sqrt{1+4\alpha\left(\frac{\mu}{r^{d}}-\frac{q^2}{r^{2(d-1)}}-\frac{1}{\ell^2}\right)}\right) \ .
\end{equation}
Inserting this $f(r)$ into eq.~\eqref{eq:BH_lambda2} for $k=1$ returns a rather long expression for the frequencies $\lambda$ in terms of $\r$, and the other bulk variables.~Ultimately, this formula should be written in terms of the boundary data $\mathcal{P}$.~Instead of reporting that formula, we provide plots of $\lambda$ in  figure~\ref{fiG:dynamics_AdS}.\footnote{In all the figures of this section, $\lambda$ is plotted in units of $\ell$.} 

From that figure one can observe that $\lambda^2$ is positive very close to the horizon and diverges as $\mathfrak{r}\to r_h$, while $\lambda^2\to0$ as we approach the asymptotic boundary of AdS, and there does not exist any intermediate region where it becomes negative.~Exploring, more generally, the parameter space $(\mu,q,\tigma)$ we find that: in the uncharged case ($q=0$), for fixed mass $\mu$ we have $\lambda^2>0$ for all positive values of the GB coupling $\tigma$; this behavior persists when $\tigma$ is fixed for small black holes (sufficiently small $\mu<\ell$); similarly, when $\tigma<0$, $\lambda^2$ is always positive.~In the charged case ($q\neq0$) a similar picture holds for all values of $\tigma$ and $q$ below extremality, namely there are solely unstable modes.~For comparison, we mention that in the instance of Einstein-Maxwell theory all the spherically symmetric modes are unstable. 

One could also perform an analysis of the behavior of eq.~\eqref{eq:BH_lambda2} for charged black holes in Minkowski space, the results of which are presented in figure~\ref{fiG:RN_dynamics_flat}.~The general theme is that there are no stable modes, irrespective of the value of the coupling $\alpha$, as in pure Einstein-Maxwell theory.~Note that the near-horizon growing modes have a Rindler origin, as we discuss shortly.~These conclusions hold true for any value of the charge $q$ (including the uncharged case $q=0$).

In all cases for spherical horizons, we found that the boundary dynamics behavior of EGB is similar to the one found in Einstein gravity, where at the linearized level around the static solution, perturbations grow exponentially. It would be interesting to understand the end-point of this behavior, which requires a non-linear analysis.

\paragraph{Planar horizons}  One could also study non-linearly modes that preserve the background symmetries in the case of planar horizons.
\begin{figure}[t]
    \centering   
    \begin{subfigure}[t]{0.48\textwidth}
        \centering
        \includegraphics[width=\textwidth]{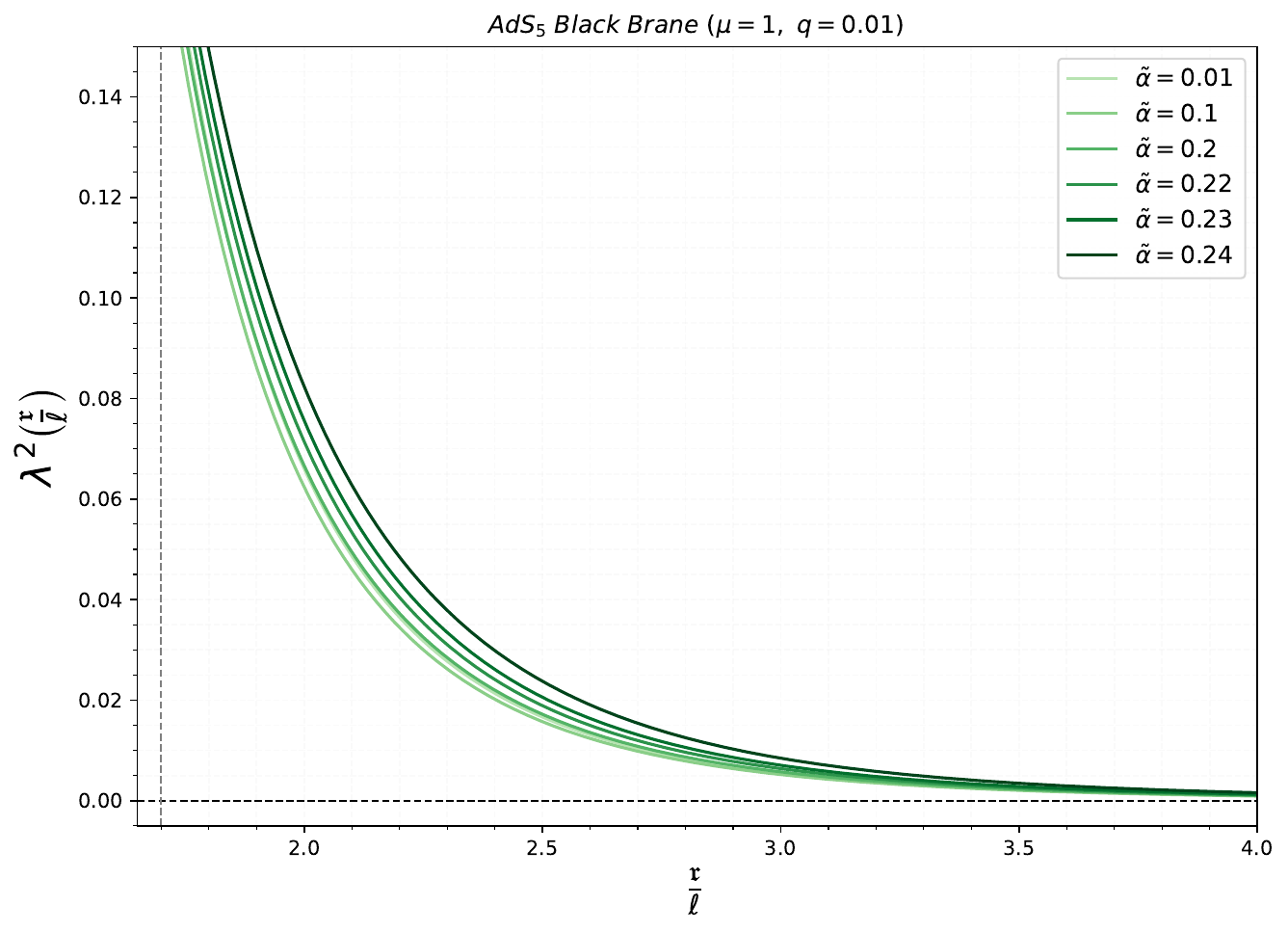}
        \caption{$q=0.01$}
        \label{fig:BD_BH_beta_d=5a}
    \end{subfigure}
    \hfill
    \begin{subfigure}[t]{0.48\textwidth}
        \centering
        \includegraphics[width=\textwidth]{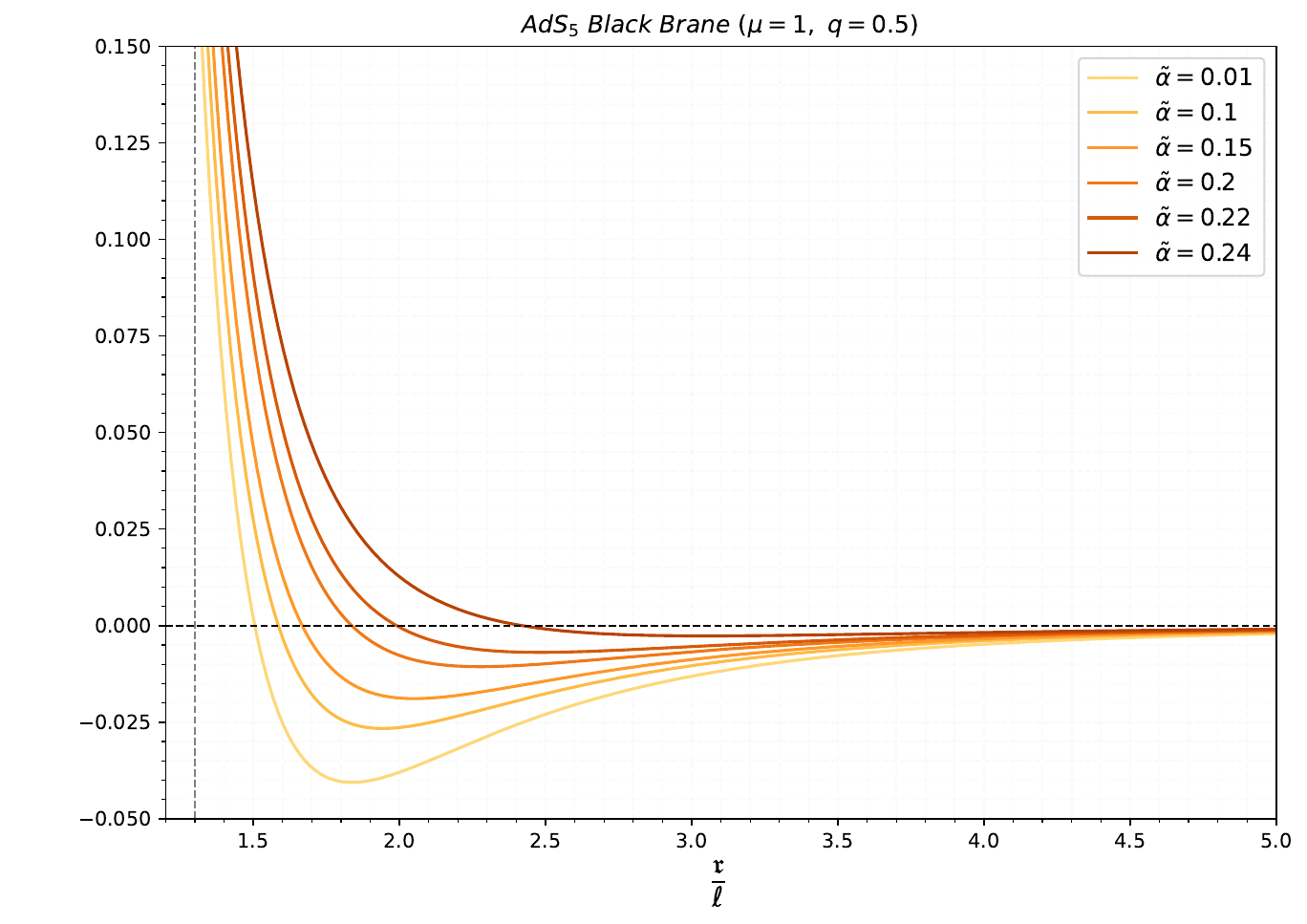}
        \caption{$q=0.5$}
        \label{fig:BD_BH_beta_d=5b}
    \end{subfigure}
    \caption{Plots of $\lambda^2(\tfrac{\mathfrak{r}}{\ell})$~\eqref{eq:BH_lambda2} for the charged AdS$_5$ black brane, when $ \mu=1, \ q=0.01 \ \text{(\textit{left})} \ \text{and} \  q=0.5 \ \text{(\textit{right})}$, for the various indicated values of the coupling $\tigma$. The gray vertical dashed line indicates the radius of the outer event horizon when $\tigma=0.01$, while the curves in all cases blow-up at the position of the horizon.~Again, the general picture is the same in higher dimensions.}
    \label{fiG:black_brane_dynamics_AdS}
\end{figure}In the example of a charged AdS black brane with the blackening factor given by~\eqref{eq:f(r)_EMGB},
\begin{equation}
    f(r)=\frac{1}{2\alpha}\left(1-\sqrt{1+4\alpha\left(\frac{\mu}{r^{d}}-\frac{q^2}{r^{2(d-1)}}-\frac{1}{\ell^2}\right)}\right) \ ,
\end{equation}
the story exhibits notable differences compared to the black hole case, especially for sufficiently large values of the charge below extremality.~In this regime, as illustrated in figure~\ref{fiG:black_brane_dynamics_AdS}, for all values of $\tigma$ in the interval $0\leq\tigma\leq \tfrac{1}{4}$ (recall that $\tfrac{1}{4}$ is the upper bound for the existence of AdS vacua) there only exist stable spherically symmetric linearized modes -- except close to the horizon where $\lambda^2$ becomes positive.~This near-horizon behavior is generically expected from eq.~\eqref{eq:BH_lambda2}, since near an outer and non-extremal horizon (\ie $f(r_h)=0$ and $f'(r_h)\neq0$), $\lambda$ is very large and positive signaling exponentially growing modes, and as we will see below that has a Rindler origin.~Going back to figure~\ref{fiG:black_brane_dynamics_AdS}, for small values of the charge, we observe that there are only unstable modes.~A parallel discussion to the small-$q$ case also holds when the black brane is uncharged.~For negative values of $\tigma$ a similar pattern appears, namely for large enough $q<m$ we have stable modes away from the near-horizon region and only unstable ones when $\mathfrak{r}\rightarrow r_h$.~The small charge and uncharged sectors again do not exhibit any stable modes.

We remark that interestingly, in pure Einstein-Maxwell theory subject to CBC \eqref{eq:CBCs}, when the charge is large enough the behavior of $\lambda^2(\mathfrak{r})$ closely resembles that of the curves in the right plot of figure~\ref{fiG:black_brane_dynamics_AdS}.~This is demonstrated in appendix~\ref{app:pert_GR}.~Thus, all the stable black brane modes discussed within EMGB exist also in EM, with the effect of the Gauss-Bonnet correction being that now stable modes start appearing when the boundary is placed further away from the horizon.~In addition, the near-horizon growing modes can be associated to a Rindler-like behavior \cite{Anninos:2025zgr}.~In particular, taking the planar Rindler line element to be, 
\begin{equation}
    ds^2=-\left(c\rho\right)^2 dt^2+d\rho^2+\frac{r_h^2}{\ell^2}d\vec{x}^2_{d-1} \ ,
\end{equation}
for some constant $c$ and performing a similar analysis subject to the extended conformal boundary conditions~\eqref{eq:dynamics_sphere_bcs}, one observes that the linearized dynamics of the corresponding conformal factor has frequencies
\begin{equation}\label{eq:near_Rindler}
    \lambda \ell=\pm 1 +\mathcal{O}\Big(\frac{\tigma}{\mathfrak{r}/\ell^2}\Big) \ ,
\end{equation}
which is the Einstein gravity behavior, and higher-derivative effects kick in once we move slightly away from the strict Rindler limit.~Consequently, all the near-horizon growing modes observed in the context of EMGB can be traced back to such a behavior.~We further remark that the expansion in eq.~\eqref{eq:near_Rindler}, including all the sub-leading terms, mimics structurally the (Rindler-like) stretched horizon limit explored in ref.~\cite{Anninos:2025zgr}, see \eg eq.~(5.2) there.~The observation that exponentially growing modes appear to be a generic feature of near-horizon physics subject to CBC was already made in the context of Einstein gravity, \eg in ref.~\cite{Anninos:2024wpy}.

\begin{figure}[t] 
  \centering
  \includegraphics[width=.65\textwidth,height=7cm]{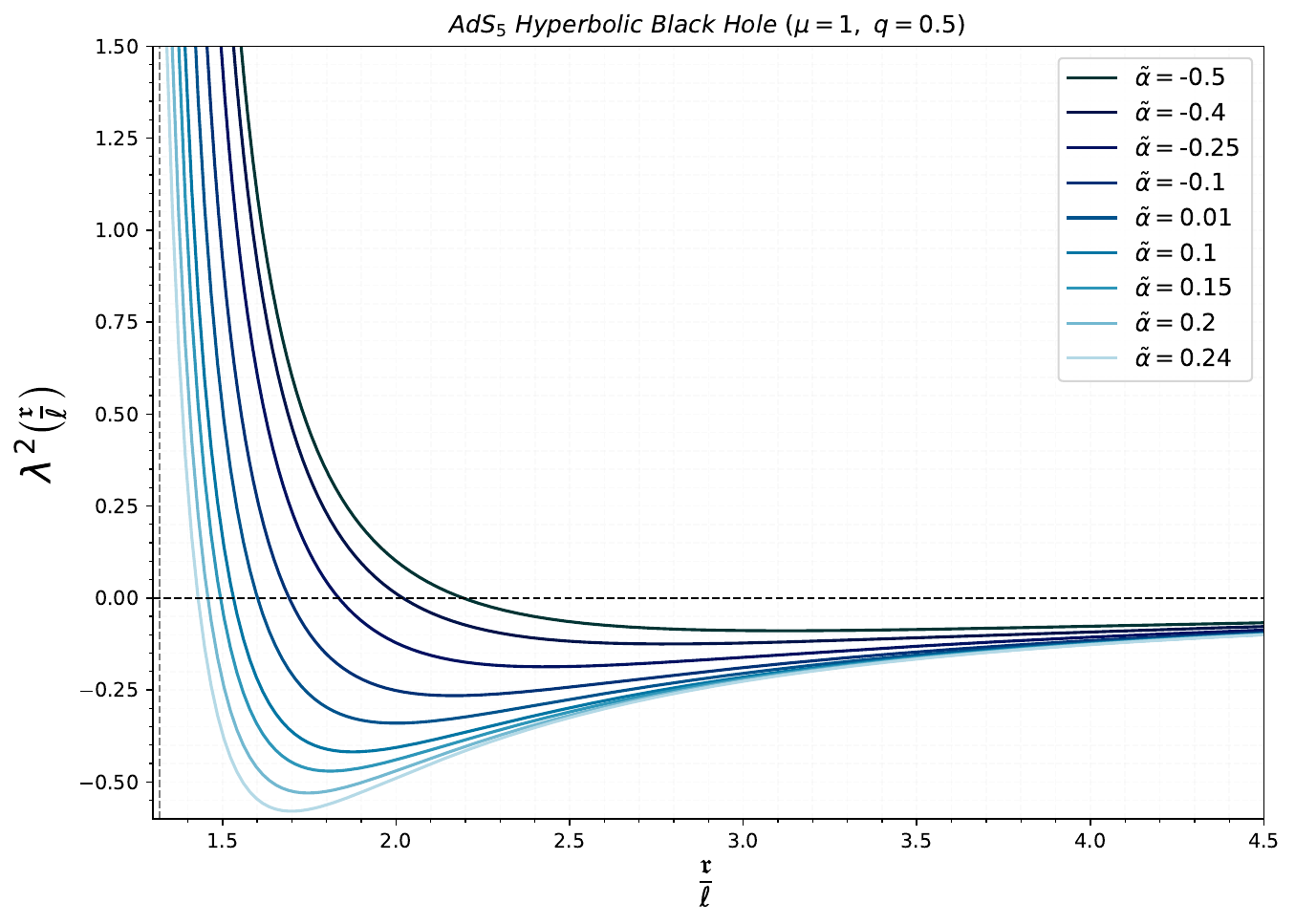}
  \caption{Plots of $\lambda^2(\tfrac{\mathfrak{r}}{\ell})$~\eqref{eq:f(r)_dynamics} for the charged AdS$_5$ black hole with hyperbolic horizon, where $\mu=1 \ \text{and} \  q=0.5$, for the various indicated values of the coupling $\tigma$. The gray vertical dashed line corresponds to the radius of the outer event horizon when $\tigma=0.24$.~The qualitative features of the graphs are independent of the dimension $d$.}
  \label{fiG:Hyper_black_brane_dynamics_AdS}
\end{figure}
\paragraph{Hyperbolic horizons} Another interesting example concerns charged AdS black holes with hyperbolic horizon topology, and here we briefly report on the results of such an analysis.~Our line element is of the form of eq.~\eqref{eq:dynamics_sphere_bcs} with $k=-1$, and the blackening factor reads~\eqref{eq:f(r)_EMGB}
\begin{equation}
    f(r)=-1+\frac{r^2}{2\alpha}\left(1-\sqrt{1+4\alpha\left(\frac{\mu}{r^{d}}-\frac{q^2}{r^{2(d-1)}}-\frac{1}{\ell^2}\right)}\right) \ .
\end{equation}
Figure~\ref{fiG:Hyper_black_brane_dynamics_AdS} contains the plot of $\lambda^2$, that governs the behavior of the linearized solutions, as in eq.~\eqref{eq:BH_lambda2}.~One observes that, for any value of the charge $q$ below extremality (including $q=0$), there exist stable modes except very close to the horizon (once again, the near-horizon growing modes can be found in a corresponding Rindler analysis).~This is also true in Einstein-Maxwell theory with hyperbolic horizons, as illustrated in appendix~\ref{app:pert_GR}.~The curious feature here is that this is the first example where stable modes exist for all negative values of the GB coupling $\tigma$.~The general trend is that hyperbolic horizons appear to stabilize the majority of exponentially growing modes in the linearized regime, save the ones of Rindler origin localized near $r=r_h$, even in Einstein gravity.

\section{Discussion}\label{sec:discuss}
Quasi-local thermodynamics subject to conformal boundary conditions in Einstein gravity exhibit a behavior that, in the high-temperature limit, mimics that of a conformal field theory in one lower dimension. Given that we expect Einstein gravity will receive higher curvature corrections when considering short distance scales, in this work, we initiated the study of conformal boundary conditions in higher curvature theories of gravity by analyzing the case of Einstein-Maxwell-Gauss-Bonnet. We showed that CBC can be appropriately generalized and the functional form of the high-temperature expansion in this generalized setting is not disturbed. Below we discuss some implications of our results, as well as potential future directions.

\paragraph{Linearized Lorentzian dynamics beyond spherical symmetry}
 
In Einstein gravity, spherical cavities around maximally symmetric spacetimes allow for exponentially growing modes, that grow faster for large angular momentum ~\cite{Anninos:2023epi,Anninos:2024wpy,Liu:2024ymn,Anninos:2024xhc, Liu:2025xij, Anninos:2025zgr}. These large angular momentum modes have radial profiles that localize exponentially close to the boundary. One might imagine at these scales higher curvature (or quantum) effects might affect the results obtained in Einstein gravity. In this work we studied the behavior of the spherically symmetric modes, but a systematic study of the large angular momentum modes is also possible. Allowed frequencies will be now obtained by imposing the generalized conformal boundary conditions for EGB~\eqref{eq:ECBCs}. It would be interesting to see how the corrections in the GB coupling affect the behavior of the large angular momentum modes, found in Einstein gravity. The Kodama-Ishibashi formalism has been extended to LL gravity \cite{Ishibashi:2011ws}, which might serve useful for this purpose.

\paragraph{A one-parameter family of boundary conditions in EGB} In ref.~\cite{Liu:2024ymn}, a one-parameter family of boundary conditions within Einstein gravity was introduced.~In particular, the authors considered putting an arbitrary numerical coefficient $\Theta$ in front of the Gibbons-Hawking-York boundary term, \ie $-\frac{\Theta}{8\pi G_N}\int_{\partial \mathcal{M}}d^{d}y \sqrt{h} \mathcal{K} \ ,$ such that the boundary conditions fix the conformal class of the metric at the boundary, and some combination of the determinant of the induced metric and the trace of the extrinsic curvature, \ie $\lbrace [h_{mn}]|_{\partial\mathcal{M}},h^p\mathcal{K}|_{\partial\mathcal{M}}\rbrace$, such that $p$ is related to $\Theta$ through $\Theta=1+\tfrac{d-1}{d(2p-1)}$.\footnote{This modified action is also stationary under the following generalized umbilic boundary conditions $\lbrace \tilde{\pi}^{\mu\nu}|_{\partial\mathcal{M}}=0,h^p\mathcal{K}|_{\partial\mathcal{M}}= \text{fixed} \rbrace$ . } The $p=0$ case reduces to the conformal boundary conditions of Einstein gravity.~One can generalize the extended boundary conditions discussed in this paper by allowing a general coefficient in front of the GB boundary term, $ -\frac{\Phi}{16\pi G_N}\int_{\partial\mathcal{M}} d^{d}y\sqrt{h}\mathcal{Q}$. This action has a well-defined variational principal provided the following quantities are kept fixed at the boundary, $\lbrace [h_{mn}]|_{\partial\mathcal{M}},h^p\mathcal{P}|_{\partial\mathcal{M}}\rbrace$, where $\Theta$ is chosen as above and $\Phi$ is given by $\Phi=1+\tfrac{d-3}{d(2p-1)}$. The detailed derivation is given in appendix~\ref{app:action_variations}. It would be interesting to analyze the thermodynamics stemming from this generalized action and the corresponding generalized boundary conditions. Note that at high temperatures, the behavior will not be extensive unless $p=0$, following the argument of ref.~\cite{Banihashemi:2024yye} given in the context of Einstein gravity.

\paragraph{Higher curvature generalizations} To obtain the results presented in this paper we had to make use of the natural generalization of the conformal boundary conditions~\eqref{eq:CBCs} to Gauss-Bonnet gravity, namely the extended conformal boundary conditions~\eqref{eq:ECBCs}.~The  difference is in the scalar quantity that is fixed at the boundary (along with the conformal class of the induced metric) is a higher derivative extension of the trace of the extrinsic curvature. Similarly, for more general Lanczos-Lovelock theories, this scalar quantity will receive further higher curvature contributions. In line with what was observed here for GB gravity, we expect these higher curvature corrections will be proportional to the boundary terms constructed in \cite{Myers:1987yn} for the LL theory subject to Dirichlet boundary conditions. More specifically, recall that the action of the $2n$-th LL theory corresponds to the $2n$-dimensional Euler density.\footnote{For example, the Einstein-Hilbert term is obtained for $n=1$, Gauss-Bonnet for $n=2$ and so on.} The boundary term is then the contribution to the Euler density if the $2n$-dimensional manifold has a boundary.

One may also consider more general higher–curvature interactions, \eg the
$R^{4}$ terms appearing in type~II superstring theory
\cite{Gross:1986iv,Grisaru:1986vi,Myers:1987qx}.  In general, however, such corrections do not preserve the property that the gravitational equations of motion remain second order in time derivatives.  Consequently, the underlying classical system of non-linear partial differential equations requires an augmented initial value formulation in which higher (time) derivatives of the metric, or equivalently the associated higher conjugate momenta, must be specified \cite{Ostrogradsky:1850fid}. An alternative viewpoint is that generic higher–curvature terms introduce additional propagating degrees of freedom (typically ghosts) at the scale set by their couplings. In effective field theory, one does not wish to retain these modes but only to capture how
the long-wavelength `Einsteinian' modes are perturbed by the higher–curvature interactions. Projecting out the unwanted excitations demands a more refined treatment, \eg see~\cite{Simon:1990ic,Parker:1993dk}, but would not substantially alter the qualitative structure of the boundary data. The special virtue of EGB and, more generally, Lovelock theories is that the gravitational equations of motion remain second order in time derivatives. Accordingly, constructing the extended conformal boundary conditions~\eqref{eq:ECBCs} was comparatively straightforward. For generic higher–curvature
corrections treated perturbatively around Einstein gravity, we  expect that an analogous set of ECBC can be defined in which, as in eq.~\eqref{eq:paste}, the boundary data is the conformal class of the boundary metric and a scalar built from the extrinsic curvature augmented by the appropriate higher–curvature corrections.

Additionally, from an effective field theory point of view, it is natural to consider higher derivative terms involving the gauge field (and possibly mixed terms combining the gauge field strength and the curvature tensor).~Black hole solutions with four derivative terms in the Maxwell field were studied in GB gravity \cite{Wiltshire:1988uq}.~Another natural extension of the present work might be the dimensional reduction of GB gravity, which in the lower dimensional theory, extends the GB term is a special family of higher derivative terms involving the Kaluza-Klein gauge field while preserving the second-order character of the equations of motion, \eg see \cite{Mueller-Hoissen:1985www,Huang:1988mw}. It would be interesting to consider conformal thermodynamics for black holes in these theories, though we expect that the general high-temperature extensivity~\eqref{eq:S_conf} persists. 

We would also like to comment here on the following:\footnote{We thank D.~Anninos for
illuminating discussions on this point.}~the gravitational path integral must, in principle, be invariant under field redefinitions. Under such transformations, however, the effective couplings appearing in the action~\eqref{eq:full_action} are generally reshuffled.  It is therefore necessary to specify what one means, in a covariant sense, by a particular coupling such as the Gauss--Bonnet parameter~$\alpha$. The natural prescription is to identify its distinct contribution to an appropriate physical observable, \eg a boundary graviton correlator. For instance, in the analysis of EGB gravity within the AdS/CFT framework~\cite{Buchel:2009sk}, the Gauss--Bonnet coupling controls a unique tensor structure in the three-point function of the boundary stress tensor. A context in which such a covariant definition of higher–curvature couplings plays an essential role is the computation of the
sphere path integral~\cite{Anninos:2020hfj}.

In the present work, we have not introduced a corresponding covariant definition of~$\alpha$, and so one should exercise some caution when interpreting the $\alpha$-dependent corrections to $\mathcal{N}_{\mathrm{dof}}(\mathcal{P})$
in eq.~\eqref{eq:house_on_fire} for nonvanishing $\Lambda$.  Nonetheless, we expect that any residual ambiguity can be resolved through further computations of boundary correlation functions of the graviton, \ie stress tensor correlators in the putative dual CFT.

\paragraph{A universal dimensionless parameter for flat space quantum gravity} As demonstrated in section~\ref{sec:entropies}, the effective number of degrees of freedom $\mathcal{N}_\text{dof}$ for Minkowski space black holes is not affected by the GB correction, taking the same form as in Einstein gravity,
\begin{equation}\label{eq:flat_Ndof_disc}
    \mathcal{N}_{\mathrm{dof}}(\mathcal{P}) =
\dfrac{\Sigma_{k,d-1}(2\pi)^{d-1}}{4 G_N\, \mathcal{P}^{d-1}} \ .
\end{equation}
Here, we would like to argue that this universality persists in arbitrary LL gravity theories.

Note the overall scaling behavior of the flat space conformal entropy~\eqref{eq:Ndof_Mink} can be understood through dimensional analysis.~In particular, in the case of Einstein gravity, the only possible dimensionless combination is $G_N \mathcal{K}^{d-1}$ and therefore it seems that all the coefficients of the thermal effective action~\eqref{eq:I_EFT} will have the form $c_i(d)/G_N \mathcal{K}^{d-1}$, with some dimension-dependent numerical coefficient $c_i(d)$.~That is true for the first three coefficients, as computed in \cite{Banihashemi:2025qqi}, and our argument suggests that it is true for every order in the small-$\tilde{\beta}$ expansion.

Let us now provide an argument for EMGB gravity, perturbatively in the coupling $\alpha$.~In the context of ECBC, again based on dimensional analysis to leading order in $\tilde{\beta}$, we have that $r_h\propto \tfrac{1}{\mathcal{P}\tilde{\beta}}$ and therefore the GB correction can only enter in the entropy~\eqref{eq:Swald} at $\mathcal{O}(\tilde{\beta}^{3-d})$ without ruining the scaling structure of the expansion~\eqref{eq:I_EFT}.~This is exactly what one observes in eq.~\eqref{eq:Ndof_Mink}.~This also implies that any other higher curvature correction to the action~\eqref{eq:full_action} with parametrically small couplings will not affect the leading GR contribution to $\mathcal{N}_{\text{dof}}(\mathcal{P})$~\eqref{eq:N_dof(K)_GR_cases}, which hence appears to be rather universal in flat space.

Continuing in this line of reasoning, since in EGB gravity the only dimensionless combination involving $\alpha$ (when $\Lambda=0$) is $\alpha \mathcal{P}^2$, we expect the overall structure of the conformal entropy to be 
\begin{equation}
    \mathcal{S}_{\text{conf}}^{(\text{EGB})} \propto\frac{1}{G_N\mathcal{P}^{d-1}\tilde{\beta}^{d-1}}\left(\sum_{n\geq0}c_{2n}(d)\tilde{\beta}^{2n}+\sum_{m\geq1}\tilde{c}_{2m}(d)(\alpha\mathcal{P}^2)^m\tilde{\beta}^{2m}\right) \ ,
\end{equation}
with again $c_i(d)$ and $\tilde{c}_j(d)$ being numerical coefficients that depend on the dimension.~Of course, the above arguments are based on the fact that in flat space, in contrast to (A)dS, there is no additional dimensionful parameter to combine with $\alpha$ and hence only the combination $\alpha\mathcal{P}^2$ can appear in the various expressions. 

In principle, the story could change when additional scales are present, as in the case of charged black holes.~That is because in the presence of \eg electric charge, there now exist possible dimensionless combinations involving the GB coupling and there does not seem to be an a priori reason for why these should not appear.~Nevertheless, we explicitly computed $\mathcal{N}_\text{dof}$ in the charged case in section~\ref{subsec:charged_conformal_thermo} and confirmed that it does not receive corrections from $\tilde\alpha$ in the flat space limit.

The arguments presented thus far regarding the universality of the leading EGB contribution are valid in a perturbative expansion in $\alpha$.~A more elaborate argument is required non-perturbatively, since \eg $r_h$ can be a general function of $\alpha\mathcal{P}^2$.~Generically that is indeed the case, however in the regime that we are studying one further has to take a small-$\tilde{\beta}$ limit.~This in turn effectively reduces the dependence of $r_h$ on $\alpha$ to its perturbative form, see \eg eq.~\eqref{eq:rh_flat_non-pert}.~One can also understand this universal behavior through a more physical argument.~In particular, recall that the flat space $\mathcal{N}_\text{dof}$ can be obtained as the $\mathcal{P}\ell\to\infty$ limit of the corresponding AdS expression.~This type of flat limit essentially amounts to zooming on a region deep into the AdS bulk, that looks like Minkowski space.~However, as our non-perturbative curves in figure~\ref{fig:non-pert_AdS} indicate, the GB correction is important only near the AdS boundary and becomes indistinguishable from Einstein gravity as we move radially inwards.~This implies that deep into the bulk the physics of conformal thermodynamics should be universal, coinciding with that of Einstein gravity.~This argument is general and thus we expect it to hold for arbitrary Lanczos-Lovelock gravity theories in general dimensions.

Overall, studying flat space with (E)CBC provides an additional scale \ie $\mathcal{P}$, which can be leveraged to construct a dimensionless combination \ie $G_N\mathcal{P}^{d-1}$.~This provides a natural parameter to expand around, that is otherwise absent in Minkowski space semiclassical quantum gravity.~Given its universality, the flat space $\mathcal{N}_\text{dof}$ of eq.~\eqref{eq:flat_Ndof_disc} might have interesting implications for a microscopic description of flat space,\footnote{Note that it is still possible to take the limit $G_N \P^{d-1} \to 0$ of the flat space conformal thermodynamics, in which case the cavity will fill the entire space.~That may provide contact with other approaches to flat space holography, based on null infinity.} and thus deserves further investigation.

\paragraph{Holography for finite regions} The extensivity of the high-temperature conformal thermodynamics subject to (E)CBC is hinting toward a dual field theoretic description that is local, in contrast to the $T\bar{T}$-deformed field theories dual to finite Dirichlet boundaries in AdS/CFT duality~\cite{Freidel:2008sh,McGough:2016lol,Zamolodchikov:2004ce,Smirnov:2016lqw,Hartman:2018tkw}.~Moreover, the extensive behavior is universal and does not depend on the sign of the cosmological constant. This curious observation might serve as a guiding principle towards finding microscopic realizations of these conformal boundaries. For example, ref.~\cite{Allameh:2025gsa} recently proposed a dual theory for $d=2$ with negative cosmological constant.~To gain intuition for the higher dimensional case, one could attempt to compute stress tensor and/or matter correlation functions inside the worldtube and interpret them in terms of boundary data. Note that, as discussed in \cite{Anninos:2024xhc}, the presence of the scale $\mathcal{P}$ might induce some sort of conformal symmetry breaking, at least when the boundary is sufficiently close to the conformal boundary of AdS. 

In addition, it would be interesting to study whether the familiar notions of causal \cite{Hamilton:2005ju} and entanglement \cite{Dong:2016eik} wedge reconstruction from AdS/CFT generalize to conformal boundaries placed at finite distance.~This could be achieved by leveraging the proposal for generalized entanglement wedges for gravitating spacetime regions, that was put forward in \cite{Bousso:2022hlz,Bousso:2023sya}. Ultimately, one would wish to understand quantum aspects of conformal timelike boundaries and their role within string/M-theory.~Liouville walls in the string target space discussed in \eg \cite{Silverstein:2022dfj}, seem to be an intriguing starting point for progress in this direction.

\section*{Acknowledgements}

It is a pleasure to thank Michael Anderson, Dionysios Anninos, Pau Figueras, Davide Gaiotto, Chawakorn Maneerat, Edgar Shaghoulian and Eva Silverstein  for useful discussions. TZ especially thanks Sergio Sanjurjo for illuminating discussions and comments on the manuscript. We would also like to thank all the participants of the workshop \href{https://scgp.stonybrook.edu/archives/45387}{Timelike Boundaries in Classical and Quantum Gravity} for stimulating discussions and the Simons Center for Geometry and Physics at Stony Brook University for its hospitality, during which this work was finalized. Research at Perimeter Institute is supported in part by the Government of Canada through the Department of Innovation, Science and Economic Development Canada and by the Province of Ontario through the Ministry of Colleges and Universities. The work of DAG is funded by UKRI Stephen Hawking Fellowship EP/W005530/1 ``Quantum Emergence of an Expanding Universe". DAG is further supported by STFC consolidated grant ST/X000753/1. RCM is also supported in part by a Discovery Grant from the Natural Sciences and Engineering Research Council of Canada, and by funding from the BMO Financial Group.

\appendix

\section{Boundary Terms for EMGB Gravity}\label{app:action_variations}
In this appendix we discuss various boundary conditions that yield a well-posed variational problem for EMGB gravity, once the bulk part of the action~\eqref{eq:full_action} is supplemented with appropriate boundary terms.
\subsection{Variation of the Einstein-Hilbert Action}

The Einstein-Hilbert (EH) action in Lorentzian signature takes the form\footnote{We work in Lorentzian signature, so that we can have general formulas both for timelike and spacelike hypersurfaces.~As usual, there is an overall minus sign in all the action expressions that follow if one wishes to work in Euclidean signature.}
\begin{equation}\label{eq:EH_action}
    \mathcal{I}_{\text{EH}}=\frac{1}{16\pi G_N}\int_{\mathcal{M}} d^{d+1}x \sqrt{-g}\left(-2\Lambda+R\right) \ .
\end{equation}
Varying this action with respect to the inverse metric $g^{\mu\nu}$ we obtain\footnote{In general, the action variation also contains a corner contribution uncovered by Hayward~\cite{Hayward:1993my}, whose coefficient was shown to vanish for conformal boundary conditions in~\cite{Odak:2021axr}.~We are going to ignore corner terms in what follows, since they are irrelevant for the Euclidean computations we are after in this work.}
\begin{equation}\label{eq:dI_EH}
   \delta \mathcal{I}_{\text{EH}}\hat{=} \frac{\epsilon}{16\pi G_N} \left( \int_{\partial\mathcal{M}}d^{d}y \sqrt{|h|} \left(\mathcal{K}_{\mu\nu}-\mathcal{K} h_{\mu\nu}\right) \delta h ^{\mu\nu} -2 \delta\left( \int_{\partial\mathcal{M}}d^{d}y \sqrt{|h|} \mathcal{K}\right) \right) \ ,
\end{equation}
where $\epsilon=1 (-1)$ if the boundary is timelike (spacelike)  and the overhats on equal signs denote on-shell variations, as usual.~Therefore, in order to have a well-posed variational principle, we have to impose boundary conditions, and add the corresponding boundary terms in the action. In the following, we examine the cases of interest.~As is standard in such manipulations, the following relations will be useful
\begin{equation}
    \begin{cases}
        \delta{h}_{\mu \nu}= - h_{\mu\sigma} h_{\nu\rho}   \delta h^{\rho\sigma} \ ,\\
        \delta \sqrt{|h|}=-\frac{1}{2}\sqrt{|h|} h_{\mu\nu}\delta h^{\mu\nu} \ .
    \end{cases}
\end{equation}

\paragraph{Dirichlet boundary conditions} The natural choice is to fix the induced metric along the boundary, \ie~impose Dirichlet boundary conditions $\delta h_{\mu\nu}|_{\partial\mathcal{M}}=0$. This makes the first term in the on-shell variation~\eqref{eq:dI_EH} to vanish and further instructs us to supplement the bulk EH action with the following Gibbons-Hawking-York (GHY) term~\cite{York:1972sj,Gibbons:1976ue},
\begin{equation}\label{eq:York}
    \mathcal{I}_{\text{GHY}}=\frac{\epsilon}{8\pi G_N} \int_{\partial\mathcal{M}} d^{d}y \sqrt{|h|}\mathcal{K} \ .
\end{equation}
Note that the numerical coefficient of this term is independent of the spacetime dimension.

\paragraph{Umbilic boundary conditions} From eq.~\eqref{eq:dI_EH} we see that specifying the boundary condition $\mathcal{K}_{\mu\nu}-\mathcal{K}h_{\mu\nu}|_{\partial\mathcal{M}}=0$ and supplementing the EH action with the boundary term~\eqref{eq:York} also yields a well-posed variational principle. Again, the coefficient of the York term is universal for all dimensions.

\paragraph{Neumann boundary conditions} Another possibility would be to fix the conjugate momentum on the boundary, which in ADM terms reads,
\begin{equation}
    \pi^{\mu\nu}:=\frac{\delta}{\delta h_{\mu\nu}}\left(\mathcal{I}_{\text{EH}}+\mathcal{I}_{\text{GHY}}\right)=-\frac{\epsilon \sqrt{|h|}}{16\pi G_N}\left(\mathcal{K}^{\mu\nu}-\mathcal{K}h^{\mu\nu}\right) \ .
\end{equation}
We can then massage the first term of eq.~\eqref{eq:dI_EH} as follows,
\begin{equation}
    \frac{\epsilon}{16\pi G_N} \int_{\partial\mathcal{M}}d^{d}y \sqrt{|h|} \left(\mathcal{K}_{\mu\nu}-\mathcal{K} h_{\mu\nu}\right) \delta h ^{\mu\nu} = -\frac{\epsilon\left(1-d\right)}{16\pi G_N}\delta\left(\int_{\partial\mathcal{M}}d^{d}y \sqrt{|h|} K\right)-\int_{\partial\mathcal{M}}d^{d}y\   h_{\mu\nu}\delta\pi^{\mu\nu} \ ,   
\end{equation}
and substituting back into eq.~\eqref{eq:dI_EH} we obtain,
\begin{equation}
    \delta\mathcal{I}_{\text{EH}}\hat{=}\frac{\epsilon\left(d-3\right)}{16\pi G_N}\delta\left(\int_{\partial\mathcal{M}} d^{d}y \sqrt{|h|}\mathcal{K}\right)-\int_{\partial\mathcal{M}}d^{d}y\   h_{\mu\nu}\delta\pi^{\mu\nu} \ ,   
\end{equation}
where we observe that subject to fixing $\delta\pi^{\mu\nu}|_{\partial\mathcal{M}}=0$ the boundary term of the EH action becomes
\begin{equation}\label{eq:York_NBCs}
   \mathcal{I}_{\text{NBCs}}^{(\text{GR})}= -\frac{\epsilon\left(d-3\right)}{16\pi G_N}\int_{\partial\mathcal{M}} d^{d}y \sqrt{|h|}\mathcal{K} \ .
\end{equation}
Unlike the previous cases, the coefficient of the GHY term now depends on the dimension and, interestingly, in $d=3$ vanishes altogether, while in $d=1$ coincides with the boundary term of the Dirichlet problem.

\paragraph{Conformal boundary conditions} In the case of CBC~\eqref{eq:CBCs}, one fixes the conformal class of the boundary metric and the trace of the extrinsic curvature; that is, one imposes $\delta h^{\mu\nu}|_{\partial \mathcal{M}}=\Omega^2(y)h^{\mu\nu}$ where $\Omega(y)$ is a boundary Weyl factor and fixes $\mathcal{K}$, respectively.~It is more useful, though, to work with $\tilde{h}^{\mu\nu}:= h^{-\frac{1}{d}}h_{\mu\nu}$ such that the requirement of fixing the conformal class reduces to $\delta\tilde{h}^{\mu\nu}|_{\partial \mathcal{M}}=0$.~Under a Weyl rescaling,
\begin{equation}\label{eq:Weyl_rescaling}
    h_{\mu\nu}\rightarrow \Omega^2(y)h_{\mu\nu} \ ,
\end{equation}
$\tilde{h}_{\mu\nu}$ remains invariant.~These boundary conditions can thus be equivalently phrased as fixing $\tilde{h}_{\mu\nu}|_{\partial \mathcal{M}}, \mathcal{K}|_{\partial\mathcal{M}}$.\footnote{Note that in this formulation we do not specify additional data than before, since by definition $\tilde{h}_{\mu\nu}$ is required to have unit determinant.} The variations of $h_{\mu\nu}$ and $\tilde{h}_{\mu\nu}$ are related via
\begin{equation}\label{eq:tilde_var}
\delta h_{\mu\nu}=h^{\frac{1}{d}}\delta \tilde{h}_{\mu\nu}+\frac{2}{d\sqrt{|h|}}h_{\mu\nu}\delta\sqrt{|h|} 
  \ .    
\end{equation}
Hence, the first term of eq.~\eqref{eq:dI_EH} takes the form,
\begin{equation}\label{eq:new_page}
\begin{aligned}
    \int_{\partial\mathcal{M}} d^{d}y \ \pi^{\mu\nu}\delta h_{\mu\nu}&=\int_{\partial\mathcal{M}} d^{d}y \ h^{\frac{1}{d}} \pi^{\mu\nu}\delta \tilde{h}_{\mu\nu}+\frac{1}{d}\int_{\partial\mathcal{M}} d^{d}y \ \pi^{\mu}_\mu  h^{\rho\sigma}\delta  h_{\rho\sigma} \,, \\
    &=\int_{\partial\mathcal{M}} d^{d}y \ h^{\frac{1}{d}} \pi^{\mu\nu}\delta \tilde{h}_{\mu\nu}+\frac{\epsilon(d-1)}{d8\pi G_N}\int_{\partial\mathcal{M}} d^{d}y \ \delta \sqrt{|h|} \mathcal{K} \ ,
\end{aligned}    
\end{equation}
and by virtue of a functional partial integration in field space, it becomes,
\begin{equation}\label{eq:conf_var}
  \int_{\partial\mathcal{M}}d^{d}y \  \tilde{\pi}^{\mu\nu}\delta \tilde{h}_{\mu\nu}+\int_{\partial\mathcal{M}}d^{d}y  \ \pi_{\mathcal{K}}\delta\mathcal{K} + \frac{\epsilon(d-1)}{d8\pi G_N}\delta\left(\int_{\partial\mathcal{M}} d^{d}y  \sqrt{|h|} \mathcal{K}\right) \ ,
\end{equation}
where we defined the canonical conjugate momenta as
\begin{equation}\label{eq:def_EH}
    \begin{cases}
        \tilde{\pi}^{\mu\nu} := -\dfrac{\epsilon\sqrt{|h|}}{16\pi G_N}h^{\frac{1}{d}}\left(\mathcal{K}^{\mu\nu}-\frac{1}{d}h^{\mu\nu}\mathcal{K}\right) \ , \\
        \pi_{\mathcal{K}}:=-\dfrac{\epsilon(d-1)}{d8\pi G_N} \sqrt{|h|} \ ,
    \end{cases} \ ,
\end{equation}
and in the first line we made the conjugate momentum traceless by adding a term proportional to $h^{\mu\nu}\delta\tilde{h}_{\mu\nu}$, which is zero since
\begin{equation}
    h^{\rho\sigma}\delta h_{\rho\sigma}=h^{\rho\sigma}\left(h^{\frac{1}{d}}\delta \tilde{h}_{\rho\sigma} +\frac{1}{d}h_{\rho\sigma}h^{\alpha\beta}\delta h_{\alpha\beta}\right)=h^{\frac{1}{d}}h^{\mu\nu}\delta \tilde{h}_{\mu\nu}+h^{\rho\sigma}\delta h_{\rho\sigma} \ .
\end{equation}
In other words, $\tilde{\pi}^{\mu\nu}$ is contracted with $\delta\tilde{h}_{\mu\nu}$ which projects to the traceless part of the corresponding tensor.~Inserting~\eqref{eq:conf_var} back in the variation~\eqref{eq:dI_EH} and imposing conformal boundary conditions, we obtain that
\begin{equation}
    \delta\mathcal{I}_{\text{EH}}\hat{=}-\frac{\epsilon}{8d\pi G_N}\delta\left(\int_{\partial\mathcal{M}} d^{d}y  \sqrt{|h|} \mathcal{K}\right) \ .
\end{equation}
Consequently, the variation of the bulk EH action has to be compensated by the variation of the following boundary term,
\begin{equation}
    \mathcal{I}_{\text{CBC}}=\frac{\epsilon}{8d\pi G_N}\int_{\partial\mathcal{M}} d^{d}y  \sqrt{|h|} \mathcal{K} \ ,
\end{equation}
where the coefficient depends on the spacetime dimension.

\paragraph{Generalized boundary conditions} In ref.~\cite{Liu:2024ymn} the authors derived the necessary boundary conditions, so that the EH action has a well-posed variational problem when it is accompanied by a GHY term with an arbitrary numerical pre-factor $\Theta$, \ie
\begin{equation}\label{eq:York_arbitrary}
    \mathcal{I}_{\text{GCBC}}^{(\text{GR})}=\frac{\epsilon\Theta}{8\pi G_N}\int_{\partial \mathcal{M}}d^{d}y \sqrt{|h|} \mathcal{K} \ .
\end{equation}
Adding such a term, the on-shell action variation~\eqref{eq:dI_EH} becomes
\begin{equation}
\begin{aligned}
    \delta \mathcal{I}_{\text{EH}} & \hat{=}  \frac{\epsilon}{16\pi G_N} \left( \int_{\partial\mathcal{M}}d^{d}y \sqrt{|h|} \left(\mathcal{K}_{\mu\nu}-\mathcal{K} h_{\mu\nu}\right) \delta h ^{\mu\nu} -2(1-\Theta) \delta\left( \int_{\partial\mathcal{M}}d^{d}y \sqrt{|h|} \mathcal{K}\right) \right) \\
     & \hat{=} -\frac{\epsilon}{16\pi G_N} \left( \int_{\partial\mathcal{M}}d^{d}y\, \tilde{\pi}^{\mu\nu} \delta \tilde{h}_{\mu\nu}+2\left(1-\Theta-\frac{d-1}{d}\right)\mathcal{K}\delta\sqrt{|h|}+2(1-\Theta)\sqrt{|h|}\delta \mathcal{K} \right) \ .
\end{aligned}
\end{equation}
The main idea is then to recognize that the last two terms can be written as
\begin{equation}
    \frac{1-\Theta}{8\pi G_N}\int_{\partial \mathcal{M}} d^{d}y \sqrt{|h|}h^{-p}\delta(h^p\mathcal{K}) \ ,
\end{equation}
for $p=\frac{1}{2(1-\Theta)}(1-\Theta-\frac{d-1}{d})$.~Thus, the generalized boundary conditions (GBCs) amount to
\begin{equation}\label{eq:GBCs}
    \text{generalized boundary conditions}: \ \lbrace [h_{mn}]|_{\partial\mathcal{M}},h^p\mathcal{K}|_{\partial\mathcal{M}}\rbrace \ ,
\end{equation}
or alternatively, 
\begin{equation}\label{eq:UGBCs}
    \text{umbilic generalized boundary conditions}: \ \lbrace \tilde{\pi}^{\mu\nu}|_{\partial\mathcal{M}}=0,h^p\mathcal{K}|_{\partial\mathcal{M}}\rbrace \ ,
\end{equation}
for $\Theta=1+\frac{d-1}{d(2p-1)}$.~One recovers Dirichlet and conformal boundary conditions, respectively, in the limits $p\rightarrow\infty$ and $p\rightarrow0$.~In~\cite{Liu:2024ymn} GBCs were shown to be elliptic in Euclidean signature, for any finite $p$.

\subsection{Variation of the Gauss-Bonnet Action}

Suppose now that we add in~\eqref{eq:EH_action} a Gauss-Bonnet term,
\begin{equation}\label{eq:I_grav}
\mathcal{I}_{\text{GB}}=\frac{\alpha}{(d-2)(d-3)16\pi G_N}\int_{\mathcal{M}} d^{d+1}x \sqrt{-g}\left(R^{\mu\nu\rho\sigma}R_{\mu\nu\rho\sigma}-4R^{\mu\nu}R_{\mu\nu}+R^2\right)\ .
\end{equation}
Recall that the coupling $\alpha$ is dimensionful and appropriately normalized by a factor of $1/32\pi^2$ so that $\mathcal{I}_{\text{GB}}$ reduces to the Euler character in $4$-dimensions. The variation of the Gauss-Bonnet term turns out to be \cite{Bunch1981,Myers:1987yn,Davis:2002gn,Deruelle:2017xel},\footnote{We again ignore corner terms, for a treatment see ref.~\cite{Chakraborty:2017zep}.}
\begin{equation}\label{eq:dI_GB}
\begin{aligned}
\frac{(d-2)(d-3)16\pi G_N}{\alpha}\delta\mathcal{I}_{\text{GB}}=\int_\mathcal{M}d^{d+1}x\sqrt{-g} \ \mathcal{E}_{\mu\nu}\delta g^{\mu\nu} &+\epsilon\int_{\partial\mathcal{M}}d^{d}y\sqrt{|h|} \ \mathcal{B}_{\mu\nu}\delta h^{\mu\nu}\\
&-\epsilon\delta\left(\int_{\partial\mathcal{M}}d^{d}y\sqrt{|h|} \ \mathcal{Q}\right) \ ,
\end{aligned}
\end{equation}
where:
\begin{equation}\label{eq:GB_relations}
\begin{cases}
    \mathcal{E}_{\mu\nu}=2\left(R_\mu^{\alpha\beta\gamma}R_{\nu\alpha\beta\gamma}-2R^{\alpha\beta}R_{\mu\alpha\nu\beta}-2R_\mu^\alpha R_{\nu\alpha}+RR_{\mu\nu}\right)-\frac{1}{2}g_{\mu\nu}\left(R^{\mu\nu\rho\sigma}R_{\mu\nu\rho\sigma}-4R^{\mu\nu}R_{\mu\nu}+R^2\right) \\
    \ \ \ \ \  \equiv2R_\mu^{\alpha\beta\gamma}P_{\nu\alpha\beta\gamma}-\frac{1}{2}g_{\mu\nu}R^{\alpha\beta\gamma\delta}P_{\alpha\beta\gamma\delta} \ ,\\
   \mathcal{B}_{\mu\nu}=2\left(3J_{\mu\nu}-Jh_{\mu\nu}-2\tilde{P}_{\nu\alpha\mu\beta}\mathcal{K}^{\alpha\beta}\right) \ \ \ \text{or} \ \ \ \mathcal{B}^i_j =2\delta^{i\mu\nu\rho}_{j\alpha\beta\gamma} \mathcal{K}^\alpha_\mu\left(\frac{1}{2}\tilde{R}^{\beta\gamma}_{\nu\rho}-\frac{\epsilon}{3}\mathcal{K}^\beta_\nu\mathcal{K}^\gamma_\rho\right)\ ,\\
   \mathcal{Q}=4\left(J-2\tilde{G}^{\mu\nu}\mathcal{K}_{\mu\nu}\right)= 4 \delta^{\mu\nu\rho}_{\alpha\beta\gamma} \mathcal{K}^\alpha_\mu\left(\frac{1}{2}\tilde{R}^{\beta\gamma}_{\nu\rho}-\frac{\epsilon}{3}\mathcal{K}^\beta_\nu\mathcal{K}^\gamma_\rho\right) \ ,\\
   P_{\alpha\beta\gamma\delta}=R_{\alpha\beta\gamma\delta}+2R_{\beta[\gamma}g_{\delta]\alpha}-2R_{\alpha[\gamma}g_{\delta]\beta}+Rg_{\alpha[\gamma}g_{\delta]\beta} \ ,\\
   J_{\mu\nu}=\epsilon\left(\frac{2}{3}\mathcal{K}\mathcal{K}_{\mu\alpha}\mathcal{K}^\alpha_\nu+\frac{1}{3}\mathcal{K}_{\mu\nu}\left(\mathcal{K}^{\alpha\beta}\mathcal{K}_{
\alpha\beta}-\mathcal{K}^2\right)-\frac{2}{3}\mathcal{K}_{\mu\alpha}\mathcal{K}^{\alpha\beta}\mathcal{K}_{\beta\nu}\right) \ ,
\end{cases} 
\end{equation}
with the tildes again denoting that the corresponding tensor is computed for the induced metric $h_{\mu\nu}$ and $\delta^{ij\dots}_{k\ell\dots}$ being the generalized Kronecker delta defined as
\begin{equation}
    \delta^{\mu_1 \dots \mu_p}_{\nu_1 \dots \nu_p} =
\sum_{\sigma \in S_p} \operatorname{sgn}(\sigma)\,
\delta^{\mu_1}_{\nu_{\sigma(1)}} \cdots \delta^{\mu_p}_{\nu_{\sigma(p)}} =
\begin{vmatrix}
\delta^{\mu_1}_{\nu_1} & \cdots & \delta^{\mu_1}_{\nu_p} \\
\vdots & \ddots & \vdots \\
\delta^{\mu_p}_{\nu_1} & \cdots & \delta^{\mu_p}_{\nu_p}
\end{vmatrix} \ .
\end{equation}
As in the pure GR case, one can explore a variety of possible boundary conditions.

\paragraph{Dirichlet boundary conditions} We can again impose Dirichlet boundary conditions $\delta h_{\mu\nu}|_{\partial\mathcal{M}}=0$, such that the action has to be supplemented by the Bunch-Myers (BM) term~\cite{Bunch1981,Myers:1987yn},
\begin{equation}\label{eq:bdyRob}
    \mathcal{I}_{\text{BM}}=\frac{\epsilon\alpha}{(d-2)(d-3)16\pi G_N} \int_{\partial\mathcal{M}}d^{d}y \sqrt{|h|}\mathcal{Q} \ .
\end{equation}
As in GR, the coefficient of this boundary term is the same irrespective of dimension, apart from the particular dimension-dependent normalization of the GB coupling we have chosen.

\paragraph{Umbilic boundary conditions} Merging eqs.~\eqref{eq:dI_EH} and~\eqref{eq:dI_GB} one observes that fixing the combination
\begin{equation}
    \mathcal{K}_{\mu\nu}-\mathcal{K}h_{\mu\nu}+\frac{\alpha}{(d-2)(d-3)}\mathcal{B}_{\mu\nu}\Big|_{\partial\mathcal{M}}=0 \ ,
\end{equation}
while choosing again eqs.~\eqref{eq:York} and~\eqref{eq:bdyRob} as the boundary terms, the variational principle is well-posed.

\paragraph{Neumann boundary conditions} As in the instance of pure GR, one could try to fix the corresponding canonically conjugate momentum, which in the GB case reads
\begin{equation}
    \slashed{\mathcal{B}}^{\mu\nu}:=\frac{\delta}{\delta h_{\mu\nu}}\left(\mathcal{I}_{\text{GB}}+\mathcal{I}_{\text{BM}}\right)=-\frac{\epsilon \alpha\sqrt{|h|}}{(d-2)(d-3)16\pi G_N}\mathcal{B}^{\mu\nu} \ .
\end{equation}
Therefore, eq.~\eqref{eq:dI_GB} becomes ($\mathcal{B}:=h^{\mu\nu}\mathcal{B}_{\mu\nu}$),
\begin{equation}\label{eq:Neum_GB}
\begin{aligned}
    \delta\mathcal{I}_{\text{GB}}+\frac{\epsilon\alpha}{(d-2)(d-3)16\pi G_N}\delta\left(\int_{\partial\mathcal{M}}d^{d}y \sqrt{|h|}\mathcal{Q}\right)&=-\epsilon\int_{\partial\mathcal{M}}d^{d}y \slashed{\mathcal{B}}^{\mu\nu}\delta h_{\mu\nu}\\
    &=-\epsilon\delta\int_{\partial\mathcal{M}}d^{d}y \slashed{\mathcal{B}}+\epsilon\int_{\partial\mathcal{M}}d^{d}y h_{\mu\nu}\delta\slashed{\mathcal{B}}^{\mu\nu} \ ,
\end{aligned}    
\end{equation}
but,
\begin{equation}
    \mathcal{B}=2\left(3J-dJ-2\tilde{P}^\mu_{\alpha\mu\beta}\mathcal{K}^{\alpha\beta}\right)\equiv\frac{3-d}{2}\mathcal{Q} \ ,
\end{equation}
since
\begin{equation}
\begin{aligned}    \tilde{P}^\mu_{\alpha\mu\beta}&=\tilde{R}^\mu_{\alpha\mu\beta}+2\tilde{R}_{\alpha[\mu}h_{\beta]}^\mu-2\tilde{R}^\mu_{[\mu}h_{\beta]\alpha}+\tilde{R}h^\mu_{[\mu}h_{\beta]\alpha}\\
&=\tilde{R}_{\alpha\beta}+\left(\tilde{R}_{\alpha\mu}\delta^\mu_\beta-\tilde{R}_{\alpha\beta}\delta^\mu_\mu\right)-\left(\tilde{R}^\mu_{\mu}h_{\beta\alpha}-\tilde{R}^\mu_\beta h_{\mu\alpha}\right)+\frac{\tilde{R}}{2}\left(\delta^\mu_\mu h_{\beta\alpha}-\delta^\mu_\beta h_{\mu\alpha}\right)\\
&=(3-d)\left(\tilde{R}_{\alpha\beta}-\frac{1}{2}h_{\alpha\beta}\tilde{R}\right) \ .
\end{aligned}
\end{equation}
Consequently, the trace of the GB momentum is,
\begin{equation}\label{eq:B_trace}
    \slashed{\mathcal{B}}=\frac{\epsilon \alpha\sqrt{|h|}}{(d-2)32\pi G_N}\mathcal{Q} \ ,
\end{equation}
and thus the on-shell variation~\eqref{eq:Neum_GB} becomes,
\begin{equation}
\begin{aligned}
    \delta\mathcal{I}_{\text{GB}}+\frac{\epsilon\alpha}{(d-2)(d-3)16\pi G_N}\delta\left(\int_{\partial\mathcal{M}}d^{d}y \sqrt{|h|}\mathcal{Q}\right)\hat{=}&-\frac{\epsilon\alpha}{(2-d)32\pi G_N}\delta\left(\int_{\partial\mathcal{M}}d^{d}y\sqrt{|h|}\mathcal{Q}\right)\\
    &-\epsilon\int_{\partial\mathcal{M}}d^{d}y h_{\mu\nu}\delta\slashed{\mathcal{B}}^{\mu\nu} \ .
    \end{aligned}
    \end{equation}
Hence,
\begin{equation}
    \delta\mathcal{I}_{\text{GB}}\hat{=}-\frac{\epsilon(5-d)\alpha}{(d-2)(d-3)32\pi G_N}\delta\left(\int_{\partial\mathcal{M}}d^{d}y\sqrt{|h|}\mathcal{Q}\right) \ ,
\end{equation}
which implies that the appropriate addition to the bulk action for Neumann boundary conditions, is,
\begin{equation}
    \mathcal{I}_{\text{NBCs}}^{(\text{GB})}=-\frac{\epsilon(d-5)\alpha}{(d-2)(d-3)32\pi G_N}\int_{\partial\mathcal{M}}d^{d}y\sqrt{|h|}\mathcal{Q} \ .
\end{equation}
The boundary term does not appear in $d=5$, as did the York term for NBCs in GR~\eqref{eq:York_NBCs} in $d=3$.~This is a general pattern for Lovelock gravity theories, namely the coefficient of the boundary term for the generalized Neumann boundary conditions vanishes in $d=2n+1,\,n\in\mathbb{N}^\star$, when the Lovelock action contains the Euler density of $d=2n-1$.

\paragraph{Extended conformal boundary conditions} Focusing on the conformal case, the second term of the variation~\eqref{eq:dI_GB} becomes (using eq.~\eqref{eq:tilde_var}),
\begin{equation}\label{eq:some_comp}
    \begin{aligned}
        \int_{\partial\mathcal{M}}d^{d}y \ \slashed{\mathcal{B}}^{\mu\nu}\delta h_{\mu\nu} &=\int_{\partial\mathcal{M}}d^{d}y \ h^{\frac{1}{d}} \slashed{\mathcal{B}}^{\mu\nu}\delta\tilde{h}_{\mu\nu}+\int_{\partial\mathcal{M}}d^{d}y\frac{2}{d\sqrt{|h|}} \slashed{\mathcal{B}}\delta\sqrt{|h|}\\
        &=\int_{\partial\mathcal{M}}d^{d}y \ h^{\frac{1}{d}} \slashed{\mathcal{B}}^{\mu\nu}\delta\tilde{h}_{\mu\nu}+\frac{\alpha}{d(d-2)16\pi G_N}\int_{\partial\mathcal{M}}d^{d}y \ \mathcal{Q}\delta\sqrt{|h|}\\
        =\int_{\partial\mathcal{M}}d^{d}y & \ h^{\frac{1}{d}} \slashed{\mathcal{B}}^{\mu\nu}\delta\tilde{h}_{\mu\nu}+\frac{\alpha}{d(d-2)16\pi G_N}\left[\delta\left(\int_{\partial\mathcal{M}}d^{d}y \ \mathcal{Q}\sqrt{|h|}\right)-\int_{\partial\mathcal{M}}d^{d}y \sqrt{|h|}\delta \mathcal{Q}\right]\ .
    \end{aligned}
\end{equation}
Consequently, from eq.~\eqref{eq:Neum_GB} we have (omitting for a moment the terms fixed by the boundary condition $\delta\tilde{h}_{\mu\nu}|_{\partial\mathcal{M}}=0$),
\begin{equation}
    \delta\mathcal{I}_{\text{EGB}}+\frac{\epsilon\alpha}{(d-2)(d-3)16\pi G_N}\delta\left(\int_{\partial\mathcal{M}}d^{d}y \sqrt{|h|}\mathcal{Q}\right)\hat{=}\frac{\epsilon\alpha}{d(d-2)16\pi G_N}\delta\left(\int_{\partial\mathcal{M}}d^{d}y \ \mathcal{Q}\sqrt{|h|}\right) \ ,
    \end{equation}
which, in turn, implies
\begin{equation}
\delta\mathcal{I}_{\text{GB}}\hat{=}-\frac{3\epsilon\alpha}{d(d-2)(d-3)16\pi G_N}\delta\left(\int_{\partial\mathcal{M}}d^{d}y \sqrt{|h|}\mathcal{Q}\right)\ .
\end{equation}
Re-instating all the terms
\begin{equation}\label{eq:ap_variations_diff}
    \delta\mathcal{I}_{\text{EGB}}\hat{=}\int_{\partial\mathcal{M}}d^{d}y  \tilde{\slashed{\mathcal{B}}}^{\mu\nu}\delta\tilde{h}_{\mu\nu} +\int_{\partial\mathcal{M}}d^{d}y \ \pi_{\mathcal{Q}}\delta\mathcal{Q} -\frac{3\epsilon\alpha}{d(d-2)(d-3)16\pi G_N}\delta\left(\int_{\partial\mathcal{M}}d^{d}y \sqrt{|h|}\mathcal{Q}\right) \ ,
\end{equation}
where we defined
\begin{equation}\label{eq:def_GB}
    \begin{cases}
        \tilde{\slashed{\mathcal{B}}}^{\mu\nu}:=h^{\frac{1}{d}} \left(\slashed{\mathcal{B}}^{\mu\nu}-\dfrac{1}{d}h^{\mu\nu}\slashed{\mathcal{B}}\right) \ , \\
        \pi_\mathcal{Q}:=- \dfrac{\epsilon\alpha\sqrt{|h|}}{d(d-2)16\pi G_N} \ .
    \end{cases} 
\end{equation}
Therefore, the overall on-shell variation of the EH and GB actions reads,
\begin{equation}
\begin{aligned}
    \delta\mathcal{I}_{\text{EGB}}&\hat{=}\int_{\partial\mathcal{M}}d^{d}y  \left(\tilde{\pi}^{\mu\nu}+\tilde{\slashed{\mathcal{B}}}^{\mu\nu}\right)\delta\tilde{h}_{\mu\nu}+\int_{\partial\mathcal{M}}d^{d}y \ \left(\pi_\mathcal{K}\delta\mathcal{K}+ \pi_{\mathcal{Q}}\delta\mathcal{Q}\right)\\
    &-\frac{\epsilon}{d8\pi G_N}\delta\left(\int_{\partial\mathcal{M}} d^{d}y  \sqrt{|h|} \mathcal{K}\right)-\frac{3\epsilon\alpha}{d(d-2)(d-3)16\pi G_N}\delta\left(\int_{\partial\mathcal{M}}d^{d}y \sqrt{|h|}\mathcal{Q}\right) \ ,
\end{aligned}    
\end{equation}
and using eqs.~\eqref{eq:def_EH}, \eqref{eq:def_GB},
\begin{equation}
\begin{aligned}
    \delta\mathcal{I}_{\text{EGB}}\hat{=}&\int_{\partial\mathcal{M}}d^{d}y  \left(\tilde{\pi}^{\mu\nu}-\tilde{\slashed{\mathcal{B}}}^{\mu\nu}\right)\delta\tilde{h}_{\mu\nu}-\frac{\epsilon}{d16\pi G_N}\int_{\partial\mathcal{M}}d^{d}y\sqrt{|h|} \ \left(2(d-1)\delta\mathcal{K}+ \frac{\alpha}{d-2}\delta\mathcal{Q}\right)\\
    &-\frac{\epsilon}{d8\pi G_N}\delta\left(\int_{\partial\mathcal{M}} d^{d}y  \sqrt{|h|} \mathcal{K}\right)-\frac{3\epsilon\alpha}{d(d-2)(d-3)16\pi G_N}\delta\left(\int_{\partial\mathcal{M}}d^{d}y \sqrt{|h|}\mathcal{Q}\right) \ .
\end{aligned}    
\end{equation}
This last relation implies that we have to modify the GR conformal boundary conditions~\eqref{eq:CBCs}, where we again fix the conformal class of the induced metric along the boundary but now we fix the following scalar quantity,
\begin{equation}
   \mathcal{P}:= \mathcal{K}+ \frac{\alpha}{2(d-1)(d-2)}\mathcal{Q} \ ,
\end{equation}
which generalizes the trace of the extrinsic curvature, by adding a suitable higher curvature correction, \ie
\begin{equation}
    \mathcal{Q}=4\left(\mathcal{K}\mathcal{K}^{\mu\nu}\mathcal{K}_{\mu\nu}-\frac{2}{3}\mathcal{K}^{\mu\nu}\mathcal{K}_{\nu\rho}\mathcal{K}^{\rho}_{\mu}-\frac{1}{3}\mathcal{K}^3-2\tilde{G}^{\mu\nu}\mathcal{K}_{\mu\nu}\right) \ ,
\end{equation}
leading to the ECBC~\eqref{eq:ECBCs}.~Then, apart from the boundary term for CBC, one has to further add the following boundary term
\begin{equation}
    \mathcal{I}_{\text{ECBC}}=\frac{3\epsilon\alpha}{d(d-2)(d-3)16\pi G_N}\int_{\partial\mathcal{M}}d^{d}y \sqrt{|h|}\mathcal{Q} \ ,
\end{equation}
to the bulk EGB action, in order to ensure a well-posed variational problem subject to the ECBC.

\paragraph{Extended generalized boundary conditions} There exists a GB generalization of the  one parameter family of GBCs of ref.~\cite{Liu:2024ymn}.~In particular, plugging an arbitrary coefficient $\Phi$ in front of the BM term
\begin{equation}
    \mathcal{I}_{\text{GCBC}}^{(\text{GB})}=\frac{\epsilon\Phi}{16\pi G_N}\int_{\partial\mathcal{M}} d^{d}y\sqrt{|h|}\mathcal{Q} \ ,
\end{equation}
and also in the York term~\eqref{eq:York_arbitrary}, the full on-shell variation of EGB gravity takes the form:
\begin{equation}\label{eq:Big_variation}
   \begin{aligned}
    \delta\mathcal{I}_{\text{EGB}}\hat{=}-\frac{\epsilon}{16\pi G_N}&\int_{\partial\mathcal{M}}d^{d}y \Bigg[\tilde{\Pi}^{\mu\nu}\delta\tilde{h}_{\mu\nu}
    +2\left(1-\Theta-\frac{d-1}{d}\right)\mathcal{K}\delta\sqrt{|h|}+2(1-\Theta)\sqrt{|h|}\delta\mathcal{K}\\
    &+\frac{\alpha}{(d-2)(d-3)}\left(\left(1-\Phi-\frac{d-3}{d}\right)\mathcal{Q}\delta\sqrt{|h|}+(1-\Phi)\sqrt{|h|}\delta\mathcal{Q} \right)  \Bigg] \ ,
    \end{aligned}
\end{equation}
where $\tilde{\Pi}^{\mu\nu}=\tilde{\pi}^{\mu\nu}+\tfrac{\alpha}{(d-2)(d-3)}\tilde{\slashed{\mathcal{B}}}^{\mu\nu}$.~Upon setting
\begin{equation}
    \Phi=\frac{\Theta(d-3)+2}{d-1} \ ,
\end{equation}
and fixing the conformal class $\delta\tilde{h}_{\mu\nu}|_{\partial\mathcal{M}}=0$, the variation~\eqref{eq:Big_variation} can be written as
\begin{equation}
    \delta\mathcal{I}_{\text{EGB}}\hat{=}-\epsilon\frac{1-\Theta}{8\pi G_N}\int_{\partial \mathcal{M}} d^{d}y \sqrt{|h|}h^{-p}\delta(h^p\mathcal{P}) \ ,
\end{equation}
again with $p=\frac{1}{2(1-\Theta)}(1-\Theta-\frac{d-1}{d})$.~Therefore, the extended generalized boundary conditions (EGBCs) have the following quantities fixed,
\begin{equation}\label{eq:EGBCs}
    \text{extended generalized boundary conditions}: \ \lbrace [h_{mn}]|_{\partial\mathcal{M}},h^p\mathcal{P}|_{\partial\mathcal{M}}\rbrace \ ,
\end{equation}
or 
\begin{equation}\label{eq:EUGBCs}
    \text{extended umbilic generalized boundary conditions}: \ \lbrace \tilde{\Pi}^{\mu\nu}|_{\partial\mathcal{M}}=0,h^p\mathcal{P}|_{\partial\mathcal{M}}\rbrace \ ,
\end{equation}
when the total boundary term supplementing the EGB bulk action is given by
\begin{equation}
    \mathcal{I}_{\text{EGBCs}}=\frac{\epsilon}{16\pi G_N}\int_{\partial \mathcal{M}}d^{d}y \sqrt{|h|}\left( 2\Theta \mathcal{K}+\frac{\alpha}{(d-2)(d-3)}\Phi \mathcal{Q}\right)  \ ,
\end{equation}
with
\begin{equation}
    \begin{cases}
        \Theta=1+\dfrac{d-1}{d(2p-1)} \ ,\\
        \Phi=1+\dfrac{d-3}{d(2p-1)} \ .
    \end{cases}
\end{equation}
For $p=0$ one recovers the ECBC~\eqref{eq:ECBCs}; for $p\rightarrow\infty$, the Dirichlet ~\eqref{eq:York} and~\eqref{eq:bdyRob} terms. In the limit $\alpha \rightarrow0$ one goes back to the boundary conditions~\eqref{eq:GBCs} and~\eqref{eq:UGBCs}.

\subsection{Variation of the Maxwell Action} \label{MaxApp}
In our conventions~\eqref{eq:full_action}, the Lorentzian Maxwell action reads
\begin{equation}
    \mathcal{I}_{\text{M}}=-\frac{1}{64\pi G_N}\int_\mathcal{M}d^{d+1}x\sqrt{-g}F^{\mu\nu}F_{\mu\nu} \ .
\end{equation}
For completeness, let us derive the boundary terms.~We are going to use form notation, in which the curvature is $F=dA$ and the above action takes the form
\begin{equation}
    \mathcal{I}_{\text{M}}=-\frac{1}{32\pi G_N}\int_\mathcal{M}F\wedge\star F \ .
\end{equation}
Looking at gauge field variations of the form $A\rightarrow A+\delta A$, the variation of the action to leading order becomes
\begin{equation}\label{eq:Maxwell_var}
    \delta\mathcal{I}_\text{M}=-\frac{1}{16\pi G_N}\int_\mathcal{M}d \delta A\wedge\star F=-\frac{1}{16\pi G_N}\left(\epsilon\int_{\partial\mathcal{M}}\delta A\wedge\star F - \int_\mathcal{M} \delta A\wedge d\star F\right) \ ,
\end{equation}
where we used the standard graded Leibniz rule ($a$ is a $p$-form and $b$ a $q$-form)
\begin{equation}
    d(a\wedge\star b)=da\wedge\star b+(-1)^p a\wedge d\star b \ .
\end{equation}
On-shell of the equations of motion $d\star F=0$, the variation~\eqref{eq:Maxwell_var} becomes
\begin{equation}\label{eq:Max_var_part}
    \delta\mathcal{I}_\text{M}\hat{=-}\frac{\epsilon}{16\pi G_N}\int_{\partial\mathcal{M}}\delta A\wedge\star F \ ,
\end{equation}
which implies that imposing Dirichlet boundary conditions on the gauge field, \ie $\delta A|_{\partial\mathcal{M}}=0$, yields a well-posed variational problem without the need to add any boundary terms to the bulk action.~In the context of black hole thermodynamics, this corresponds to working in the grand-canonical ensemble, since the boundary potential is fixed.~However, one might wish to fix the charge along $\partial\mathcal{M}$, in which case the thermodynamic ensemble is the canonical one.~That can be accomplished by doing a functional integration by parts on eq.~\eqref{eq:Max_var_part}, as follows
\begin{equation}\label{eq:Max_bdy_var}
    \delta\mathcal{I}_\text{M}\hat{=}-\frac{\epsilon}{16\pi G_N}\delta\left(\int_{\partial\mathcal{M}} A\wedge\star F\right)+\frac{\epsilon}{16\pi G_N}\int_{\partial\mathcal{M}}A\wedge\delta(\star F) \ .
\end{equation}
Hence, if one supplements the Maxwell action with the following boundary term
\begin{equation}
    \mathcal{I}_\text{M}^{(\partial\mathcal{M})}=\frac{\epsilon}{16\pi G_N}\int_{\partial\mathcal{M}} A\wedge\star F \equiv  \frac{\epsilon}{16\pi G_N}\int_{\partial\mathcal{M}}d^{d}y\sqrt{|h|}F^{\mu\nu}n_\mu A_\nu \ ,
\end{equation}
then from the remaining term of eq.~\eqref{eq:Max_bdy_var}, the required boundary condition for a good variational problem is $\delta(\sqrt{|h|}F^{\mu\nu} n_\mu)|_{\partial\mathcal{M}}=0$.~This corresponds to fixing the electric charge, since
\begin{equation}
    Q_e=\int_{\partial\mathcal{M}} \star F \ .
\end{equation}
From this derivation it is also evident that boundary conditions on the metric and the gauge field can be imposed independently.

\section{Local and Covariant Boundary Terms for CBC}\label{app:no_K_bdy_term}

In this appendix we demonstrate that there does not exist a local and covariant boundary term which yields a well-defined variational principle for the bulk action~\eqref{eq:full_action} subject to the CBCs~\eqref{eq:CBCs}.~For simplicity, we restrict to Euclidean signature.

To begin, note that after taking into account the Gauss-Codazzi equations, there exist nine independent local and covariant boundary terms one can write down, whose integrands take the form~\cite{Bunch1981} (again curvature tensors with overhead tilde are evaluated for the boundary metric)
\begin{equation}\label{eq:nine_int}
    \mathcal{K}^3\ , \  \ \mathcal{K} \mathcal{K}^{ij}\mathcal{K}_{ij} \ , \ \mathcal{K}^{ij}\mathcal{K}_{jk}\mathcal{K}^k_i \ , \ \tilde{R}\mathcal{K} \ , \ \mathcal{K}^{ij}\tilde{R}_{ij} \ ,  \  D^iD_i\mathcal{K} \ , \ D_iD_j\mathcal{K}^{ij} \ , \ \mathcal{K}\mathcal{L}_n \mathcal{K} \ , \ \mathcal{K}^{ij} \mathcal{L}_n \mathcal{K}_{ij} \ , \\
\end{equation}
where $\mathcal{L}_n$ is the Lie derivative with respect to the normal $n$ to the boundary $\partial\mathcal{M}$.~As noted \eg in ref~\cite{Bunch1981}, the variations of $D^iD_i\mathcal{K} ,\, D_iD_j\mathcal{K}^{ij}$ yield third order derivatives of the metric variation, and there is no linear combination of the two or some contribution from the variations of any of the other of the seven terms of eq.~\eqref{eq:nine_int} that can cancel them.~Hence, these cannot be included in the boundary term.~A similar argument shows that the variations of the last two terms in eq.~\eqref{eq:nine_int} cannot be canceled either.

The above imply that we are left with five possible local and covariant curvature quantities, and we will consider the general boundary term constructed out of them, which takes the following form:
\begin{equation}\label{eq:gen_bdy_term}
    \mathcal{I}_{\partial\mathcal{M}}=\frac{\alpha}{(d-2)(d-3)16\pi G_N}\int_{\partial\mathcal{M}} d^dy\sqrt{h}\Big(c_1\mathcal{K}^3 +c_2\mathcal{K} \mathcal{K}^{ij}\mathcal{K}_{ij}+ c_3\mathcal{K}^{ij}\mathcal{K}_{jk}\mathcal{K}^k_i+ c_4\tilde{R}\mathcal{K}+ c_5\mathcal{K}^{ij}\tilde{R}_{ij}\Big) \ .
\end{equation}
Here, $c_i$ are arbitrary numerical coefficients, that possibly depend on the dimension $d$.

We would now like to examine whether there exists a combination of the coefficients $c_i$, such that the variation of eq.~\eqref{eq:gen_bdy_term} cancels the boundary pieces of the on-shell variation of the bulk E(M)GB action after imposing CBC~\eqref{eq:CBCs}.~Since, for CBC, we fix the conformal class of the induced metric, it will be convenient to decompose the various variations in terms of the variation of the conformal class and the variation of the Weyl factor.~To accomplish that, we recall that for two $d$-dimensional metrics that are related via a Weyl rescaling $\tilde{h}_{ij}=\Omega^2(y^m)h_{ij}$, the extrinsic curvature, Ricci tensor and Ricci scalar are related by (see \eg appendix D in \cite{Wald:1984rg}),
\begin{equation}\label{eq:KRRij_eqs}
    \begin{cases}
        \tilde{\mathcal{K}}_{ij}  = \Omega\big(\mathcal{K}_{ij}+h_{ij}n^kD_k\log{\Omega}\big)  \ , \\
        \tilde{R}_{ij}   = R_{ij}-(d-2)D_iD_j\log{\Omega}-h_{ij}h^{pq}D_pD_q\log{\Omega}\\
         \qquad\quad +(d-2)D_i\log{\Omega}D_j\log{\Omega}-(d-2)h_{ij}h^{pq}D_p\log{\Omega}D_q\log{\Omega}
        \ ,\\
        \tilde{R}  = \Omega^{-2}\Big(R-2(d-1)h^{ij}D_iD_j\log{\Omega}-(d-1)(d-2)h^{ij}D_i\log{\Omega}D_j\log{\Omega}\Big) \ ,
    \end{cases}
\end{equation}
where the curvature tensors with an overhead tilde are computed for $\tilde{h}_{ij}$, and the ones without for $h_{ij}$.~In our case we have that $\Omega=h^{\frac{1}{2d}}$, see definition above eq.~\eqref{eq:Weyl_rescaling}.

Given eqs.~\eqref{eq:KRRij_eqs}, one can decompose the variations of the boundary quantities into terms that are fixed by the CBC and terms proportional to the variation of the Weyl factor, in our case $h$.~Doing so and imposing CBC~\eqref{eq:CBCs}, \ie setting $\delta\tilde{h}_{ij}|_{\partial\mathcal{M}}=\delta\mathcal{K}|_{\partial\mathcal{M}}=0$, the variations of the terms in eq.~\eqref{eq:gen_bdy_term} read
\begin{align}
    &\delta \int_{\partial\mathcal{M}}d^dy\sqrt{h}\mathcal{K}^3=\int_{\partial\mathcal{M}}d^dy\,\mathcal{K}^3\delta\sqrt{h} \ , \nonumber \\
    &\delta \int_{\partial\mathcal{M}}d^dy\sqrt{h}\mathcal{K}\mathcal{K}^{ij}\mathcal{K}_{ij}=\frac{1}{d}\int_{\partial\mathcal{M}}d^dy \Big((d-2)\mathcal{K}^{ij}\mathcal{K}_{ij}\delta\sqrt{h}+2\sqrt{h}h^{-\frac{1}{2d}}\mathcal{K}^2\Big)\mathcal{K}\delta\sqrt{h}\ , \nonumber \\
    &\delta \int_{\partial\mathcal{M}}d^dy\sqrt{h}\mathcal{K}^{ij}\mathcal{K}_{jk}\mathcal{K}^k_i=\frac{1}{d}\int_{\partial\mathcal{M}}d^dy\Big((d-3)\mathcal{K}^{ij}\mathcal{K}_{jk}\mathcal{K}^k_i+3\sqrt{h}h^{-\frac{1}{d}}\mathcal{K}\mathcal{K}^{ij}\mathcal{K}_{ij}\Big)\delta\sqrt{h} \ , \nonumber \\
    &\begin{aligned}\delta \int_{\partial\mathcal{M}}d^dy\sqrt{h}\tilde{R}\mathcal{K}=&\frac{1}{d}\int_{\partial\mathcal{M}}d^dy \Big[(d-2)\tilde{R}\mathcal{K}-2(d-1)\big(D^iD_i(h^{-\frac{1}{d}}\mathcal{K}+\mathcal{K}h^{-\frac{d+2}{2d}}D^iD_i\sqrt{h}\big)\\
    &-\frac{2(d-1)(d-2)}{d}\big(D_i(h^{-\frac{1}{d}}\mathcal{K}D^i\log{\sqrt{h}})+h^{-\frac{1}{d}}\mathcal{K}D^i\log{\sqrt{h}}D_i\log{\sqrt{h}}\big)\Big]\delta\sqrt{h}\ , 
    \end{aligned}\nonumber \\
    &\begin{aligned}\delta \int_{\partial\mathcal{M}}d^dy\sqrt{h}\mathcal{K}^{ij}\tilde{R}_{ij}=&\frac{1}{d}\int_{\partial\mathcal{M}}d^dy\Big[(d+1)\mathcal{K}^{ij}\tilde{R}_{ij}+\sqrt{h}h^{\frac{2d+1}{2d}}\tilde{R}\mathcal{K}\\
    &-\frac{(d+2)(d-2)}{d}\big(\mathcal{K}^{ij}D_i\log{\sqrt{h}}D_j\log{\sqrt{h}}+D_j(\mathcal{K}^{ij}D_i\log{\sqrt{h}})\big)\\
    &-(d-2)\big(D_i \mathcal{K}^{ij}D_j\log{\sqrt{h}}+D_jD_i\mathcal{K}^{ij}\big)-\big(D^iD_i(h^{-\frac{1}{d}}\mathcal{K})+h^{-\frac{d+2}{2d}}D^iD_i\sqrt{h}\big)\\
    &+\frac{2(d-2)}{d}\big(D_i(h^{-\frac{1}{d}}\mathcal{K}D^i\log{\sqrt{h}})+h^{-\frac{1}{d}}\mathcal{K}D^i\log{\sqrt{h}}D_i\log{\sqrt{h}}\big)\Big]\delta\sqrt{h} \ . \end{aligned}\
\end{align}
Combining, then, the variation of eq.~\eqref{eq:gen_bdy_term} with the on-shell variation of the bulk action~\eqref{eq:ap_variations_diff}, and imposing CBC, we arrive at
\begin{equation}\label{eq:full_full_variation}
\begin{aligned}
   \kappa\big(\delta\mathcal{I}_{\text{EGB}}-\delta\mathcal{I}_{\partial\mathcal{M}}\big)&\hat{=} \frac{1}{d}\int_{\partial\mathcal{M}}d^dy\Bigg[\left(1+(d-2)(1-c_2)-3\left(\frac{2}{3}+c_3\right)h^{\frac{d-2}{2d}}\right)\mathcal{K} \mathcal{K}^{ij}\mathcal{K}_{ij}\delta\sqrt{h}\\
   &-(d-3)c_3\mathcal{K}^{ij}\mathcal{K}_{jk}\mathcal{K}^k_i\delta\sqrt{h}+\left(-3-dc_1+2(1-c_2)h^{\frac{d-1}{2d}}\right)\mathcal{K}^3\delta\sqrt{h}\\
   &-\left(8+(d+1)c_5\right)\mathcal{K}^{ij}\tilde{R}_{ij}\delta\sqrt{h}+\left(1-(d-2)c_4-(2+c_5)h^{\frac{3d+1}{d}}\right)\tilde{R}\mathcal{K}\delta\sqrt{h}\\
   &+\Bigg(\frac{(d+2)(d-2)}{d}(2+c_5)\big(\mathcal{K}^{ij}D_i\log{\sqrt{h}}D_j\log{\sqrt{h}}+D_j(\mathcal{K}^{ij}D_i\log{\sqrt{h}})\big)\\
   &+(d-2)(2+c_5)\big(D_i \mathcal{K}^{ij}D_j\log{\sqrt{h}}+D_jD_i\mathcal{K}^{ij}\big)\\
   &+(2+c_5-2(d-1)(1-c_4))\big(D^iD_i(h^{-\frac{1}{d}}\mathcal{K})+h^{-\frac{d+2}{2d}}D^iD_i\sqrt{h}\big)\\
   &-\frac{2(d-2)}{d}(2+c_5-(d-1)(1-c_4))\big(D_i(h^{-\frac{1}{d}}\mathcal{K}D^i\log{\sqrt{h}})\\
   &+h^{-\frac{1}{d}}\mathcal{K}D^i\log{\sqrt{h}}D_i\log{\sqrt{h}}\big)\Bigg)\delta\sqrt{h}\Bigg] \,,
\end{aligned}    
\end{equation}
where $\kappa:= \tfrac{(d-2)(d-3)16\pi G_N}{\alpha}$.~This total variation has to vanish, by appropriately tuning $c_i$.~Inspecting, however, \eg the second line of eq.~\eqref{eq:full_full_variation} one is forced to choose $c_3=0$, but then nothing can cancel the contribution proportional to $h^{\frac{d-2}{2d}}$ in the first line, and so on.~Therefore, there does not exist a choice of $c_i$ in eq.~\eqref{eq:gen_bdy_term} such that the variation of the bulk EGB action is stationary on-shell, and hence there does not exist a local and covariant boundary term for EGB gravity with CBC.

\section{On-Shell Actions for EMGB Solutions}\label{app:on-shell_actions}

In this appendix, we briefly describe how to compute the on-shell actions~\eqref{eq:full_action} for the solutions ~\eqref{eq:f(r)_EMGB}. In computing the on-shell actions, the following curvature quantities for the general metric of the form~\eqref{eq:line_element} will be useful~\cite{Deser:2005pc,Haroon:2020vpr},
\begin{equation}\label{eq:useful_relations}
    \begin{cases}
        &R=\frac{(d-1)(d-2)(k-f(r))}{r^2}-\frac{2(d-1)f'(r)}{r}-f''(r) \ ,\\
        &R^{\mu\nu}R_{\mu\nu}=\frac{1}{2}\left(f''(r)+\frac{(d-1)f'(r)}{r}\right)^2+(d-1)\left(\frac{(d-2)(k-f(r))}{r^2}-\frac{f'(r)}{r}\right) \ , \\
        & R^{\mu\nu\rho\sigma}R_{\mu\nu\rho\sigma}=\left(f''(r)\right)^2+2(d-1)\left(\frac{f'(r)}{r}\right)^2+2(d-1)(d-2)\left(\frac{k-f(r)}{r^2}\right)^2 \ ,\\
        &\mathcal{K}=a(r)+\lambda b(r) \ ,\\
        &\mathcal{K}^{\mu\nu}\mathcal{K}_{\mu\nu}=a^2(r)+\lambda b^2(r) \ , \\
        &\mathcal{K}^{\mu\nu}\mathcal{K}_{\nu\rho}\mathcal{K}^{\rho}_\mu=a^3(r)+\lambda b^3(r)\ ,\\
        &\tilde{G}^{\mu\nu}\mathcal{K}_{\mu\nu}=-\frac{\lambda(\lambda-1)k}{2r^2}\left(a(r)+(\lambda-2)b(r)\right)  \ ,
    \end{cases}    
\end{equation}
where we defined $\lambda=d-1,\  a(r)=f'(r)/2\sqrt{f(r)} \ \text{and} \  b(r)=\sqrt{f(r)}/r$.~Moreover, the trace of the EGB field equations is (recall the definition of $\mathcal{E}_{\mu\nu}$ from eq.~\eqref{eq:GB_relations} and also that $\mathcal{L}_{\text{GB}}=R^{\mu\nu\rho\sigma}R_{\mu\nu\rho\sigma}-4R^{\mu\nu}R_{\mu\nu}+R^2$),
\begin{equation}
g^{\mu\nu}\mathcal{E}^{\text{EGB}}_{\mu\nu}=g^{\mu\nu}\left(G_{\mu\nu}+\Lambda g_{\mu\nu}+\frac{\alpha}{(d-2)(d-3)}\mathcal{E}_{\mu\nu}\right)=\frac{1-d}{2}R+(d+1)\Lambda-\frac{\alpha}{2(d-2)}\mathcal{L}_{\text{GB}} \ ,   
\end{equation}
which has to be zero on-shell, and thus,
\begin{equation}
    \frac{\alpha}{(d-2)(d-3)}\left(R^{\mu\nu\rho\sigma}R_{\mu\nu\rho\sigma}-4R^{\mu\nu}R_{\mu\nu}+R^2\right)\hat{=}\frac{1}{d-3}\left(2(d+1)\Lambda-(d-1)R\right) \ .
\end{equation}
Using this, we can re-express the bulk term of the action~\eqref{eq:full_action} as
\begin{equation}
\begin{aligned}
    \mathcal{I}_{\text{bulk}}&\hat{=}\frac{1}{(d-3)8\pi G_N}\int_{\mathcal{M}} d^{d+1}x \sqrt{g} \left(R-4\Lambda\right)\\
    &\hat{=} -\frac{\beta\Sigma_{k,d-1}}{(d-3)8\pi G_N}\left[\frac{4\Lambda}{d}r^{d}+r^{d-1}f'(r)+(d-1)r^{d-2}(f(r)-k)\right] \Bigg|_{r_h}^{e^\bomega\mathfrak{r}} \ ,
\end{aligned}   
\end{equation}
since the $r$ integral reduces to a boundary term by successive integrations by parts,
\begin{equation}
    \begin{aligned}
        \int dr r^{d-1}R&= \int dr r^{d-1}\left[-\left(\frac{k(d-1)(d-2)}{r}+f'(r)\right) '-\frac{(d-1)(d-2)f(r)+2(d-1)rf'(r)}{r^2}\right]\\
            &=-\left[r^{d-1}f'(r)+(d-1)r^{d-2}\left(f(r)-k\right)\right]_{r_h}^{e^\bomega\mathfrak{r}} \ .
    \end{aligned}
\end{equation}
A useful identity is given by (here $q=0$),
\begin{equation}\label{eq:action_identity}
\begin{aligned}[t]
    r^{d-1}f'(r)\left(1-\frac{2(d-1)\alpha}{d-3}\frac{f(r)-k}{r^2}\right)= -\frac{2}{d-3}\left[r^{d-1} f'(r) + (d-1)r^{d-2}\left(f(r)-k\right) - 2(d-1)\,\frac{r^{d}}{\ell^{2}}\right]\\ +\frac{(d-1)(d-4)}{d-3}\mu \ .
\end{aligned}    
\end{equation}
Using eq.~\eqref{eq:action_identity}, the bulk on-shell action simplifies to,
\begin{equation}\label{eq:EGB_bulk_action}
\begin{aligned}
    \mathcal{I}_{\text{bulk}}&\hat{=}\frac{\beta \Sigma_{k,d-1}}{16\pi G_N}\left[-\frac{(d-1)(d-4)}{d-3}\mu+r^{d-1}f'(r)\left(1-\frac{2(d-1)\alpha}{d-3}\frac{f(r)-k}{r^2}\right)\right]_{r_h}^{e^\bomega\mathfrak{r}}\\
    &\hat{=}\frac{\beta \Sigma_{k,d-1}(e^\bomega\mathfrak{r})^{d-1}f'(e^\bomega\mathfrak{r})}{16\pi G_N}\left(1-\frac{2(d-1)\alpha}{d-3}\frac{f(e^\bomega\mathfrak{r})-k}{(e^\bomega\mathfrak{r})^2}\right) - \frac{\Sigma_{k,d-1}r_h^{d-1}}{4G_N}\left(1+2k\frac{d-1}{d-3}\frac{\alpha}{r_h^2}\right)\\
    &\equiv \frac{\beta \Sigma_{k,d-1}(e^\bomega\mathfrak{r})^{d-1}f'(e^\bomega\mathfrak{r})}{16\pi G_N}\left(1-\frac{2(d-1)\alpha}{d-3}\frac{f(e^\bomega\mathfrak{r})-k}{(e^\bomega\mathfrak{r})^2}\right) -\mathcal{S}_{\text{EGB}} \ ,
\end{aligned}    
\end{equation}
where we used that $\beta=4\pi/f'(r_h)$, as well as the definition of the Wald entropy $\mathcal{S}_{\text{EGB}}$~\eqref{eq:Swald}.~Using eqs.~\eqref{eq:useful_relations} we can express the trace of the $J_{\mu\nu}$ tensor~\eqref{eq:GB_relations} as
\begin{equation}
        J=\mathcal{K}\mathcal{K}^{\mu\nu}\mathcal{K}_{\mu\nu}-\frac{1}{3}\mathcal{K}^3-\frac{2}{3}\mathcal{K}^{\mu}_{\alpha}\mathcal{K}^{\alpha\beta}\mathcal{K}_{\beta\mu}=-\lambda(\lambda-1)a(r)b^2(r)-\frac{\lambda(\lambda-1)(\lambda-2)}{3}b^3(r) \ .
\end{equation}
Therefore, the GB boundary term~\eqref{eq:bdyRob} can be collectively expressed as
\begin{equation}\label{eq:bdyEGB_subt}
    \mathcal{I}_{BM}=\frac{b_d\alpha(d-1)}{(d-2)4\pi G_N}\int_{\partial\mathcal{M}}d^{d}y\sqrt{h}\left[b^2(r)\left(a(r)+\frac{(\lambda-2)}{3}b(r)\right)-\frac{k}{r^2}\left(a(r)+(d-4)b(r)\right)\right] \ ,
\end{equation}
while the corresponding GR boundary term~\eqref{eq:York} reads,
\begin{equation}\label{eq:bdyGR_subt}
    \mathcal{I}_{YGH}=-\frac{a_d}{8\pi G_N}\int_{\partial \mathcal{M}}d^{d}y \sqrt{h} \left[a(r)+\lambda b(r)\right] \ .
\end{equation}
Putting eqs.~\eqref{eq:EGB_bulk_action}, \eqref{eq:bdyEGB_subt}, \eqref{eq:bdyGR_subt} together we have the following full on-shell action (here we specialize to $k=a_d=b_d=1$ and $q=0$),
\begin{equation}\label{eq:EGB_Dir_on-shell_total}
\begin{aligned}
\mathcal{I}_{\text{grav}}\hat{=} &\frac{\beta_{\partial\mathcal{M}}\Omega_{d-1}(d-1)(e^\bomega\mathfrak{r})^{d-4}\sqrt{f(e^\bomega\mathfrak{r})}\left(2\alpha(f(e^\bomega\mathfrak{r})-3) - 3 (e^\bomega\mathfrak{r})^{2}\right)}{24\pi G_{N}}\\
&\qquad\qquad\qquad\qquad\qquad\qquad -\frac{\Omega_{d-1} r_h^{d-1}}{4 G_{N}} \left(1 + \dfrac{2(d-1)}{(d-3)}\frac{\alpha}{r_h^2}\right) \ ,
\end{aligned}
\end{equation}
where $\beta_{\partial \mathcal{M}}:=\sqrt{f(e^\bomega\mathfrak{r})}\beta$. 

\section{Flat EMGB black hole} \label{app: flat EGB bh}
Here we provide the computations for the high-temperature limit of the conformal entropy discussion in section~\ref{sec:entropies}, directly in flat space.

\paragraph{Flat EGB Black Hole} We consider first the case of an asymptotically flat BD black hole with spherical horizon ($q=0,\Lambda=0,k=1$ in eq.~\eqref{eq:f(r)_EMGB}), with line element
\begin{equation}\label{eq:flat_metric}
    ds^2_{\text{BD}}=f(r)d\tau^2+\frac{dr^2}{f(r)}+r^2d\Omega^2_{d-1}\ , \  \ f(r)=1+\frac{r^2}{2\alpha}\left(1-\sqrt{1+\frac{4\alpha\mu}{r^{d}}}\right) \ ,
\end{equation}
The boundary is at $r=e^\bomega \mathfrak{r}$ and we consider thermodynamics for the region between the black hole horizon at $r=r_h$ and the boundary $\partial\mathcal{M}$.~As in (A)dS, inverting the boundary data eqs.~\eqref{eq:ECBCs} for the metric~\eqref{eq:flat_metric} to obtain $r_h,e^\bomega\mathfrak{r}$ as a function of $\lbrace\tilde{\beta},\mathcal{P}\rbrace$ is not possible analytically, even for the simplest case of $d=4$.~Nonetheless, perturbatively in the GB coupling, we get (recall that $y=r_h/e^\bomega\mathfrak{r}$),
\begin{equation}\label{eq:EGB_betaK-highB}
    \begin{cases}
        \tilde{\beta}=\dfrac{2\pi\sqrt{1-y^{d-2}}}{d-2}\left(2y+\dfrac{\alpha}{r_h^2}\dfrac{(d-2)y^{2d-1}+(2-3d)y^{d-1}+2dy}{(d-2)(1-y^{d-2})}\right) +\mathcal{O}(\alpha^2)\ ,\\
        \begin{aligned}
        \mathcal{P}&=
            \dfrac{2(d-1)y-dy^{d-1}}{2r_h\sqrt{1-y^{d-2}}}\\
            &+\dfrac{\alpha}{r_h^2}\dfrac{-d y^{3 (d-1)}+16 (d-3) y^3+3 \left(2 (d-3) y^2+d\right) y^{2 d-3}-6 \left(4 (d-3) + y^{-2}\right) y^{d+1}}{12r_h(1-y^{d-2})^{3/2}}+\mathcal{O}(\alpha^2) \ .
        \end{aligned}
    \end{cases}
\end{equation}
Due to the absence of additional scales in flat space, the high-temperature limit corresponds to sending $y\rightarrow1$ (as in GR).~However we have to keep the following dimensionless quantity also fixed, $\rho:=\alpha/r_h^2$, whose necessity can be understood by observing the second term of the eqs.~\eqref{eq:EGB_betaK-highB}.~This is somewhat peculiar, since it essentially requires scaling $G_N$ as we rescale the coordinates with the fluctuating boundary conformal factor.~This problem originates from the fact that flat space lacks an additional dimensionful parameter, and $G_N$ is the only available quantity that comes in to play that role.~This step can be justified by the fact that one recovers the same results from the flat limit ($\mathcal{P}\ell\rightarrow\infty$) of (A)dS, as shown in section~\ref{sec:entropies}. We remark, lastly, that our results remain unchanged whichever of the following three dimensionless parameters we keep fixed in the high-temperature limit: $\alpha/r_h^2,\ \alpha/e^\bomega\mathfrak{r}r_h\ \text{or} \   \alpha/e^{2\bomega}\mathfrak{r}^2$. In the $y\rightarrow1$ limit, eqs.~\eqref{eq:EGB_betaK-highB} become
\begin{equation}
    \begin{cases}
        \tilde{\beta}=\dfrac{2\pi\sqrt{1-y}}{\sqrt{d-2}}\left(2+\frac{\alpha}{r_h^2}\frac{d}{d-2}\right)+\dots \ ,\\
        \mathcal{P}=\dfrac{1}{2r_h \sqrt{(d-2)(1-y)}}\left(d-2+\frac{\alpha}{r_h^2}\frac{3d-5}{2}\right)+\dots \ ,
    \end{cases}
\end{equation}
where the dots indicate subleading corrections.~Using these, the Gauss-Bonnet leading correction to the entropy takes the form
\begin{equation}\label{eq:leadinga_Sconflat}
\begin{aligned}
    \mathcal{S}_\text{conf}^{(\text{GB})}&=\frac{(2\pi)^{d-1}\Omega_{d-1}}{4G_N\mathcal{P}^{d-1}}\frac{1}{\tilde{\beta}^{d-1}}+\mathcal{O}\left(\frac{1}{\tilde{\beta}^{d-5}}\right) \ ,
\end{aligned}    
\end{equation}
Note that, as in eq.~\eqref{eq:Ndof_Mink}, to leading order in $\alpha$, the GB correction solely affects the subleading term in the high-temperature expansion.

\paragraph{Flat EMGB Black Hole}
We now consider Einstein-Maxwell-Gauss-Bonnet gravity in flat space with line element ($\ell\rightarrow\infty,k=1$ in eq.~\eqref{eq:f(r)_EMGB}),
\begin{equation}
    ds^2_{\text{BD}}=f(r)d\tau^2+\frac{dr^2}{f(r)}+r^2d\Omega^2_{d-1} \ , \  \ f(r)=1+\frac{r^2}{2\alpha}\left(1-\sqrt{1+\frac{4\alpha\mu}{r^{d}}-\frac{4\alpha q^2}{r^{2(d-1)}}}\right)  \ .
\end{equation}
In the following, it will be useful to express our quantities in terms of the dimensionless ratios $\rho=\alpha/r_h^2, \ \gamma=q^2/r_h^{2(d-2)}$, which we keep fixed as we take the high-temperature limit.~For this metric then, the ECBC~\eqref{eq:ECBCs} at $\mathcal{O}(\alpha)$ and in the $y\rightarrow1$ limit are
\begin{equation}
    \begin{cases}
        \tilde{\beta}=  \dfrac{
2\pi\sqrt{(d-2)(1-y)(1-\gamma)}
\big[2(d-2)(1-\gamma) + \rho\big(6\gamma-1+(d+1)(1-2\gamma)\big)\big]}{(d - 2)^{2}(1-\gamma)^2} +\dots \  ,\\
        \mathcal{P}= \dfrac{2(d-2)(1-\gamma)+\rho\big(3d-5-2\gamma(d-2)\big)}{4r_h\sqrt{(d-2)(1-y)(1-\gamma)}}+\dots \ .
    \end{cases}
\end{equation}
Inverting these equations in this regime, we obtain,
\begin{equation}
    r_h=\frac{2\pi}{\tilde{\beta}\mathcal{P}} +\mathcal{O}(\tilde{\beta}) \ ,
\end{equation}
and therefore the conformal entropy, to leading order in the high-temperature expansion, coincides with the uncharged case~\eqref{eq:Ndof_Mink}, \ie
\begin{equation}
    \mathcal{S}_\text{conf}^{(\text{EMGB})}=\frac{(2\pi)^{d-1}\Omega_{d-1}}{4G_N\mathcal{P}^{d-1}}\frac{1}{\tilde{\beta}^{d-1}}+\mathcal{O}\left(\frac{1}{\tilde{\beta}^{d-3}}\right)  \ .
\end{equation}
Again, a better justification of the prescription we implemented for the high-temperature limit here, is that the answer can be recovered from the flat limit of (A)dS, see section~\ref{subsec:charged_conformal_thermo}.

\section{Near the AdS Boundary in EMGB Gravity}\label{app:GB_AdS_bdy}

In this appendix we discuss aspects of timelike boundaries in Einstein-Maxwell-Gauss-Bonnet gravity, placed near the conformal boundary of AdS. As uncovered by Fefferman and Graham (FG)~\cite{FeffermanGraham1985} and generalized to arbitrary Lovelock metrics in ref.~\cite{albin2020poincare}, the metric of every asymptotically AdS$_{d+1}$ spacetime admits an expansion near the conformal boundary of AdS as (notice that $\ell_{\mt{AdS}}$ is the AdS curvature radius, which does not coincide with the $\ell$ appearing in the action for higher curvature gravity solutions),
\begin{equation}
    \frac{ds^2}{\ell_{\mt{AdS}}^2}=d\rho^2+\frac{e^{2\rho}}{\ell_{\mt{AdS}}^2}\left(g^{(0)}_{ij}+\ell_{\mt{AdS}}^2e^{-2\rho}g^{(2)}_{ij}+\dots+\ell_{\mt{AdS}}^{d-1}e^{-d\rho}g^{(d)}_{ij}+\dots\right)dx^idx^j \ ,
\end{equation}
where $\rho$ is a radial coordinate and, again, Latin indices correspond to boundary directions.~As in Einstein gravity, $g_{ij}^{(0)}$ constitutes the conformal class of the boundary metric that has to be specified freely and $g^{(d)}_{ij}$ is related to the boundary stress-tensor.~Placing our timelike boundary $\partial\mathcal{M}$ at a large value $\rho=\mathfrak{r}$, the induced metric reads
\begin{equation}
    g_{ij}|_{\partial\mathcal{M}}=e^{2\mathfrak{r}}\left(g^{(0)}_{ij}+\dots\right) \ ,
\end{equation}
while the extrinsic curvature,
\begin{equation}
    \mathcal{K}_{ij}|_{\partial\mathcal{M}}=\frac{e^{2\mathfrak{r}}}{\ell_{\mt{AdS}}}\left(g^{(0)}_{ij}+\dots\right) \ ,
\end{equation}
and hence its trace is:
\begin{equation}
    \mathcal{K}|_{\partial\mathcal{M}}=g^{ij}\mathcal{K}_{ij}|_{\partial\mathcal{M}}=\frac{d}{\ell_{\mt{AdS}}}+\dots \ .
\end{equation}
Other curvature invariants constructed out of $\mathcal{K}_{ij}$ can be computed to leading order in a similar fashion, \eg
\begin{equation}
    \mathcal{K}^{ij}\mathcal{K}_{ij}|_{\partial\mathcal{M}}=\frac{d}{\ell_{\mt{AdS}}^2}+\dots \ , \qquad \mathcal{K}^{ij}\mathcal{K}_{jk}\mathcal{K}^{k}_i|_{\partial\mathcal{M}}=\frac{d}{\ell_{\mt{AdS}}^3}+\dots \ .
\end{equation}
Hence, the extended boundary condition near the AdS boundary takes the form
\begin{equation}
    \begin{aligned}
         \mathcal{P}\ell_{\mt{AdS}}|_{\partial\mt{AdS}}&=\mathcal{K}\ell_{\mt{AdS}}|_{\partial\mt{AdS}}+\frac{\alpha}{2(d-1)(d-2)}\mathcal{Q}\ell_{\mt{AdS}}|_{\partial\mt{AdS}} \,,\\
         &=d\left(1-\frac{2}{3}\tigma g^{-2}(\tigma)\right)+\frac{2\tigma \ell^2}{d-1}\tilde{R}(g^{(0)}_{ij})e^{-2\mathfrak{r}}+\mathcal{O}\left(e^{-2\mathfrak{r}}\right)\ ,
        \end{aligned}
\end{equation}
for $g(\tigma)$ as in eq.~\eqref{eq:f(sigma)}.~Therefore, the boundary condition at the AdS boundary is
\begin{equation}
    \mathcal{P}\ell_{\mt{AdS}}|_{\partial\mt{AdS}}=\frac{d}{3}\left(2+\sqrt{1-4\tigma}\right) \ .
\end{equation}

\section{Entropy Universality in Lovelock Gravity, and Beyond}\label{ap:entropy_univ}

As emphasized in the main body of this work, entropy in Lanczos-Lovelock gravity is independent of the kind of boundary conditions one wishes to impose.~Rather, it is an intrinsic feature of the horizon, a fact that can be linked to the Wald formula~\cite{Wald:1993nt,PhysRevLett.70.3684,Jacobson:1993vj,Jacobson:1994qe,Dong:2013qoa}.~However, within the framework of Euclidean thermodynamics there exists a simpler computation, see \eg refs.~\cite{Carlip:1993sa,Banados:1993qp}, demonstrating directly that the only contribution to the entropy comes exclusively from the horizon.

Let us begin by discussing this computation in pure Einstein gravity. The Euclidean action is just:
\begin{equation}\label{eq:GR_action}
    \mathcal{I}_{\text{GR}}=-\frac{1}{16\pi G_N}\int_{\mathcal{M}}d^{d+1}x\sqrt{g}\left(-2\Lambda+R\right)-\frac{a_d}{8\pi G_N}\int_{\partial\mathcal{M}}d^{d}y\sqrt{h}\mathcal{K} \ ,
\end{equation}
where again $a_d$ is determined by the boundary conditions, so that the action~\eqref{eq:GR_action} has a good variational problem.~In fact, the precise form of the boundary action is not important for the following arguments.~All that matters is that there exists a local boundary term that makes the action stationary under first order variations of the metric.~We then want to compute the entropy though the usual formula,
\begin{equation}
    \mathcal{S}_{\text{GR}}=-\left(1-\beta\partial_{\beta}\right)\mathcal{I}_{\text{GR}} \ ,
\end{equation}
where $\beta$ is the proper boundary temperature.\footnote{Or the conformal temperature $\tilde{\beta}$ used for (E)CBC or whichever temperature-like quantity defines the thermal ensemble.} We are, therefore, interested in how the action changes to first order in $\beta$. To that extend, we can vary $\beta$ infinitesimally while keeping the bulk solution fixed.\footnote{This is the opposite of what we usually do, where we compute the on-shell action for any $\beta$, appropriately adjusted so that we satisfy the equations of motion everywhere in the interior, and then differentiate to get the entropy. The two approaches coincide for first order variations of the metric, because the action~\eqref{eq:GR_action} is stationary under those.} Recall that, near the horizon $r_h$, metrics of the general form~\eqref{eq:line_element} become
\begin{equation}
    ds^2=\rho^2 B^2 d\tau^2+d\rho^2+r_h^2d\Sigma_{k,d-2}^2 \ ,
\end{equation}
where $\rho=2 \sqrt{\frac{r-r_h}{f'(r_h)}}$ and $B=\frac{f'(r_h)}{2\sqrt{f(e^\bomega\mathfrak{r})}}$, with the boundary being placed at $r=e^\bomega\mathfrak{r}$. As usual, in order to avoid a conical singularity at the horizon, one identifies
\begin{equation}
    B\tau \sim B \tau+2\pi \ .
\end{equation}
Here, however, we want to vary $\beta$ infinitesimally away from the value $2\pi/B$, namely take $\tau \sim \tau +\beta$. This introduces a conical singularity at $r=r_h$, with angle opening $2\pi+\epsilon$ or, equivalently, $\epsilon=\beta B-2\pi$.~Because of this singularity, the Ricci scalar at $r=r_h$ takes the form
\begin{equation}\label{eq:GR_Ricci}
    R=-2\epsilon\delta_h(\tau,r_h)=\left(4\pi-\beta B\right)\delta_h(\tau,r_h) \ ,
\end{equation}
where $\delta_h(\tau,r_h)$ is a delta function that gives one when integrated against the transverse coordinates to the horizon.~We are now ready to compute the entropy.~The boundary action, being a local function of the boundary geometry, will be proportional to $\beta$ and therefore will not contribute to the entropy.~The same goes on for the bulk action, except at the position of the conical excess, where there also exists a constant piece~\eqref{eq:GR_Ricci}.~Hence, we get
\begin{equation}
    \mathcal{S}_{\text{GR}}=\frac{1}{16\pi G_N}\int_h d\Sigma_{k,d-1} \, 4\pi \, \delta_h(\tau,r_h) =\frac{A_k}{4G_N} \ ,
\end{equation}
$A_k$ being the area of the horizon of constant curvature $k$.~Because the operator $1-\beta\partial_\beta$ annihilates contributions proportional to $\beta$, this calculation makes manifest that the entropy is controlled solely by the horizon.~Other thermodynamic quantities, such as the energy that involves only a first-$\beta$ derivative, will not have this feature and indeed they do typically receive contributions from the bulk and boundary actions (away from the horizon).~Notice that these arguments also hold for the case of Einstein-Maxwell theory.

This derivation can also be readily generalized to arbitrary Lovelock gravity theories, for ensembles that are defined by boundary conditions that possess a stationary action under first order metric variations, possibly after the addition of a local boundary term.~The only essential difference from before is the contribution of the bulk action at the position of the conical excess, which is just the sum of general Lovelock densities $\chi_{2n}(g)$ evaluated on the manifold $(\mathcal{M},g)$,
\begin{equation}
    \mathcal{I}_{\text{LL}}=-\frac{1}{16\pi G_N}\int_{\mathcal{M}}d^{d+1}x\sqrt{g}\left(\sum_{n\geq1}\lambda_n \chi_{2n}(g)\right)+\text{local boundary terms} \ ,
\end{equation}
with some dimensionful coefficients $\lambda_n$.~For an opening angle of $\epsilon$ away from $2\pi$, the contribution of the density $\chi_{2n}(g)$ at the excess is given by (generalizing eq.~\eqref{eq:GR_Ricci})~\cite{Fursaev:1995ef},
\begin{equation}\label{eq:Love_Ricci} 
    \chi_{2n}(g)|_h=-2(n-1)\epsilon\chi_{2(n-1)}(\gamma) \delta_h(\tau,r_h) \ ,
\end{equation}
where the density $\chi_{2(n-1)}(\gamma)$ is evaluated for the codimension-$2$ horizon metric $\gamma$ with curvature quantities \textit{intrinsic} to the horizon.~The calculation proceeds as previously, yielding
\begin{equation}
    \mathcal{S}_{\text{LL}}=\frac{A_k}{4G_N}\left(1+\sum_{n\geq3} (n-1)\lambda_n\chi_{2(n-1)}(\gamma)\right)\ .
\end{equation}
Rather remarkably, this reproduces the Jacobson-Myers entropy~\cite{Jacobson:1994qe}, the generalization of the Wald entropy~\cite{Wald:1993nt,Iyer:1995kg}.\footnote{This is slightly ambiguous, since in the examples of Killing horizons considered here the additional extrinsic curvature terms vanish and the Jacobson-Myers entropy reduces to the Wald entropy.}~For Einstein-Gauss-Bonnet gravity, we get
\begin{equation}
    \mathcal{S}_{\text{EGB}}= \frac{1}{4 G_N} \int_hd\Sigma_{k,d-1} \left(1+\frac{2\,\alpha}{(d-2)(d-3)} R(\gamma) \right) \delta_h(\tau,r_h) =\frac{A_k}{4G_N}\left(1+2k\frac{d-1}{d-3}\frac{\alpha}{r_h^2}\right)  \ ,
\end{equation}
which is the familiar expression~\eqref{eq:Swald}.~Again, the discussion is unaltered by the addition of $U(1)$ gauge fields.~A similar reasoning can also be followed for more general actions that are covariant functionals of the Riemann tensor, in the spirit of ref.~\cite{Wald:1993nt}.

\section{Lorentzian Linearized Dynamics in Einstein-Maxwell Gravity}\label{app:pert_GR}
Here we briefly discuss the behavior of linearized modes subject to conformal boundary conditions in Einstein-Maxwell gravity, about backgrounds with planar and hyperbolic horizons.\footnote{The case of uncharged spherical horizons has been analyzed previously in refs.~\cite{Anninos:2023epi,Anninos:2024wpy,Anninos:2024xhc,Liu:2024ymn}.~The addition of charge does not affect their conclusions, at least for the $l=0$ modes.}~Overall, these modes exhibit better stability properties than the corresponding modes of spherical horizons.

\begin{figure}[t]
    \centering   
    \begin{subfigure}[t]{0.48\textwidth}
        \centering
        \includegraphics[width=\textwidth]{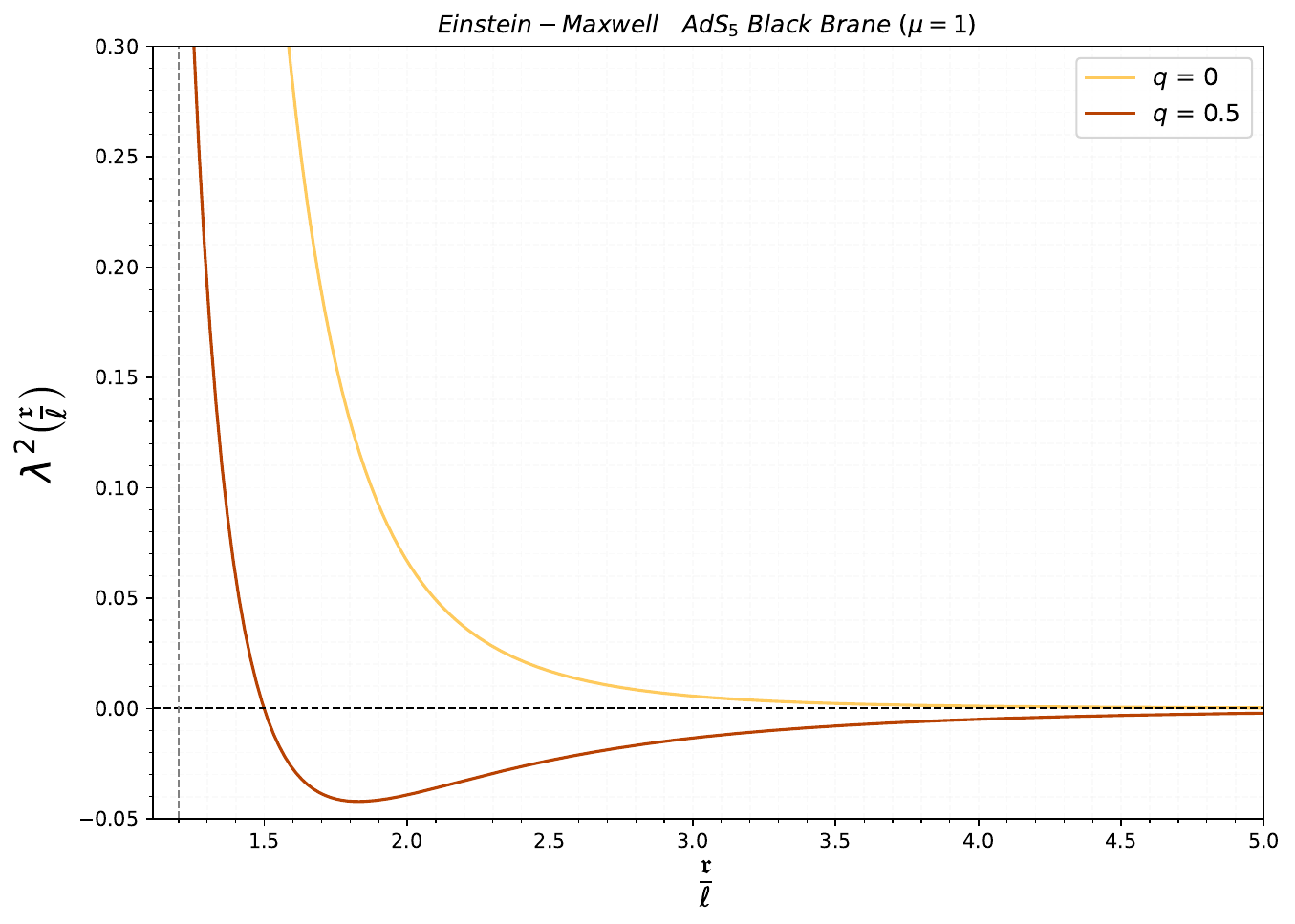}
        \caption{Planar Horizon}
        \label{fig:BD_BH_planar_beta_d=5a}
    \end{subfigure}
    \hfill
    \begin{subfigure}[t]{0.48\textwidth}
        \centering
        \includegraphics[width=\textwidth]{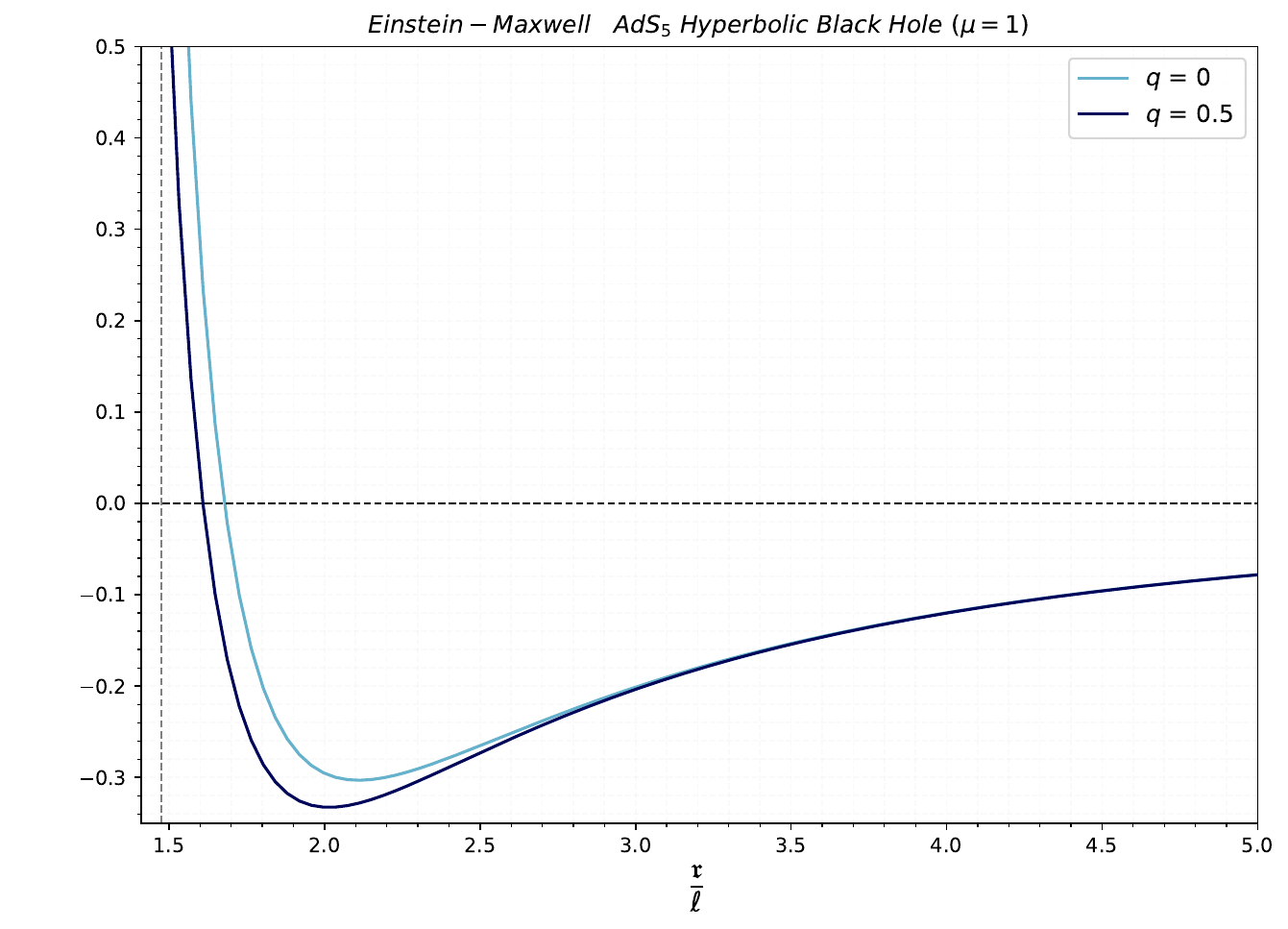}
        \caption{Hyperbolic Horizon}
        \label{fig:BD_BH_planar_beta_d=5b}
    \end{subfigure}
     \caption{Plots of $\lambda^2(\tfrac{\mathfrak{r}}{\ell})$ for~\textit{left}: the ADS$_5$ black brane~\eqref{eq:GR_planarBH} and~\textit{right}: the AdS$_5$ hyperbolic black hole~\eqref{eq:GR_hyperBH}, for $q=0,\,0.5$ and $\mu=1$.~In both plots, the gray vertical dashed line corresponds to the radius of the outer event horizon when $q=0.5$, and their general characteristics do not depend on the spacetime dimension.}
  \label{fig:GR_Lorentzian_dynamcis}
\end{figure}

\paragraph{Planar horizons} We again have the general black brane line element
\begin{equation}
    ds^2=\frac{r^2}{\ell^2}\left(-f(r)dt^2+d\vec{x}^2_{d-1}\right)+\frac{\ell^2}{r^2}\frac{dr^2}{f(r)} \ ,
\end{equation}
and our boundary placed at $r=e^{\bomega(u)}\mathfrak{r}$.~The CBC~\eqref{eq:CBCs} reads\footnote{Note that here we are using a slightly different metric ansatz for the black brane compared to the one in eq.~\eqref{eq:f(r)_gen_dynamics}, to make contact with the known literature, \eg ref.~\cite{Anninos:2024xhc}.}
\begin{equation}\label{eq:L_planar_CBCs}
    \begin{cases}
        ds^2|_{\partial\mathcal{M}}=\dfrac{e^{2\bomega(u)}\mathfrak{r}^2}{\ell^2}\left(-du^2+d\vec{x}^2_{d-1}\right) \ ,\\
        \mathcal{K}|_{\partial\mathcal{M}}=\text{const.} \ .
    \end{cases}
\end{equation}
Requiring $\mathcal{K}$ to be constant yields a differential equation for the Weyl factor $\bomega(u)$,
\begin{equation}\label{eq:GR_Weyl_black_brane dynamics}
\begin{aligned}
   \ell^4\partial_u^2\bomega(u)=-(d-1)(\ell^2\partial_u{\bomega}(u))^{2}
&-
(e^{\bomega(u)}\mathfrak{r})^2\Big(
df\big(e^{\bomega(u)}\mathfrak{r}\big)
+\frac{1}{2}e^{\bomega(u)}\mathfrak{r}f'\big(e^{\bomega(u)}\mathfrak{r}\big)
\Big)\\
&+\mathcal{K}\ell e^{\bomega(u)}\mathfrak{r}
\sqrt{
(e^{\bomega(u)}\mathfrak{r})^{2}f\big(e^{\bomega(u)}\mathfrak{r}\big)
+(\ell^{2}\partial_u{\bomega}(u))^{2}
} \ .
\end{aligned}
\end{equation}
Linearizing that equation around $\bomega(u)\rightarrow0$, we get
\begin{equation}
   \partial_u^2\delta\bomega(u)= \frac{
\mathfrak{r}^{3}\left[
\mathfrak{r}(f'(\mathfrak{r}))^{2}
-2f(\mathfrak{r})\left((d+1)f'(\mathfrak{r})+\mathfrak{r}f''(\mathfrak{r})\right)\right]}{4\ell^{4}f(\mathfrak{r})}\, \delta\bomega(u)  \ ,
\end{equation}
which in turn admits solutions of the form,
\begin{equation}\label{eq:GR_black_brane_lambda2}
   \delta \bomega(u)\propto e^{\pm \lambda u} \ , \quad \ \text{with} \  \lambda^2=-\frac{\mathfrak{r}^3\sqrt{f(\mathfrak{r})}\partial_r \mathcal{K}|_{r=\mathfrak{r}}}{\ell^3} \ ,
\end{equation}
where $\mathcal{K}$ here corresponds to the extrinsic curvature of $\partial\mathcal{M}$ for the conformal representative metric with $\bomega(u)=0$,
\begin{equation}
    \mathcal{K}=\frac{1}{2\ell\sqrt{f(\mathfrak{r})}}\left(\mathfrak{r}f'(\mathfrak{r})+2df(\mathfrak{r})\right) \ .
\end{equation}
Evaluating eq.~\eqref{eq:GR_black_brane_lambda2} for the planar Reissner–Nordström blackening factor (eq.~\eqref{eq:f(r)_GR} with $k=0$),    
\begin{equation}\label{eq:GR_planarBH}
    f(r)=\frac{r^2}{\ell^2}-\frac{\mu}{r^{d-2}}+\frac{q^2}{r^{2(d-2)}} \ ,
\end{equation}
one arrives at the left plot of figure~\ref{fig:GR_Lorentzian_dynamcis}.~In the uncharged and small charged sectors, only unstable modes exist.~In contrast, when the charge $q$ is of the same order of magnitude as the mass $\mu$ then the boundary modes are stable, except when $\partial\mathcal{M}$ is placed near the (outer) horizon.

\paragraph{Hyperbolic Horizons} For the line element with hyperbolic horizon
\begin{equation}
    ds^2=-f(r)dt^2+\frac{dr^2}{f(r)}+r^2dH_{d-1}^2
\end{equation}
the conformal boundary conditions along $\partial\mathcal{M}$ are simply,
\begin{equation}
       \begin{cases}
        ds^2|_{\partial\mathcal{M}}=e^{2\bomega(u)}\left(-du^2+\mathfrak{r}^2dH_{d-1}^2\right) \ ,\\
        \mathcal{K}|_{\partial\mathcal{M}}=\text{const.} \ .
    \end{cases}
\end{equation}
The dynamics of the conformal factor are governed by the following differential equation,
\begin{equation}\label{eq:GR_Weyl_brane dynamics}
\small
  \mathfrak{r}^2 \partial_u^2\bomega(u)= -(d-1)\left(
f(e^{\bomega(u)}\mathfrak{r})+(\mathfrak{r}\partial_u\bomega(u))^2\right)
-\frac{1}{2} e^{\bomega(u)}\mathfrak{r} f'(e^{\bomega(u)}\mathfrak{r}) +e^{\bomega(u)}\mathfrak{r}\mathcal{K}\sqrt{
f(e^{\bomega(u)}\mathfrak{r})
+  (\mathfrak{r}\partial_u\bomega(u))^2} \ ,
\end{equation}
which upon linearization reduces to\footnote{Note that this equation is identical to the one for spherical horizons~\cite{Liu:2024ymn,Galante:2025tnt} and therefore the only information about the (spherical or hyperbolic) nature of the horizon comes through $f(r)$.}
\begin{equation}
   \mathfrak{r}^2\partial^2_u\delta\bomega(u)= \frac{4(d-1) f^{2}(\mathfrak{r})
+  \left(\mathfrak{r}f'(\mathfrak{r})\right)^{2}
- 2 \mathfrak{r} f(\mathfrak{r}) \left(
(d-1)f'(\mathfrak{r})
+ \mathfrak{r} f''(\mathfrak{r})
\right)}{4 f(\mathfrak{r})}\,\delta\bomega(u) \ , 
\end{equation}
and thus admits solutions of the form
\begin{equation}
   \delta \bomega(u)\propto e^{\pm \lambda u} \ , \quad \ \text{with} \  \lambda^2=-\sqrt{f(\mathfrak{r})}\partial_r\mathcal{K}|_{r=\mathfrak{r}}
\end{equation}
for $\mathcal{K}$ as in eq.~\eqref{eq:invert_devil_b,K} when $\alpha=0$.~For the hyperbolic Reissner–Nordström blackening factor (eq.~\eqref{eq:f(r)_GR} with $k=-1$),    
\begin{equation}\label{eq:GR_hyperBH}
    f(r)=-1+\frac{r^2}{\ell^2}-\frac{\mu}{r^{d-2}}+\frac{q^2}{r^{2(d-2)}} \ ,
\end{equation}
the quantity $\lambda^2(\mathfrak{r})$ is given in the right plot of figure~\ref{fig:GR_Lorentzian_dynamcis}.~This is the first case where the modes are stable in both the uncharged and charged sectors, except the Rindler-like growing modes when $\mathfrak{r}\rightarrow r_h$.

\bibliographystyle{JHEP}
\bibliography{mono}

@article{Anninos:2025zgr,
    author = "Anninos, Dionysios and Galante, Dami{\'a}n A. and Georgescu, Silvia and Maneerat, Chawakorn and Svesko, Andrew",
    title = "{The Stretched Horizon Limit}",
    eprint = "2512.16738",
    archivePrefix = "arXiv",
    primaryClass = "hep-th",
    month = "12",
    year = "2025"
}

@article{Coleman:2021nor,
    author = "Coleman, Evan and Mazenc, Edward A. and Shyam, Vasudev and Silverstein, Eva and Soni, Ronak M. and Torroba, Gonzalo and Yang, Sungyeon",
    title = "{De Sitter microstates from T$ \overline{T} $ + {\ensuremath{\Lambda}}$_{2}$ and the Hawking-Page transition}",
    eprint = "2110.14670",
    archivePrefix = "arXiv",
    primaryClass = "hep-th",
    doi = "10.1007/JHEP07(2022)140",
    journal = "JHEP",
    volume = "07",
    pages = "140",
    year = "2022"
}

@article{Silverstein:2024xnr,
    author = "Silverstein, Eva and Torroba, Gonzalo",
    title = "{Timelike-bounded dS$_{4}$ holography from a solvable sector of the T$^{2}$ deformation}",
    eprint = "2409.08709",
    archivePrefix = "arXiv",
    primaryClass = "hep-th",
    doi = "10.1007/JHEP03(2025)156",
    journal = "JHEP",
    volume = "03",
    pages = "156",
    year = "2025"
}

@article{Taylor:2018xcy,
    author = "Taylor, Marika",
    title = "{$T \bar{T}$ deformations in general dimensions}",
    eprint = "1805.10287",
    archivePrefix = "arXiv",
    primaryClass = "hep-th",
    doi = "10.4310/ATMP.2023.v27.n1.a2",
    journal = "Adv. Theor. Math. Phys.",
    volume = "27",
    number = "1",
    pages = "37--63",
    year = "2023"
}

@article{Figueras:2024bba,
    author = "Figueras, Pau and Held, Aaron and Kov{\'a}cs, {\'A}ron D.",
    title = "{Well-posed initial value formulation of general effective field theories of gravity}",
    eprint = "2407.08775",
    archivePrefix = "arXiv",
    primaryClass = "gr-qc",
    month = "7",
    year = "2024"
}

@article{Kovacs:2020pns,
    author = "Kov{\'a}cs, {\'A}ron D. and Reall, Harvey S.",
    title = "{Well-Posed Formulation of Scalar-Tensor Effective Field Theory}",
    eprint = "2003.04327",
    archivePrefix = "arXiv",
    primaryClass = "gr-qc",
    doi = "10.1103/PhysRevLett.124.221101",
    journal = "Phys. Rev. Lett.",
    volume = "124",
    number = "22",
    pages = "221101",
    year = "2020"
}

@article{Kovacs:2020ywu,
    author = "Kov{\'a}cs, {\'A}ron D. and Reall, Harvey S.",
    title = "{Well-posed formulation of Lovelock and Horndeski theories}",
    eprint = "2003.08398",
    archivePrefix = "arXiv",
    primaryClass = "gr-qc",
    doi = "10.1103/PhysRevD.101.124003",
    journal = "Phys. Rev. D",
    volume = "101",
    number = "12",
    pages = "124003",
    year = "2020"
}

@article{Fournodavlos:2021eye,
    author = "Fournodavlos, Grigorios and Smulevici, Jacques",
    title = "{The Initial Boundary Value Problem in General Relativity: The Umbilic Case}",
    eprint = "2104.08851",
    archivePrefix = "arXiv",
    primaryClass = "gr-qc",
    doi = "10.1093/imrn/rnab359",
    journal = "Int. Math. Res. Not.",
    volume = "2023",
    number = "5",
    pages = "3790--3807",
    year = "2023"
}

@article{Gibbons:1976ue,
    author = "Gibbons, G. W. and Hawking, S. W.",
    title = "{Action Integrals and Partition Functions in Quantum Gravity}",
    reportNumber = "PRINT-76-0995 (CAMBRIDGE)",
    doi = "10.1103/PhysRevD.15.2752",
    journal = "Phys. Rev. D",
    volume = "15",
    pages = "2752--2756",
    year = "1977"
}

@article{McGough:2016lol,
    author = "McGough, Lauren and Mezei, M{\'a}rk and Verlinde, Herman",
    title = "{Moving the CFT into the bulk with $ T\overline{T} $}",
    eprint = "1611.03470",
    archivePrefix = "arXiv",
    primaryClass = "hep-th",
    doi = "10.1007/JHEP04(2018)010",
    journal = "JHEP",
    volume = "04",
    pages = "010",
    year = "2018"
}

@article{Zamolodchikov:2004ce,
    author = "Zamolodchikov, Alexander B.",
    title = "{Expectation value of composite field T anti-T in two-dimensional quantum field theory}",
    eprint = "hep-th/0401146",
    archivePrefix = "arXiv",
    reportNumber = "BONN-TH-2004-02",
    month = "1",
    year = "2004"
}

@article{Smirnov:2016lqw,
    author = "Smirnov, F. A. and Zamolodchikov, A. B.",
    title = "{On space of integrable quantum field theories}",
    eprint = "1608.05499",
    archivePrefix = "arXiv",
    primaryClass = "hep-th",
    doi = "10.1016/j.nuclphysb.2016.12.014",
    journal = "Nucl. Phys. B",
    volume = "915",
    pages = "363--383",
    year = "2017"
}

@article{Hartman:2018tkw,
    author = "Hartman, Thomas and Kruthoff, Jorrit and Shaghoulian, Edgar and Tajdini, Amirhossein",
    title = "{Holography at finite cutoff with a $T^2$ deformation}",
    eprint = "1807.11401",
    archivePrefix = "arXiv",
    primaryClass = "hep-th",
    doi = "10.1007/JHEP03(2019)004",
    journal = "JHEP",
    volume = "03",
    pages = "004",
    year = "2019"
}

@article{anderson2008boundary,
  title={On boundary value problems for Einstein metrics},
  author={Anderson, Michael T},
  journal={Geometry \& Topology},
  volume={12},
  number={4},
  pages={2009--2045},
  year={2008},
  publisher={Mathematical Sciences Publishers}
}

@article{anderson2010extension,
  title={Extension of symmetries on Einstein manifolds with boundary},
  author={Anderson, Michael T},
  journal={Selecta Mathematica},
  volume={16},
  number={3},
  pages={343--375},
  year={2010},
  publisher={Springer}
}

@article{Avramidi:1997sh,
    author = "Avramidi, Ivan G. and Esposito, Giampiero",
    title = "{Lack of strong ellipticity in Euclidean quantum gravity}",
    eprint = "hep-th/9708163",
    archivePrefix = "arXiv",
    reportNumber = "DSF-97-43",
    doi = "10.1088/0264-9381/15/5/006",
    journal = "Class. Quant. Grav.",
    volume = "15",
    pages = "1141--1152",
    year = "1998"
}

@article{An:2020nfw,
    author = "An, Zhongshan and Anderson, Michael T.",
    title = "{On the initial boundary value problem for the vacuum Einstein equations and geometric uniqueness}",
    eprint = "2005.01623",
    archivePrefix = "arXiv",
    primaryClass = "math.AP",
    month = "5",
    year = "2020"
}

@article{An:2021fcq,
    author = "An, Zhongshan and Anderson, Michael T.",
    title = "{The initial boundary value problem and quasi-local Hamiltonians in General Relativity}",
    eprint = "2103.15673",
    archivePrefix = "arXiv",
    primaryClass = "gr-qc",
    doi = "10.1088/1361-6382/ac0a86",
    month = "3",
    year = "2021"
}

@article{An:2025rlw,
    author = "An, Zhongshan and Anderson, Michael T.",
    title = "{Well-posed geometric boundary data in General Relativity, I: Conformal-mean curvature boundary data}",
    eprint = "2503.12599",
    archivePrefix = "arXiv",
    primaryClass = "math.AP",
    month = "3",
    year = "2025"
}

@article{An:2025gvr,
    author = "An, Zhongshan and Anderson, Michael T.",
    title = "{Well-posed geometric boundary data in General Relativity, II: Dirichlet boundary data}",
    eprint = "2505.07128",
    archivePrefix = "arXiv",
    primaryClass = "math.AP",
    month = "5",
    year = "2025"
}

@article{An:2025cbs,
    author = "An, Zhongshan and Anderson, Michael T.",
    title = "{Well-posed geometric boundary data in General Relativity, III: conformal-volume boundary data}",
    eprint = "2507.15567",
    archivePrefix = "arXiv",
    primaryClass = "gr-qc",
    month = "7",
    year = "2025"
}

@article{Liu:2025xij,
    author = "Liu, Xiaoyi and Reall, Harvey S. and Santos, Jorge E. and Wiseman, Toby",
    title = "{Ill-posedness of the Cauchy problem for linearized gravity in a cavity with conformal boundary conditions}",
    eprint = "2505.20410",
    archivePrefix = "arXiv",
    primaryClass = "gr-qc",
    month = "5",
    year = "2025"
}

@article{York:1972sj,
    author = "York, Jr., James W.",
    title = "{Role of conformal three geometry in the dynamics of gravitation}",
    doi = "10.1103/PhysRevLett.28.1082",
    journal = "Phys. Rev. Lett.",
    volume = "28",
    pages = "1082--1085",
    year = "1972"
}

@article{Anninos:2023epi,
    author = "Anninos, Dionysios and Galante, Dami{\'a}n A. and Maneerat, Chawakorn",
    title = "{Gravitational observatories}",
    eprint = "2310.08648",
    archivePrefix = "arXiv",
    primaryClass = "hep-th",
    doi = "10.1007/JHEP12(2023)024",
    journal = "JHEP",
    volume = "12",
    pages = "024",
    year = "2023"
}

@article{Anninos:2024wpy,
    author = "Anninos, Dionysios and Galante, Dami{\'a}n A. and Maneerat, Chawakorn",
    title = "{Cosmological observatories}",
    eprint = "2402.04305",
    archivePrefix = "arXiv",
    primaryClass = "hep-th",
    doi = "10.1088/1361-6382/ad5824",
    journal = "Class. Quant. Grav.",
    volume = "41",
    number = "16",
    pages = "165009",
    year = "2024"
}

@article{Galante:2025tnt,
    author = "Galante, Dami{\'a}n A. and Maneerat, Chawakorn and Svesko, Andrew",
    title = "{Conformal boundaries near extremal black holes}",
    eprint = "2504.14003",
    archivePrefix = "arXiv",
    primaryClass = "hep-th",
    doi = "10.1088/1361-6382/ae0408",
    journal = "Class. Quant. Grav.",
    volume = "42",
    number = "19",
    pages = "195003",
    year = "2025"
}

@article{Anninos:2024xhc,
    author = "Anninos, Dionysios and Arias, Ra{\'u}l and Galante, Dami{\'a}n A. and Maneerat, Chawakorn",
    title = "{Gravitational observatories in AdS$_{4}$}",
    eprint = "2412.16305",
    archivePrefix = "arXiv",
    primaryClass = "hep-th",
    doi = "10.1007/JHEP07(2025)234",
    journal = "JHEP",
    volume = "07",
    pages = "234",
    year = "2025"
}

@article{Witten:2018lgb,
    author = "Witten, Edward",
    title = "{A note on boundary conditions in Euclidean gravity}",
    eprint = "1805.11559",
    archivePrefix = "arXiv",
    primaryClass = "hep-th",
    doi = "10.1142/S0129055X21400043",
    journal = "Rev. Math. Phys.",
    volume = "33",
    number = "10",
    pages = "2140004",
    year = "2021"
}

@article{Bunch1981,
  author  = "T. S. Bunch",
  title   = "Surface terms in higher derivative gravity",
  journal = "Journal of Physics A: Mathematical and General",
  volume  = "14",
  number  = "5",
  pages   = "L139--L143",
  year    = "1981",
  doi     = "10.1088/0305-4470/14/5/008",
  publisher = "IOP Publishing"
}

@article{Myers:1987yn,
    author = "Myers, Robert C.",
    title = "{Higher Derivative Gravity, Surface Terms and String Theory}",
    reportNumber = "NSF-ITP-87-16",
    doi = "10.1103/PhysRevD.36.392",
    journal = "Phys. Rev. D",
    volume = "36",
    pages = "392",
    year = "1987"
}

@article{Davis:2002gn,
    author = "Davis, Stephen C.",
    title = "{Generalized Israel junction conditions for a Gauss-Bonnet brane world}",
    eprint = "hep-th/0208205",
    archivePrefix = "arXiv",
    doi = "10.1103/PhysRevD.67.024030",
    journal = "Phys. Rev. D",
    volume = "67",
    pages = "024030",
    year = "2003"
}

@article{Deruelle:2017xel,
    author = "Deruelle, Nathalie and Merino, Nelson and Olea, Rodrigo",
    title = "{Einstein-Gauss-Bonnet theory of gravity: The Gauss-Bonnet-Katz boundary term}",
    eprint = "1709.06478",
    archivePrefix = "arXiv",
    primaryClass = "gr-qc",
    doi = "10.1103/PhysRevD.97.104009",
    journal = "Phys. Rev. D",
    volume = "97",
    number = "10",
    pages = "104009",
    year = "2018"
}

@article{Chakraborty:2017zep,
    author = "Chakraborty, Sumanta and Parattu, Krishnamohan and Padmanabhan, T.",
    title = "{A Novel Derivation of the Boundary Term for the Action in Lanczos-Lovelock Gravity}",
    eprint = "1703.00624",
    archivePrefix = "arXiv",
    primaryClass = "gr-qc",
    doi = "10.1007/s10714-017-2289-5",
    journal = "Gen. Rel. Grav.",
    volume = "49",
    number = "9",
    pages = "121",
    year = "2017"
}

@article{Cai:2001dz,
    author = "Cai, Rong-Gen",
    title = "{Gauss-Bonnet black holes in AdS spaces}",
    eprint = "hep-th/0109133",
    archivePrefix = "arXiv",
    doi = "10.1103/PhysRevD.65.084014",
    journal = "Phys. Rev. D",
    volume = "65",
    pages = "084014",
    year = "2002"
}

@article{Boulware:1985wk,
    author = "Boulware, David G. and Deser, Stanley",
    title = "{String Generated Gravity Models}",
    reportNumber = "DOE-ER-40048-27 P5",
    doi = "10.1103/PhysRevLett.55.2656",
    journal = "Phys. Rev. Lett.",
    volume = "55",
    pages = "2656",
    year = "1985"
}

@article{WILTSHIRE198636,
title = "{Spherically symmetric solutions of Einstein-Maxwell theory with a Gauss-Bonnet term}",
journal = {Physics Letters B},
volume = {169},
number = {1},
pages = {36-40},
year = {1986},
issn = {0370-2693},
doi = {https://doi.org/10.1016/0370-2693(86)90681-7},
url = {https://www.sciencedirect.com/science/article/pii/0370269386906817},
author = {D.L. Wiltshire}}

@article{Padmanabhan:2013xyr,
    author = "Padmanabhan, T. and Kothawala, D.",
    title = "{Lanczos-Lovelock models of gravity}",
    eprint = "1302.2151",
    archivePrefix = "arXiv",
    primaryClass = "gr-qc",
    doi = "10.1016/j.physrep.2013.05.007",
    journal = "Phys. Rept.",
    volume = "531",
    pages = "115--171",
    year = "2013"
}

@article{Deser:2005pc,
    author = "Deser, Stanley and Ryzhov, A. V.",
    title = "{Curvature invariants of static spherically symmetric geometries}",
    eprint = "gr-qc/0505039",
    archivePrefix = "arXiv",
    reportNumber = "BRX-TH-563",
    doi = "10.1088/0264-9381/22/16/012",
    journal = "Class. Quant. Grav.",
    volume = "22",
    pages = "3315--3324",
    year = "2005"
}

@article{Haroon:2020vpr,
    author = "Haroon, Sumarna and Hennigar, Robie A. and Mann, Robert B. and Simovic, Fil",
    title = "{Thermodynamics of Gauss-Bonnet-de Sitter Black Holes}",
    eprint = "2002.01567",
    archivePrefix = "arXiv",
    primaryClass = "gr-qc",
    doi = "10.1103/PhysRevD.101.084051",
    journal = "Phys. Rev. D",
    volume = "101",
    pages = "084051",
    year = "2020"
}

@article{York:1986it,
    author = "York, Jr., James W.",
    title = "{Black hole thermodynamics and the Euclidean Einstein action}",
    doi = "10.1103/PhysRevD.33.2092",
    journal = "Phys. Rev. D",
    volume = "33",
    pages = "2092--2099",
    year = "1986"
}

@article{Liu:2024ymn,
    author = "Liu, Xiaoyi and Santos, Jorge E. and Wiseman, Toby",
    title = "{New Well-Posed boundary conditions for semi-classical Euclidean gravity}",
    eprint = "2402.04308",
    archivePrefix = "arXiv",
    primaryClass = "hep-th",
    doi = "10.1007/JHEP06(2024)044",
    journal = "JHEP",
    volume = "06",
    pages = "044",
    year = "2024"
}

@article{Wald:1993nt,
    author = "Wald, Robert M.",
    title = "{Black hole entropy is the Noether charge}",
    eprint = "gr-qc/9307038",
    archivePrefix = "arXiv",
    reportNumber = "EFI-93-42",
    doi = "10.1103/PhysRevD.48.R3427",
    journal = "Phys. Rev. D",
    volume = "48",
    number = "8",
    pages = "R3427--R3431",
    year = "1993"
}

@article{PhysRevLett.70.3684,
  title = {Black hole entropy and higher curvature interactions},
  author = {Jacobson, Ted and Myers, Robert C.},
  journal = {Phys. Rev. Lett.},
  volume = {70},
  issue = {24},
  pages = {3684--3687},
  numpages = {0},
  year = {1993},
  month = {Jun},
  publisher = {American Physical Society},
  doi = {10.1103/PhysRevLett.70.3684},
  url = {https://link.aps.org/doi/10.1103/PhysRevLett.70.3684}
}

@article{Jacobson:1993vj,
    author = "Jacobson, Ted and Kang, Gungwon and Myers, Robert C.",
    title = "{On black hole entropy}",
    eprint = "gr-qc/9312023",
    archivePrefix = "arXiv",
    reportNumber = "MCGILL-93-22, NSF-ITP-93-152, UMDGR-94-75",
    doi = "10.1103/PhysRevD.49.6587",
    journal = "Phys. Rev. D",
    volume = "49",
    pages = "6587--6598",
    year = "1994"
}

@inproceedings{Jacobson:1994qe,
    author = "Jacobson, Ted and Kang, Gungwon and Myers, Robert C.",
    title = "{Black hole entropy in higher curvature gravity}",
    booktitle = "{16th Annual MRST (Montreal-Rochester-Syracuse-Toronto) Meeting on High-energy Physics: What Next? Exploring the Future of High-energy Physics}",
    eprint = "gr-qc/9502009",
    archivePrefix = "arXiv",
    reportNumber = "MCGILL-95-04, UMDGR-95-092, MCGILL-95--04, UMDGR--95-092",
    month = "5",
    year = "1994"
}

@article{Dong:2013qoa,
    author = "Dong, Xi",
    title = "{Holographic Entanglement Entropy for General Higher Derivative Gravity}",
    eprint = "1310.5713",
    archivePrefix = "arXiv",
    primaryClass = "hep-th",
    reportNumber = "SU-ITP-13-21",
    doi = "10.1007/JHEP01(2014)044",
    journal = "JHEP",
    volume = "01",
    pages = "044",
    year = "2014"
}

@article{Myers:2010ru,
    author = "Myers, Robert C. and Robinson, Brandon",
    title = "{Black Holes in Quasi-topological Gravity}",
    eprint = "1003.5357",
    archivePrefix = "arXiv",
    primaryClass = "gr-qc",
    doi = "10.1007/JHEP08(2010)067",
    journal = "JHEP",
    volume = "08",
    pages = "067",
    year = "2010"
}

@article{Banihashemi:2024yye,
    author = "Banihashemi, Batoul and Shaghoulian, Edgar and Shashi, Sanjit",
    title = "{Flat space gravity at finite cutoff}",
    eprint = "2409.07643",
    archivePrefix = "arXiv",
    primaryClass = "hep-th",
    doi = "10.1088/1361-6382/ada2d7",
    journal = "Class. Quant. Grav.",
    volume = "42",
    number = "3",
    pages = "035010",
    year = "2025"
}

@article{Banihashemi:2025qqi,
    author = "Banihashemi, Batoul and Shaghoulian, Edgar and Shashi, Sanjit",
    title = "{Thermal effective actions from conformal boundary conditions in gravity}",
    eprint = "2503.17471",
    archivePrefix = "arXiv",
    primaryClass = "hep-th",
    doi = "10.1088/1361-6382/adee72",
    journal = "Class. Quant. Grav.",
    volume = "42",
    number = "15",
    pages = "155004",
    year = "2025"
}

@article{Allameh:2024qqp,
    author = "Allameh, Kuroush and Shaghoulian, Edgar",
    title = "{Modular invariance and thermal effective field theory in CFT}",
    eprint = "2402.13337",
    archivePrefix = "arXiv",
    primaryClass = "hep-th",
    doi = "10.1007/JHEP01(2025)200",
    journal = "JHEP",
    volume = "01",
    pages = "200",
    year = "2025"
}

@article{Camanho:2009vw,
    author = "Camanho, Xian O. and Edelstein, Jose D.",
    title = "{Causality constraints in AdS/CFT from conformal collider physics and Gauss-Bonnet gravity}",
    eprint = "0911.3160",
    archivePrefix = "arXiv",
    primaryClass = "hep-th",
    doi = "10.1007/JHEP04(2010)007",
    journal = "JHEP",
    volume = "04",
    pages = "007",
    year = "2010"
}

@article{Buchel:2009sk,
    author = "Buchel, Alex and Escobedo, Jorge and Myers, Robert C. and Paulos, Miguel F. and Sinha, Aninda and Smolkin, Michael",
    title = "{Holographic GB gravity in arbitrary dimensions}",
    eprint = "0911.4257",
    archivePrefix = "arXiv",
    primaryClass = "hep-th",
    reportNumber = "UWO-TH-09-16",
    doi = "10.1007/JHEP03(2010)111",
    journal = "JHEP",
    volume = "03",
    pages = "111",
    year = "2010"
}

@article{Myers:2010tj,
    author = "Myers, Robert C. and Sinha, Aninda",
    title = "{Holographic c-theorems in arbitrary dimensions}",
    eprint = "1011.5819",
    archivePrefix = "arXiv",
    primaryClass = "hep-th",
    doi = "10.1007/JHEP01(2011)125",
    journal = "JHEP",
    volume = "01",
    pages = "125",
    year = "2011"
}

@article{Benjamin:2023qsc,
    author = "Benjamin, Nathan and Lee, Jaeha and Ooguri, Hirosi and Simmons-Duffin, David",
    title = "{Universal asymptotics for high energy CFT data}",
    eprint = "2306.08031",
    archivePrefix = "arXiv",
    primaryClass = "hep-th",
    reportNumber = "CALT-TH 2023-014, IPMU 23-0020",
    doi = "10.1007/JHEP03(2024)115",
    journal = "JHEP",
    volume = "03",
    pages = "115",
    year = "2024"
}

@article{Carlip:1993sa,
    author = "Carlip, Steven and Teitelboim, Claudio",
    title = "{The Off-shell black hole}",
    eprint = "gr-qc/9312002",
    archivePrefix = "arXiv",
    reportNumber = "IASSNS-HEP-93-84, UCD-93-34",
    doi = "10.1088/0264-9381/12/7/011",
    journal = "Class. Quant. Grav.",
    volume = "12",
    pages = "1699--1704",
    year = "1995"
}

@article{Banados:1993qp,
    author = "Banados, Maximo and Teitelboim, Claudio and Zanelli, Jorge",
    title = "{Black hole entropy and the dimensional continuation of the Gauss-Bonnet theorem}",
    eprint = "gr-qc/9309026",
    archivePrefix = "arXiv",
    reportNumber = "IASSNS-HEP-93-53",
    doi = "10.1103/PhysRevLett.72.957",
    journal = "Phys. Rev. Lett.",
    volume = "72",
    pages = "957--960",
    year = "1994"
}

@article{Iyer:1995kg,
    author = "Iyer, Vivek and Wald, Robert M.",
    title = "{A Comparison of Noether charge and Euclidean methods for computing the entropy of stationary black holes}",
    eprint = "gr-qc/9503052",
    archivePrefix = "arXiv",
    doi = "10.1103/PhysRevD.52.4430",
    journal = "Phys. Rev. D",
    volume = "52",
    pages = "4430--4439",
    year = "1995"
}

@article{Fursaev:1995ef,
    author = "Fursaev, Dmitri V. and Solodukhin, Sergey N.",
    title = "{On the description of the Riemannian geometry in the presence of conical defects}",
    eprint = "hep-th/9501127",
    archivePrefix = "arXiv",
    reportNumber = "JINR-E2-95-28, JINR, E2-95-28",
    doi = "10.1103/PhysRevD.52.2133",
    journal = "Phys. Rev. D",
    volume = "52",
    pages = "2133--2143",
    year = "1995"
}

@article{Choquet-Bruhat:1988jdt,
    author = "Choquet-Bruhat, Y.",
    title = "{The Cauchy Problem for Stringy Gravity}",
    doi = "10.1063/1.527841",
    journal = "J. Math. Phys.",
    volume = "29",
    pages = "1891--1895",
    year = "1988"
}

@inproceedings{choquet1989gravitation,
  title="{Gravitation with Gauss Bonnet terms}",
  author={Choquet-Bruhat, Yvonne},
  booktitle={Conference on Mathematical Relativity},
  volume={19},
  pages={53--73},
  year={1989},
  organization={Australian National University, Mathematical Sciences Institute}
}

@article{Freidel:2008sh,
    author = "Freidel, Laurent",
    title = "{Reconstructing AdS/CFT}",
    eprint = "0804.0632",
    archivePrefix = "arXiv",
    primaryClass = "hep-th",
    month = "4",
    year = "2008"
}

@article{Hayward:1993my,
    author = "Hayward, G.",
    title = "{Gravitational action for space-times with nonsmooth boundaries}",
    doi = "10.1103/PhysRevD.47.3275",
    journal = "Phys. Rev. D",
    volume = "47",
    pages = "3275--3280",
    year = "1993"
}

@article{Odak:2021axr,
    author = "Odak, Gloria and Speziale, Simone",
    title = "{Brown-York charges with mixed boundary conditions}",
    eprint = "2109.02883",
    archivePrefix = "arXiv",
    primaryClass = "hep-th",
    doi = "10.1007/JHEP11(2021)224",
    journal = "JHEP",
    volume = "11",
    pages = "224",
    year = "2021"
}

@article{Wiltshire:1988uq,
    author = "Wiltshire, David L.",
    title = "{Black Holes in String Generated Gravity Models}",
    reportNumber = "IC/88/56",
    doi = "10.1103/PhysRevD.38.2445",
    journal = "Phys. Rev. D",
    volume = "38",
    pages = "2445",
    year = "1988"
}

@article{Allameh:2025gsa,
    author = "Allameh, Kuroush and Shaghoulian, Edgar",
    title = "{Timelike Liouville theory and AdS$_3$ gravity at finite cutoff}",
    eprint = "2508.03236",
    archivePrefix = "arXiv",
    primaryClass = "hep-th",
    month = "8",
    year = "2025"
}

@incollection{FeffermanGraham1985,
  author    = {Fefferman, Charles and Graham, C. Robin},
  title     = {Conformal Invariants},
  booktitle = {\'Elie Cartan et les Math{\'e}matiques d'Aujourd'hui},
  series    = {Ast{\'e}risque},
  volume    = {hors s{\'e}rie},
  pages     = {95--116},
  year      = {1985}
}

@article{albin2020poincare,
  title={Poincar{\'e}-Lovelock metrics on conformally compact manifolds},
  author={Albin, Pierre},
  journal={Advances in Mathematics},
  volume={367},
  pages={107108},
  year={2020},
  publisher={Elsevier}
}

@article{Chamblin:1999tk,
    author = "Chamblin, Andrew and Emparan, Roberto and Johnson, Clifford V. and Myers, Robert C.",
    title = "{Charged AdS black holes and catastrophic holography}",
    eprint = "hep-th/9902170",
    archivePrefix = "arXiv",
    reportNumber = "DAMTP-1999-29, EHU-FT-9902, UK-99-02, MCGILL-99-07",
    doi = "10.1103/PhysRevD.60.064018",
    journal = "Phys. Rev. D",
    volume = "60",
    pages = "064018",
    year = "1999"
}

@article{Banerjee:2012iz,
    author = "Banerjee, Nabamita and Bhattacharya, Jyotirmoy and Bhattacharyya, Sayantani and Jain, Sachin and Minwalla, Shiraz and Sharma, Tarun",
    title = "{Constraints on Fluid Dynamics from Equilibrium Partition Functions}",
    eprint = "1203.3544",
    archivePrefix = "arXiv",
    primaryClass = "hep-th",
    reportNumber = "TFR-TH-12-05, IPMU12-0037",
    doi = "10.1007/JHEP09(2012)046",
    journal = "JHEP",
    volume = "09",
    pages = "046",
    year = "2012"
}

@article{Camanho:2014apa,
    author = "Camanho, Xian O. and Edelstein, Jose D. and Maldacena, Juan and Zhiboedov, Alexander",
    title = "{Causality Constraints on Corrections to the Graviton Three-Point Coupling}",
    eprint = "1407.5597",
    archivePrefix = "arXiv",
    primaryClass = "hep-th",
    doi = "10.1007/JHEP02(2016)020",
    journal = "JHEP",
    volume = "02",
    pages = "020",
    year = "2016"
}

@article{Anninos:2022ujl,
    author = {Anninos, Dionysios and Galante, Dami{\'a}n A. and M{\"u}hlmann, Beatrix},
    title = "{Finite features of quantum de Sitter space}",
    eprint = "2206.14146",
    archivePrefix = "arXiv",
    primaryClass = "hep-th",
    doi = "10.1088/1361-6382/acaba5",
    journal = "Class. Quant. Grav.",
    volume = "40",
    number = "2",
    pages = "025009",
    year = "2023"
}

@article{Camps:2013zua,
    author = "Camps, Joan",
    title = "{Generalized entropy and higher derivative Gravity}",
    eprint = "1310.6659",
    archivePrefix = "arXiv",
    primaryClass = "hep-th",
    doi = "10.1007/JHEP03(2014)070",
    journal = "JHEP",
    volume = "03",
    pages = "070",
    year = "2014"
}

@article{Maldacena:1997re,
    author = "Maldacena, Juan Martin",
    title = "{The Large $N$ limit of superconformal field theories and supergravity}",
    eprint = "hep-th/9711200",
    archivePrefix = "arXiv",
    reportNumber = "HUTP-97-A097, HUTP-98-A097",
    doi = "10.4310/ATMP.1998.v2.n2.a1",
    journal = "Adv. Theor. Math. Phys.",
    volume = "2",
    pages = "231--252",
    year = "1998"
}

@article{Gubser:1998bc,
    author = "Gubser, S. S. and Klebanov, Igor R. and Polyakov, Alexander M.",
    title = "{Gauge theory correlators from noncritical string theory}",
    eprint = "hep-th/9802109",
    archivePrefix = "arXiv",
    reportNumber = "PUPT-1767",
    doi = "10.1016/S0370-2693(98)00377-3",
    journal = "Phys. Lett. B",
    volume = "428",
    pages = "105--114",
    year = "1998"
}

@article{Witten:1998qj,
    author = "Witten, Edward",
    title = "{Anti de Sitter space and holography}",
    eprint = "hep-th/9802150",
    archivePrefix = "arXiv",
    reportNumber = "IASSNS-HEP-98-15",
    doi = "10.4310/ATMP.1998.v2.n2.a2",
    journal = "Adv. Theor. Math. Phys.",
    volume = "2",
    pages = "253--291",
    year = "1998"
}

@article{Witten:1998zw,
    author = "Witten, Edward",
    editor = "Bergstrom, L. and Lindstrom, U.",
    title = "{Anti-de Sitter space, thermal phase transition, and confinement in gauge theories}",
    eprint = "hep-th/9803131",
    archivePrefix = "arXiv",
    reportNumber = "IASSNS-HEP-98-21",
    doi = "10.4310/ATMP.1998.v2.n3.a3",
    journal = "Adv. Theor. Math. Phys.",
    volume = "2",
    pages = "505--532",
    year = "1998"
}

@article{Myers:2012ed,
    author = "Myers, Robert C. and Singh, Ajay",
    title = "{Comments on Holographic Entanglement Entropy and RG Flows}",
    eprint = "1202.2068",
    archivePrefix = "arXiv",
    primaryClass = "hep-th",
    doi = "10.1007/JHEP04(2012)122",
    journal = "JHEP",
    volume = "04",
    pages = "122",
    year = "2012"
}

@article{Freedman:1999gp,
    author = "Freedman, D. Z. and Gubser, S. S. and Pilch, K. and Warner, N. P.",
    title = "{Renormalization group flows from holography supersymmetry and a c theorem}",
    eprint = "hep-th/9904017",
    archivePrefix = "arXiv",
    reportNumber = "CERN-TH-99-86, HUTP-99-A015, USC-99-1, MIT-CTP-2846",
    doi = "10.4310/ATMP.1999.v3.n2.a7",
    journal = "Adv. Theor. Math. Phys.",
    volume = "3",
    pages = "363--417",
    year = "1999"
}

@article{Nahm:1977tg,
    author = "Nahm, W.",
    title = "{Supersymmetries and Their Representations}",
    reportNumber = "CERN-TH-2341",
    doi = "10.1201/9781482268737-2",
    journal = "Nucl. Phys. B",
    volume = "135",
    pages = "149",
    year = "1978"
}

@article{Minwalla:1997ka,
    author = "Minwalla, Shiraz",
    title = "{Restrictions imposed by superconformal invariance on quantum field theories}",
    eprint = "hep-th/9712074",
    archivePrefix = "arXiv",
    reportNumber = "PUPT-1748",
    doi = "10.4310/ATMP.1998.v2.n4.a4",
    journal = "Adv. Theor. Math. Phys.",
    volume = "2",
    pages = "783--851",
    year = "1998"
}

@article{Cordova:2016emh,
    author = "Cordova, Clay and Dumitrescu, Thomas T. and Intriligator, Kenneth",
    title = "{Multiplets of Superconformal Symmetry in Diverse Dimensions}",
    eprint = "1612.00809",
    archivePrefix = "arXiv",
    primaryClass = "hep-th",
    doi = "10.1007/JHEP03(2019)163",
    journal = "JHEP",
    volume = "03",
    pages = "163",
    year = "2019"
}

@article{Kodama:2000fa,
    author = "Kodama, Hideo and Ishibashi, Akihiro and Seto, Osamu",
    title = "{Brane world cosmology: Gauge invariant formalism for perturbation}",
    eprint = "hep-th/0004160",
    archivePrefix = "arXiv",
    doi = "10.1103/PhysRevD.62.064022",
    journal = "Phys. Rev. D",
    volume = "62",
    pages = "064022",
    year = "2000"
}

@article{Kodama:2003jz,
    author = "Kodama, Hideo and Ishibashi, Akihiro",
    title = "{A Master equation for gravitational perturbations of maximally symmetric black holes in higher dimensions}",
    eprint = "hep-th/0305147",
    archivePrefix = "arXiv",
    doi = "10.1143/PTP.110.701",
    journal = "Prog. Theor. Phys.",
    volume = "110",
    pages = "701--722",
    year = "2003"
}

@article{Lanczos:1938sf,
    author = "Lanczos, Cornelius",
    title = "{A Remarkable property of the Riemann-Christoffel tensor in four dimensions}",
    doi = "10.2307/1968467",
    journal = "Annals Math.",
    volume = "39",
    pages = "842--850",
    year = "1938"
}

@article{Lovelock:1971yv,
    author = "Lovelock, D.",
    title = "{The Einstein tensor and its generalizations}",
    doi = "10.1063/1.1665613",
    journal = "J. Math. Phys.",
    volume = "12",
    pages = "498--501",
    year = "1971"
}

@article{Ishibashi:2011ws,
    author = "Ishibashi, Akihiro and Kodama, Hideo",
    title = "{Perturbations and Stability of Static Black Holes in Higher Dimensions}",
    eprint = "1103.6148",
    archivePrefix = "arXiv",
    primaryClass = "hep-th",
    doi = "10.1143/PTPS.189.165",
    journal = "Prog. Theor. Phys. Suppl.",
    volume = "189",
    pages = "165--209",
    year = "2011"
}

@article{Silverstein:2022dfj,
    author = "Silverstein, Eva",
    title = "{Black hole to cosmic horizon microstates in string/M theory: timelike boundaries and internal averaging}",
    eprint = "2212.00588",
    archivePrefix = "arXiv",
    primaryClass = "hep-th",
    doi = "10.1007/JHEP05(2023)160",
    journal = "JHEP",
    volume = "05",
    pages = "160",
    year = "2023"
}

@article{Hamilton:2005ju,
    author = "Hamilton, Alex and Kabat, Daniel N. and Lifschytz, Gilad and Lowe, David A.",
    title = "{Local bulk operators in AdS/CFT: A Boundary view of horizons and locality}",
    eprint = "hep-th/0506118",
    archivePrefix = "arXiv",
    reportNumber = "BROWN-HET-1448, CU-TP-1130",
    doi = "10.1103/PhysRevD.73.086003",
    journal = "Phys. Rev. D",
    volume = "73",
    pages = "086003",
    year = "2006"
}

@article{Dong:2016eik,
    author = "Dong, Xi and Harlow, Daniel and Wall, Aron C.",
    title = "{Reconstruction of Bulk Operators within the Entanglement Wedge in Gauge-Gravity Duality}",
    eprint = "1601.05416",
    archivePrefix = "arXiv",
    primaryClass = "hep-th",
    reportNumber = "NSF-KITP-16-005, NSF-KITP-16-005",
    doi = "10.1103/PhysRevLett.117.021601",
    journal = "Phys. Rev. Lett.",
    volume = "117",
    number = "2",
    pages = "021601",
    year = "2016"
}

@article{Bousso:2022hlz,
    author = "Bousso, Raphael and Penington, Geoff",
    title = "{Entanglement wedges for gravitating regions}",
    eprint = "2208.04993",
    archivePrefix = "arXiv",
    primaryClass = "hep-th",
    doi = "10.1103/PhysRevD.107.086002",
    journal = "Phys. Rev. D",
    volume = "107",
    number = "8",
    pages = "086002",
    year = "2023"
}

@article{Bousso:2023sya,
    author = "Bousso, Raphael and Penington, Geoff",
    title = "{Holograms in our world}",
    eprint = "2302.07892",
    archivePrefix = "arXiv",
    primaryClass = "hep-th",
    doi = "10.1103/PhysRevD.108.046007",
    journal = "Phys. Rev. D",
    volume = "108",
    number = "4",
    pages = "046007",
    year = "2023"
}

@article{Simon:1990ic,
    author = "Simon, Jonathan Z.",
    title = "{Higher Derivative Lagrangians, Nonlocality, Problems and Solutions}",
    reportNumber = "UCSB-TH-89-50",
    doi = "10.1103/PhysRevD.41.3720",
    journal = "Phys. Rev. D",
    volume = "41",
    pages = "3720",
    year = "1990"
}

@article{Parker:1993dk,
    author = "Parker, Leonard and Simon, Jonathan Z.",
    title = "{Einstein equation with quantum corrections reduced to second order}",
    eprint = "gr-qc/9211002",
    archivePrefix = "arXiv",
    reportNumber = "WISC-MILW-92-TH-14",
    doi = "10.1103/PhysRevD.47.1339",
    journal = "Phys. Rev. D",
    volume = "47",
    pages = "1339--1355",
    year = "1993"
}

@article{Mueller-Hoissen:1985www,
    author = "Mueller-Hoissen, Folkert",
    title = "{Dimensionally Continued Euler Forms, {Kaluza-Klein} Cosmology and Dimensional Reduction}",
    reportNumber = "MPI-PAE/PTh 62/85",
    doi = "10.1088/0264-9381/3/4/020",
    journal = "Class. Quant. Grav.",
    volume = "3",
    pages = "665",
    year = "1986"
}

@article{Huang:1988mw,
    author = "Huang, Wung-Hong",
    title = "{Kaluza-Klein Reduction of Gauss-Bonnet Curvature}",
    doi = "10.1016/0370-2693(88)91579-1",
    journal = "Phys. Lett. B",
    volume = "203",
    pages = "105--108",
    year = "1988"
}

@book{Wald:1984rg,
    author = "Wald, Robert M.",
    title = "{General Relativity}",
    doi = "10.7208/chicago/9780226870373.001.0001",
    publisher = "Chicago Univ. Pr.",
    address = "Chicago, USA",
    year = "1984"
}

@article{Myers:1987qx,
    author = "Myers, Robert C.",
    title = "{Superstring Gravity and Black Holes}",
    doi = "10.1016/0550-3213(87)90402-0",
    journal = "Nucl. Phys. B",
    volume = "289",
    pages = "701--716",
    year = "1987"
}

@article{Gross:1986iv,
    author = "Gross, David J. and Witten, Edward",
    title = "{Superstring Modifications of Einstein's Equations}",
    reportNumber = "Print-86-0250 (PRINCETON)",
    doi = "10.1016/0550-3213(86)90429-3",
    journal = "Nucl. Phys. B",
    volume = "277",
    pages = "1",
    year = "1986"
}

@article{Grisaru:1986vi,
    author = "Grisaru, Marcus T. and Zanon, D.",
    title = "{$\sigma$ Model Superstring Corrections to the Einstein-hilbert Action}",
    reportNumber = "HUTP-86/A046, BRX-TH-202",
    doi = "10.1016/0370-2693(86)90765-3",
    journal = "Phys. Lett. B",
    volume = "177",
    pages = "347--351",
    year = "1986"
}

@article{Ostrogradsky:1850fid,
    author = "Ostrogradsky, M.",
    title = "{M{\'e}moires sur les {\'e}quations diff{\'e}rentielles, relatives au probl{\`e}me des isop{\'e}rim{\`e}tres}",
    journal = "Mem. Acad. St. Petersbourg",
    volume = "6",
    number = "4",
    pages = "385--517",
    year = "1850"
}

@article{Anninos:2020hfj,
    author = "Anninos, Dionysios and Denef, Frederik and Law, Y. T. Albert and Sun, Zimo",
    title = "{Quantum de Sitter horizon entropy from quasicanonical bulk, edge, sphere and topological string partition functions}",
    eprint = "2009.12464",
    archivePrefix = "arXiv",
    primaryClass = "hep-th",
    doi = "10.1007/JHEP01(2022)088",
    journal = "JHEP",
    volume = "01",
    pages = "088",
    year = "2022"
}

@article{Sarkar:2010xp,
    author = "Sarkar, Sudipta and Wall, Aron C.",
    title = "{Second Law Violations in Lovelock Gravity for Black Hole Mergers}",
    eprint = "1011.4988",
    archivePrefix = "arXiv",
    primaryClass = "gr-qc",
    doi = "10.1103/PhysRevD.83.124048",
    journal = "Phys. Rev. D",
    volume = "83",
    pages = "124048",
    year = "2011"
}

@article{Liko:2007vi,
    author = "Liko, Tomas",
    title = "{Topological deformation of isolated horizons}",
    eprint = "0705.1518",
    archivePrefix = "arXiv",
    primaryClass = "gr-qc",
    doi = "10.1103/PhysRevD.77.064004",
    journal = "Phys. Rev. D",
    volume = "77",
    pages = "064004",
    year = "2008"
}

\end{document}